\documentclass[12pt,a4paper]{report}
\usepackage[utf8]{inputenc}
\usepackage[T1]{fontenc}
\usepackage{lmodern}
\usepackage[a4paper,margin=1in]{geometry}
\usepackage{setspace}
\usepackage{graphicx}
\usepackage{amsmath,amssymb,amsfonts,amsthm}
\usepackage{bm}
\usepackage{physics}
\usepackage{hyperref}
\usepackage{url}
\usepackage{microtype}
\usepackage{booktabs}
\usepackage{caption}
\usepackage{subcaption}
\usepackage{enumitem}
\usepackage{tocloft}
\usepackage{titlesec}
\usepackage{fancyhdr}
\usepackage{titling}

\hypersetup{
    colorlinks=true,
    linkcolor=magenta,
    citecolor=magenta,
    urlcolor=magenta,
    pdftitle={An Interplay Between Fractional Calculus and Holographic Dark Energy},
    pdfauthor={Ayush Bidlan}
}

\titleformat{\chapter}[display]
  {\normalfont\bfseries\centering}
  {\Large CHAPTER \thechapter}
  {0.5em}
  {\LARGE\MakeUppercase}

\titleformat{\part}[display]
  {\normalfont\bfseries\centering}
  {\Huge}
  {PART \Roman{part}}
  {0.5em}

\begin{document}

\begin{titlepage}
    \centering
    \vspace*{1.5cm}

    {\Huge\bfseries An Interplay Between Fractional Calculus and Holographic Dark Energy \par}

    \vfill

    {\Large Ayush Bidlan\par}
    \vspace{0.8cm}

    {\large Department of Physics\par}
    {\large Sardar Vallabhbhai National Institute of Technology, Surat, India\par}

    \vspace{0.8cm}

    {\large Supervised by\par}
    {\large Dr.\ Vikash Kumar Ojha (SVNIT, Surat)\par}
    {\large Prof.\ Paulo Vargas Moniz (CMA-UBI, Portugal)\par}

    \vspace{1.2cm}

    {\large Submitted in partial fulfillment of the requirements for the degree of\par}
    {\large Master of Science in Physics\par}

    \vspace{0.8cm}

    {\large 2025 - 2026\par}

\end{titlepage}

\pagenumbering{roman}

\chapter*{Abstract}
This dissertation aims to put forth a systematic construction of a fractional-calculus extension of holographic dark energy (HDE). We show that linking late-time cosmic acceleration to non-local or memory effects encoded in a fractional (Riesz) derivative within black hole thermodynamics produces deviations from standard HDE and can address some challenges of the Hubble cutoff. In particular, a Riesz fractional spatial derivative is introduced into the Hamiltonian constraint of a Schwarzschild black hole in quantum geometrodynamics, leading to a Fractional Wheeler--DeWitt equation whose solutions yield fractionally corrected thermodynamic quantities, notably fractional Bekenstein--Hawking entropy governed by the L\'evy index \(\alpha\), with \(1<\alpha\leq2\). Using this entropy with Cohen's inequality, a new dark energy density is constructed, defining the Fractional Holographic Dark Energy (FHDE) framework.

The cosmological implications of FHDE are then investigated. Within the Hubble cutoff, its late-time evolution is analysed through cosmological observables, and the model is reconstructed using effective field descriptions with spin-\(0\) and spin-\(1\) candidates, allowing kinetic and potential terms to be extracted as functions of redshift \(z\) and \(\alpha\). The framework is then extended to BD, DGP, EPN, and Horndeski theories to derive the equation-of-state and deceleration parameters in terms of \(z\) and \(\alpha\). In addition, the fate of the Universe is studied through late-time singularities, namely the big, little, and pseudo-rip, within the Granda--Oliveros FHDE setting. In short, this dissertation proposes FHDE as a theoretically motivated extension of HDE, bridging non-locality in quantum gravity with the late-time dynamics of the Universe, and offering a route toward understanding cosmic acceleration beyond \(\Lambda\)CDM.

\noindent\textbf{Keywords:} Fractional Cosmology, Non-Local Dynamic Dark Energy, Holographic Phenomenology

\newpage
\tableofcontents

\newpage
\listoffigures
\newpage
\listoftables

\chapter*{List of Publications}
\addcontentsline{toc}{chapter}{List of Publications}

\begin{itemize}

    \item \textbf{Publications from the chapters of this thesis}
    \begin{enumerate}

        \item O.~Trivedi, \textbf{A.~Bidlan}, and \textbf{P.~Moniz}.
        ``Fractional holographic dark energy.''
        \textit{Phys.\ Lett.\ B} \textbf{858} (2024), p.~139074.
        doi:~\href{https://doi.org/10.1016/j.physletb.2024.139074}%
        {10.1016/j.physletb.2024.139074}.
        arXiv:\href{https://arxiv.org/abs/2407.16685}{2407.16685}~[gr-qc]; corresponds to Section (\ref{FHDE}) of \textbf{Chapter} (\ref{Chapter3}).

        \item \textbf{A.~Bidlan}, \textbf{P.~Moniz}, and O.~Trivedi.
        ``Reconstructing Fractional Holographic Dark Energy with scalar
        and gauge fields.''
        \textit{Eur.\ Phys.\ J.\ C} \textbf{85}, 520 (2025).
        doi:~\href{https://doi.org/10.1140/epjc/s10052-025-14238-2}%
        {10.1140/epjc/s10052-025-14238-2}.
        arXiv:\href{https://arxiv.org/abs/2502.00292}{2502.00292}~[gr-qc]; corresponds to Section (\ref{ReconFHDE}) of \textbf{Chapter} (\ref{Chapter3}).

        \item \textbf{A.~Bidlan}, \textbf{P.~Moniz}, and O.~Trivedi.
        ``Future Rip Scenarios in Fractional Holographic Dark Energy.''
        arXiv:\href{https://arxiv.org/abs/2601.04414}{2601.04414}~[gr-qc]. Under review at \textit{International Journal of Modern Physics D}; corresponds to Section (\ref{late-timecosmic}) of \textbf{Chapter} (\ref{Chapter4}).

    \end{enumerate}

    \item \textbf{Other works during this thesis period}
    \begin{enumerate}

        \item \textbf{A.~Bidlan}, D.~Dey, and P.~Bambhaniya.
        ``Gravitational Collapse and Singularity Formation in
        Brans--Dicke Gravity.''
        arXiv:\href{https://arxiv.org/abs/2601.07488}{2601.07488}~[gr-qc]. Under review at \textit{The European Journal Physical C}.

        \item B.~Patel, J.~Mistry, \textbf{A.~Bidlan},
        and P.~Bambhaniya.
        ``Causal Structure of Spacetime Singularities and Their
        Observable Signatures.''
        arXiv:\href{https://arxiv.org/abs/2603.21187}{2603.21187}~[gr-qc]. Under review at \textit{Chinese Physics C}.

    \end{enumerate}

\end{itemize}

\newpage

\clearpage
\pagenumbering{arabic}


\chapter{Introduction}

\vspace*{2cm}  

\begin{quote}
\itshape
\raggedright
\noindent
...and I say to myself. ``What A Fractional World!''---J. Tenreiro Machado
\end{quote}

\vspace{1cm}  

Fractional calculus (FC) has received increasing attention from both mathematicians and physicists as a powerful framework for extending the standard definitions of differentiation and integration to non-integer orders \cite{herrmann2011fractional}. At its foundation, FC generalises ordinary differential and integral operators to fractional-order integro-differential operators. More generally, a fractional derivative extends integer-order differentiation to rational, real, or even complex order. Although its systematic development is relatively recent and many aspects remain under active investigation, FC has already found wide application in several branches of physics, particularly in the study of stochastic processes and fractal dynamics \cite{fractal,laskin2000fractals,giona1992fractional}. For a long period, however, FC remained primarily a mathematical subject with limited direct physical application. It played an important role in the early development of Abel's theory of integral equations; see \cite{Abel1} for recent developments. Several major mathematicians, including Liouville, Riemann, Heaviside, and Hilbert, also contributed to or showed interest in FC. Historically, its practical use in real-world problems remained limited, and it was often employed only as an alternative mathematical tool to simplify otherwise \textit{difficult} calculations. A representative example is its use in obtaining simplified treatments of the diffusion and wave equations \cite{widder1976heat,courant2024methods}.

During the last two decades, the scope of FC has expanded substantially across physics, finance, computer science, chemistry, and biology \cite{herrmann2011fractional,FCDNA}. This is in marked contrast with the state of the subject a few decades ago, when genuine applications were relatively scarce. Among more recent developments, FC has proved particularly useful in polymer and materials modelling \cite{FCPolymer}. It also introduces physically relevant corrections in fluid dynamics, with important consequences for the modelling of nanofluids and related engineering systems \cite{FCFD,FCnano}. One of its most prominent applications is in the theory of \textit{fractals}\footnote{Fractals are geometric objects possessing detailed structure at arbitrarily small scales, with a fractal dimension that exceeds their topological dimension. This is elaborated further in Section~(\ref{FQM&BHT}), in the context of Fractional Quantum Mechanics (FQM).} \cite{fractal}. In particular, fractional stochastic processes are characterised by non-independent increments, so that the system retains a form of \textit{memory} of its previous evolution rather than depending only on the immediately preceding step \cite{fractal,beran2017statistics,giona1992fractional}. This memory effect follows directly from the \textit{non-local} character of fractional integro-differential operators. For example, fractional Brownian motion provides a concrete description of such behaviour through fractional modifications of the standard displacement--time relation of ordinary Brownian motion; see \cite{FCbrownian} and references therein. Several definitions to compute derivatives in FC have been introduced, including the Liouville, Riemann, Caputo, and Riesz derivatives \cite{herrmann2011fractional}. There is no single definition that is universally preferred; rather, the appropriate choice depends on the specific physical or mathematical problem under consideration.

A particularly important application of FC appears in quantum mechanics. Let us be more precise by considering the Brownian motion. In ordinary Brownian motion, particle trajectories are self-similar and non-differentiable, with a fractal dimension greater than their topological dimension. This observation motivated Feynman and Hibbs to formulate quantum mechanics as a path integral over Brownian paths, thereby placing fractal structures at the foundation of the quantum formalism \cite{feynman2010quantum}. Building on this perspective, a fractional path-integral formulation was later developed, leading to space-FQM, in which the path integral is taken over L\'evy flight trajectories characterised by the L\'evy index varying in the range $1<\alpha\leq2$. A closer look at the L\'evy distribution theory will be made in Section (\ref{LevyTheory}) of Chapter (\ref{Chapter2}), where the range of $\alpha$ will be discussed. The ordinary Gaussian or Brownian limit is recovered for \(\alpha=2\), i.e., when fractional features are not present. One of the central results of FQM is the space-fractional Schr\"odinger equation (SE), where the second-order spatial derivative in the standard SE is replaced by a fractional derivative, specifically the \textit{quantum Riesz fractional derivative} \cite{cai2019riesz,bayin2016definition}. A systematic treatment of these developments may be found in the foundational works and monograph of Nick Laskin \cite{laskin2000fractional,laskin2002fractional,laskin2000fractional1,laskin2000fractals,laskin2017time,wang2007generalized}. In a complementary direction, Naber introduced a time-fractional SE \cite{naber2004time}, in which the ordinary first-order time derivative is replaced by the Caputo fractional derivative while retaining the standard spatial structure. This then motivated the spacetime-fractional SE, where both temporal and spatial derivatives are promoted to fractional order. For broader historical background on FC, L\'evy paths, and related topics, see \cite{ross2006brief,herrmann2011fractional,samko1993aa,kyprianou2007introduction,mandelbrot1960pareto,levy1925calcul,moniz2020fractional}.

For the sake of completeness, we now introduce the mathematical definitions of the fractional integro-differential operators. We begin with the \textit{Riemann--Liouville} (RL) fractional derivative, constructed from the RL integral, which generalises repeated integration to non-integer order. For a function \(f(x)\) and real $a$, the RL fractional derivative of order $a$ is defined by
\begin{equation}\label{Fderivative1}
D^{a}f(x) = \frac{d^{n}}{dx^{n}} \left[ \frac{1}{\Gamma(n-a)} 
\int_{a}^{x} (x-t)^{n-a-1} f(t) \, dt \right],
\end{equation}
where \(n\) is the smallest integer greater than $a$, and \(\Gamma(n-a)\) denotes the gamma function. A second derivative of central importance to this thesis is the Riesz derivative, which appears naturally in fractional partial differential equations and anomalous diffusion \cite{owolabi2025modeling}. Its most convenient definition is given in Fourier space: for a function \(f(x)\) and real $\alpha$,
\begin{equation}\label{Fderivative2}
    \mathcal{F}\{D^{\alpha} f(x)\}(k) = -|k|^{\alpha} \mathcal{F}\{f(x)\}(k),
\end{equation}
where \(\mathcal{F}\{\dots\}\) denotes the Fourier transform. Both fractional operators, i.e., RL and Riesz, extend the notion of differentiation to non-integer orders, but they serve totally different purposes when it comes to their implementation in theoretical physics. The RL fractional derivative is broader in general scope and applies to functions on the real line, whereas the Riesz derivative is especially well suited to the fractional Laplacian and space-FQM \cite{cai2019riesz,bayin2016definition}. In this sense, the Riesz derivative may be viewed as a symmetric realisation of the fractional Laplacian and is related to the RL derivative through its integral representation.

More specifically, the Riesz derivative is a non-local operator (similarly any other fractional operator), since it incorporates contributions from regions of the domain beyond its immediate neighbourhood. Its fractional order generalises ordinary differentiation and makes it suitable for describing anomalous diffusion and other systems not adequately captured by local differential operators. The notion of non-locality is encoded in its integral representation via a convolution kernel that decays with distance, thereby weighting contributions from the entire domain. The Riesz derivative is also closely related to the fractional Laplacian, another non-local operator widely used in the modelling of L\'evy flights and anomalous diffusion, where particles may perform large spatial jumps \cite{cai2019riesz,giona1992fractional}. In practical terms, this non-local structure makes the operator particularly useful for modelling phenomena governed by long-range interactions. For example, in materials science, stress and strain may depend on the overall configuration of the medium rather than purely on local deformations.

Moving forward in the context of fractional operators, the next operator to be considered is the Caputo fractional derivative of order $a$, with $a$ not necessarily an integer, therefore extending the notion of differentiation beyond the standard integer hierarchy. For instance, one may formally consider a derivative of order \(0.5\). Among the most widely used definitions is the Caputo derivative, which resembles the RL derivative but reverses the order of differentiation and integration. It is defined by
\begin{equation}
    {}^C D^{a} f(x) = \frac{1}{\Gamma(n-a)} \int_a^x\frac{f^{(n)}(t)}{(x-t)^{a-n+1}} \, dt.
\end{equation}
To illustrate the practical use of this definition, consider the fractional differential equation
\begin{equation}\label{frDE}
    D^{a} y(t) + \lambda y(t) = f(t),
\end{equation}
where \(D^{a}\) is the Caputo derivative of order $a$, \(\lambda\) is a constant, and \(f(t)\) is a prescribed source term. Applying the Laplace transform to both sides gives
\begin{equation}
    \mathcal{L}\{D^{a} y(t)\} + \lambda \mathcal{L}\{y(t)\} = \mathcal{L}\{f(t)\}.
\end{equation}
Using the standard Laplace transform property of the Caputo derivative, one obtains
\begin{equation}
    s^{a} Y(s) - s^{a-1} y(0) + \lambda Y(s) = F(s),
\end{equation}
where \(Y(s)\) and \(F(s)\) are the Laplace transforms of \(y(t)\) and \(f(t)\), respectively. Solving for \(Y(s)\) yields
\begin{equation}
    Y(s) = \frac{F(s) + s^{a-1} y(0)}{s^{a} + \lambda}.
\end{equation}
The corresponding solution \(y(t)\) is then obtained by inverse Laplace transformation. When an exact inversion is not available, one may instead employ approximate methods such as finite-difference schemes, Adomian decomposition, or Mittag-Leffler function expansions \cite{herrmann2011fractional,das2011functional,baleanu2012fractional}.

Moreover, if we replace the Caputo derivative with the Riesz derivative in Eq. (\ref{frDE}), the Laplace transform of the Riesz derivative of a function \(f(t)\), for \(0<\alpha<2\), may be written as
\begin{equation}
\mathcal{L}\{D^\alpha f(t)\}(s) = s^\alpha \mathcal{L}\{f(t)\}(s) - \sum_{k=0}^{\lceil \alpha \rceil - 1} s^{\alpha - 1 - k} f^{(k)}(0),
\end{equation}
where \(f^{(k)}(0)\) denotes the \(k\)-th derivative of \(f(t)\) evaluated at \(t=0\).

Within this thesis, particular emphasis is placed on the Riesz derivative and its applications. As discussed earlier, it provides a natural framework for modelling anomalous diffusion, in which the mean-squared displacement grows non-linearly with time rather than following the standard Brownian scaling \cite{schilling2014brownian,owolabi2025modeling}. In quantum mechanics, it captures spatially non-local dynamics that lie beyond the reach of the ordinary SE formulation \cite{bayin2016definition}. In materials science, it is used to describe viscoelastic media with memory, where the stress response depends on the full deformation history rather than only the instantaneous state. In hydrology, it appears in the fractional advection--dispersion equation and yields more refined models of solute transport in heterogeneous aquifers \cite{agrawal2007fractional}. Additional applications arise in signal processing, where it is used in the analysis of fractal signals, and in mathematical finance, where it contributes to models with long-range dependence and non-Gaussian fluctuations.

From the perspective of gravity and cosmology, the Riesz derivative offers a natural mechanism for extending dynamical frameworks to non-integer derivative orders in both classical and quantum settings. In this context, fractional cosmology has emerged as a promising approach to several longstanding open problems, including the Hubble tension, the synchronisation problem, and the cosmological constant problem. More broadly, it provides a useful setting for studying the large-scale evolution of the Universe, black hole (BH) dynamics, and the propagation of gravitational waves (GWs). For representative studies, see \cite{roberts2009fractional,jamil2012fractional,torres2020quantum,gonzalez2023exact,leon2023cosmology,Marroquin:2024ddg,Micolta-Riascos:2023mqo,Landim:2021ial,Landim:2021www,Socorro:2023xmx,Fumeron:2023sqz,shchigolev2011cosmological,Junior:2023fwb,Rasouli:2022bug,canedo2025quantum}.

More specifically, classical cosmological models with fractional derivatives have generally followed one of two strategies: \((i)\) \textit{last-step modification}, in which ordinary derivatives in the field equations are replaced by fractional ones after the equations have been derived from an action, and \((ii)\) \textit{first-step modification}, in which the fractional structure is introduced directly at the level of the action. The latter is conceptually more fundamental, since it modifies the theory before the equations of motion are obtained. As an explicit example, consider fractional scalar-field cosmology in the classical regime. In this setting, we take the Friedmann–Lemaître–Robertson–Walker (FLRW) line element
\begin{equation}
    ds^{2}=-N^{2}(t)dt^{2}+a^{2}(t)\left[\frac{dr^{2}}{1-kr^{2}}+r^{2}d\Omega^{2}\right]
\end{equation}
where \(N(t)\) is the lapse function, \(a(t)\) is the scale factor, and \(k=-1,0,+1\) denotes hyperbolic, flat, and spherical spatial geometry, respectively. In what follows, \(k\) is kept generic. The corresponding action is
\begin{equation}\label{FSCaction}
    \mathcal{S}_{\text{EH}-\Phi}=\int d^{4}x\sqrt{-g}\left[\frac{R}{16\pi G}-\frac{1}{2}g^{\mu\nu}\nabla_{\mu}\Phi\nabla_{\nu}\Phi-V(\Phi)\right],
\end{equation}
where units are chosen such that \(c=\hbar=1\), and \(\Phi\) is a homogeneous scalar field minimally coupled to the Ricci scalar \(R\). Implementing a fractional integral of order \(\mu\), the action takes the form
\small{\begin{equation}
    \mathcal{S}^{(\mu)}_{\text{EH}-\Phi}=\frac{1}{\Gamma(\mu)}\int^{t}_{0}Na^{3}\Bigg\{\frac{3}{8\pi G}\left[\frac{\Ddot{a}}{aN^{2}}+\frac{k}{a^{2}}-\frac{1}{N^{2}}\left(\frac{\Dot{a}}{a}\right)^{2}-\frac{\Dot{a}\Dot{N}}{aN^{3}}\right]+\left[\frac{\Dot{\Phi}^{2}}{2N^{2}}-V(\Phi)\right]\Bigg\}(t-\tau)^{\mu-1}d\tau
\end{equation}}
where all quantities depend on the intrinsic time \(\tau\), and an overdot denotes differentiation with respect to \(\tau\). The corresponding equation of motion is
\begin{equation}
    \frac{\partial L^{\mu}_{\text{fr}}}{\partial q_{i}}-\frac{d}{d\tau}\left(\frac{\partial L^{\mu}_{\text{fr}}}{\partial\Dot{q}_{i}}\right)+\frac{d^{2}}{d\tau^{2}}\left(\frac{\partial L^{\mu}_{\text{fr}}}{\partial\Ddot{q}_{i}}\right)=0,
\end{equation}
and variation of the action with respect to the generalised variables \(q_{i}=\{N(t),a(t),\Phi(t)\}\) yields
\begin{equation}
    H^{2}+\left(1-\mu\right)\frac{H}{t}=\frac{8\pi G}{3}\rho
\end{equation}
\begin{equation}
    2\Dot{H}+3H^{2}+2(1-\mu)\frac{H}{t}+\frac{(1-\mu)(2-\mu)}{t^{2}}=-8\pi G p
\end{equation}
\begin{equation}
    \Ddot{\Phi}+3\left(H+\frac{1-\mu}{3t}\right)\Dot{\Phi}+\frac{dV(\Phi)}{d\Phi}=0.
\end{equation}

For a detailed review of the quantum treatment in this cosmological setting, see \cite{Rasouli_2024}. The purpose of presenting this construction is to illustrate the first-step modification approach, namely the implementation of fractional integrals ``directly'' at the level of the action. This procedure provides a new way to formulate familiar cosmological dynamics, with the entire structure modified by fractionally induced non-locality. It is therefore qualitatively different from an arbitrary parametric extension and instead represents a genuine interface between FC and gravitation and cosmology. Moreover, the introduction of fractional structures often leads to unfamiliar mathematical problems. In particular, Ref. \cite{Jalalzadeh_2021} showed that the fractional quantum treatment of a Schwarzschild BH gives rise to the fractional Wheeler--DeWitt (FWDW) equation. Unlike the ordinary WDW equation, no generic closed-form solution is known for the FWDW equation. As a result, one typically applies the Bohr--Sommerfeld quantisation rule in order to recover the corresponding minisuperspace wavefunction; see \cite{Jalalzadeh_2021} for a full discussion. In this way, fractional methods have been used primarily to derive the FWDW equation as an analogue of the space-fractional SE, formulated in terms of the quantum Riesz fractional derivative. One particularly interesting direction in which this construction becomes relevant is explored in Chapter (\ref{Chapter3}).

\section{Structure of this Thesis}

The subsequent chapters in this thesis are organised as follows:

\begin{itemize}
    \item Chapter (\ref{Chapter2}) introduces the conceptual foundations that connect holography and dark energy (DE) within a unified framework, together with an observational overview of the current cosmological landscape.

    \item Chapter (\ref{Chapter3}) develops the implementation of FC in the holographic dark energy (HDE) framework and presents the principal ideas and constructions underlying the fractional holographic dark energy (FHDE) model. Moreover, the FHDE model is reconstructed from several dynamical field configurations, and their kinetic and potential features are carefully examined.

    \item  Chapter (\ref{Chapter4}) then examines a range of cosmological scenarios within the classical regime. In particular, we investigate the occurrence of late-time cosmological singularities in the FHDE framework and reconstruct key cosmological parameters, including the equation-of-state (EoS) parameter and the deceleration parameter, for several modified gravity theories.

    \item Chapter (\ref{Chapter6}) presents the main conclusions of the thesis and highlights several directions for future research that remain open for further investigation.

    \item Finally, the Appendix (\ref{AppendixFC}) showcases a brief discussion on the implementation of fractional derivatives into the framework of Einstein's gravity. We present some aspects of classical and quantum fractional gravity under multiple theoretical settings.
\end{itemize}

\chapter{Holographic Description of an Evolving Dark Energy}\label{Chapter2}

DE has remained one of the most persistent open problems in theoretical cosmology since its observational discovery at the end of the twentieth century \cite{Riess_1998}. Evidence from Type Ia supernovae (SNe Ia), the Cosmic Microwave Background (CMB), and Baryon Acoustic Oscillations (BAOs)  has established that DE accounts for roughly \(70\%\) of the total energy density of the Universe. The conventional explanation of late-time cosmic acceleration is to supplement Einstein's field equations within general relativity (GR) by a cosmological constant (CC) \(\Lambda\), which acts as a homogeneous component with sufficiently negative pressure to drive accelerated expansion. When combined with pressureless cold dark matter (CDM), this yields the \(\Lambda\)CDM model, which remains the standard framework for describing the large-scale structure and evolution of the Universe \cite{turner1997caselambdacdm,1984Natur.311..517B}. In this sense, \(\Lambda\), i.e., the CC, represents the minimal modification required to account for the observed acceleration of the Universe.

Despite its observational success, the \(\Lambda\)CDM model faces a major theoretical difficulty regarding the magnitude of \(\Lambda\) itself \cite{Perivolaropoulos:2021jda,Condon:2018eqx}. At the classical level, a small value of CC is not problematic. However, quantum field theory (QFT) predicts vacuum fluctuations that contribute an enormous effective vacuum energy density, leading to a severe mismatch with the observed value. This discrepancy lies at the core of the old cosmological constant (OCC) problem \cite{Lombriser:2019jia,Weinberg:1988cp}. The OCC problem has motivated a wide class of alternative DE models formulated to produce late-time acceleration without invoking a bare CC. One natural possibility is provided by dynamical scalar fields evolving in a suitable potential, thereby generalising \(\Lambda\) into a time-dependent EoS $w(t)\neq-1$. In the limit where the field becomes stationary, the potential \(V(\varphi)\) effectively reproduces a CC \cite{COPELAND_2006} with $w(t)\rightarrow-1$.

These evolving DE candidates include not only spin-zero fields, but also higher-spin constructions such as spinor models, vector-field models, and \(p\)-form theories. Further possibilities arise in modified gravity and string-inspired approaches motivated by quantum gravity, as well as in multi-scalar cosmological models compatible with quantum-gravity considerations. Although current data do not yet provide decisive evidence for departures from \(\Lambda\)CDM, recent results from the Dark Energy Spectroscopic Instrument (DESI) Collaboration \cite{desicollaboration2024desi2024vicosmological,lodha2024desi2024constraintsphysicsfocused,Calderon_2024, DES:2026jmi} suggest possible deviations, thereby strengthening the case for a systematic investigation of evolving DE scenarios.

More specifically, data from DESI DR1 and DR2 appear to favour an evolving DE component driving the time-dependent expansion rate of the Universe \cite{Sousa-Neto:2025gpj,DESI:2024mwx,Abdul_Karim_2025,Li:2024qso,Gialamas_2025,chaudhary2025evidenceevolvingdarkenergy}. Indeed, the combined DESI DR1 and DR2 analyses indicate a non-negligible statistical preference for time-dependent DE over \(\Lambda\)CDM. The strongest quoted significance reaches \(4.2\sigma\) when DESI data are combined with Planck and DESY5 supernova observations, although more conservative reanalyses place the robust lower significance near \(3\sigma\), depending on the details of the methodology. Since no current dataset combination exceeds the \(5\sigma\) threshold, \(\Lambda\)CDM remains observationally viable, but the evidence increasingly motivates scrutiny of alternative DE dynamics. These developments open a broad theoretical landscape for dynamical DE model building.

At the phenomenological level, DE is characterised by a sufficiently negative pressure to sustain accelerated expansion. In the language of the energy conditions of GR \cite{Hawking:1973uf}, this requires a violation of the strong energy condition (SEC), conventionally written as \(\rho_{\text{DE}}+3p_{\text{DE}}<0\). If one assumes a barotropic EoS of the form \(p_{\text{DE}}=w_{\text{DE}}\rho_{\text{DE}}\), with constant \(w_{\text{DE}}\), then accelerated expansion requires \(w_{\text{DE}}<-1/3\). Some observational constraints \cite{SDSS:2003eyi}, however, allow the more extreme regime \(w_{\text{DE}}\lesssim -1\), which corresponds to a violation of the null energy condition (NEC), \(\rho_{\text{DE}}+p_{\text{DE}}<0\). In such a case, the remaining standard energy conditions are also violated. DE of this type is referred to as \textit{phantom} DE \cite{Caldwell_2003,Nojiri_2003,Hannestad_2002}. Phantom models are especially important because they allow for several types of \textit{rip singularities} \cite{Caldwell_2003,Nojiri_2003}, whose classification and possible realisation within the FHDE framework constitute a central theme of Chapter (\ref{Chapter4}); see Section (\ref{Rips}).

Before turning to the main ingredients of FHDE, it is useful to recall the developments in black hole thermodynamics (BHT) that motivated the application of holography to the DE problem. Gerard 't Hooft first formulated the holographic principle (HP) \cite{tHooft:1993dmi}\footnote{In the spirit of a modern \textit{Plato's cave}, the HP asserts that all information contained within a spatial volume can be encoded on its boundary \cite{Susskind:1994vu}.}. Shortly thereafter, Leonard Susskind provided a concrete string-theoretic interpretation of this principle \cite{Susskind:1994vu}. In 1997, Maldacena proposed the AdS/CFT correspondence, which remains the most successful and explicit realisation of the HP to date \cite{Maldacena:1997re}. The HP is now widely regarded as a foundational ingredient of any consistent theory of quantum gravity \cite{deHaroOlle:2001szl}, and its influence has extended across many areas of theoretical physics; for recent developments see \cite{Friedrich:2024wps, Friedrich:2026hjb} and references therein. In the present chapter, our interest lies in the relation between the HP and DE. In the next section, this connection will be developed into a concrete theoretical framework, namely \textit{Holographic Dark Energy} (HDE).

\section{HDE Model: An Overview}\label{HDEtheory}

Despite the remarkable observational success of the \(\Lambda\)CDM model, it is affected by two major theoretical difficulties: \((i)\) the cosmic coincidence problem and \((ii)\) the fine-tuning problem. In response to these issues, a wide range of DE models has been proposed, yet the fundamental nature of DE remains unknown. There is increasing reason to suspect that the DE problem is, at its core, a question for quantum gravity itself; see \cite{Andriot:2026lac} for a recent review. It is therefore natural to examine the DE problem through the most fundamental principle presently available in quantum gravity, namely the HP.

This was precisely the perspective adopted by Miao Li in 2004, when the HP was applied to the DE problem to formulate the HDE model \cite{Li:2004rb}\footnote{Recent field-theoretic studies suggest that the low-energy effective field theory (EFT) of HDE may be described by a \textit{massive gravity} theory \cite{deRham:2014zqa}, whose graviton carries three polarizations: one scalar mode and two tensor modes \cite{Lin:2021bxv}.}. In what follows, we discuss the conceptual features that distinguish HDE from other DE models in the literature. A summary of its observational achievements and selected shortcomings is given in Section (\ref{cosmoobs}).

In the HDE framework, the DE density is determined entirely by two physical quantities: the reduced Planck mass,
\[
M_{pl} \equiv \left(8\pi G_{\text{N}}\right)^{-1/2},
\]
where \(G_{\text{N}}\) is Newton's gravitational constant, and the cosmological infrared (IR) length scale \(L_{\text{IR}}\), identified with the future event horizon of the Universe. HDE is the first DE model simultaneously motivated by the HP and broadly consistent with cosmological observations, which makes it a particularly competitive candidate for DE. We now outline its mathematical construction.

Consider a Universe characterised by an IR length scale \(L_{\text{IR}}\). The HP implies that the physical content of such a Universe, including the DE density \(\rho_{\text{DE}}\), is encoded by degrees of freedom defined on its boundary. As noted above, the relevant quantities are \(M_{pl}\) and \(L_{\text{IR}}\). Dimensional analysis then constrains the DE density to take the form
\begin{equation}\label{HDEdensity}
    \rho_{\text{DE}} = C_{1}M^{4}_{pl} + C_{2}M^{2}_{pl}L^{-2}_{\text{IR}} 
    + C_{3}L^{-4}_{\text{IR}} + \dots,
\end{equation}
where \(C_{1}\), \(C_{2}\), and \(C_{3}\) are dimensionless constants. In principle, \(C_{2}\) and \(C_{3}\) may carry time dependence, although such dependence can be absorbed into the definition of \(L_{\text{IR}}\). By contrast, \(C_{1}\) is determined purely by ultraviolet (UV) physics and is generally expected to remain constant as a consequence of time-translation symmetry in the underlying theory. An exception arises in theories with a time-varying gravitational constant, \(G_{\text{N}}(t)\), such as Brans--Dicke (BD) gravity. The first term in Eq. (\ref{HDEdensity}) is strongly disfavoured by naturalness, since it exceeds the observed DE density by approximately \(10^{120}\) \cite{Sola:2013gha}. This discrepancy is the essence of the fine-tuning problem, also known as the OCC problem \cite{Sola:2013gha}.

In \cite{Cohen:1998zx}, Cohen, Kaplan, and Nelson (CKN) showed that the \(C_{1}\) term is also incompatible with the HP. The HP implies that local QFT cannot provide a valid description of a BH, or more generally of any state defined at its Schwarzschild radius. Accordingly, the naive QFT estimate \(\rho_{\text{DE}} \sim C_{1}M^{4}_{pl}\) is no longer admissible. Instead, local QFT must be supplemented by a non-trivial UV cutoff \(\Lambda_{\text{UV}}\). To see this, note that the total energy in a region of size \(L_{\text{IR}}=r_{s}\) is \(L^{3}_{\text{IR}}\Lambda^{4}_{\text{UV}}\). Requiring this energy not to exceed the mass of the corresponding BH leads to
\begin{equation}\label{Cohenineq}
    L^{3}_{\text{IR}}\Lambda^{4}_{\text{UV}} < L_{\text{IR}}M^{2}_{pl}.
\end{equation}

It follows that the vacuum energy density in this UV-regulated QFT satisfies \(\rho_{\text{DE}} \sim \Lambda^{4}_{\text{UV}} \lesssim M^{2}_{pl}L^{-2}_{\text{IR}}\). Consequently, the \(C_{1}\) term is eliminated, and the expansion in Eq. (\ref{HDEdensity}) begins with the second term. The third and higher-order terms are subleading, so the DE density reduces to
\begin{equation}\label{HDE}
    \rho_{\text{DE}} = 3C^{2}M^{2}_{pl}L^{-2}_{\text{IR}},
\end{equation}
where \(C\) is a dimensionless constant. It is important to emphasise that this expression is obtained from the combined input of the HP and dimensional analysis, rather than by inserting an explicit DE fluid term into the Einstein--Hilbert action. This is precisely what makes the HDE conceptually distinct from other dynamical DE models.

Before proceeding further, one must specify the IR cutoff \(L_{\text{IR}}\). The earliest proposal was the Hubble horizon, \(L_{\text{IR}}=H^{-1}\), which introduces the natural length scale set by the inverse Hubble parameter \(H\) and was initially motivated by the possibility of alleviating the fine-tuning problem. However, this choice drives the EoS parameter toward zero in the standard HDE model and fails to explain the observed accelerated expansion. Beyond the Hubble cutoff, several alternative prescriptions have been proposed, each with its own advantages and limitations. Among the most commonly used are
\begin{equation} \label{lp}
    L_{p} = a(t) \int_{0}^{t} \frac{dt}{a(t)},\quad
    L_{f} = a(t) \int_{a(t)}^{\infty} \frac{dt}{a(t)},\quad
    L_{\text{GO}} = \left(\gamma H^2 + \delta \dot{H} \right)^{-\frac{1}{2}},
\end{equation}
where \(L_{p}\), \(L_{f}\), and \(L_{\text{GO}}\) denote the \textit{particle horizon}, \textit{future event horizon}, and \textit{Granda--Oliveros} (GO) cutoff, respectively. The particle horizon typically leads to an EoS parameter greater than \(-1/3\) and therefore cannot account for present-day acceleration. The future event horizon, on the other hand, can generate accelerated expansion, but it introduces potential causality issues. Another proposal is the GO cutoff \cite{Granda:2008dk}, which depends on both \(H\) and \(\dot{H}\). This cutoff avoids the causality concerns associated with future-dependent horizons and can still support late-time acceleration together with phantom crossing of the EoS. Its drawback, however, is that the resulting model often mimics dominant-energy-like behaviour, and the squared sound speed (SSS) may fall outside observationally acceptable ranges depending on the chosen Hubble ansatz, thereby indicating possible instability in some HDE--GO realisations \cite{TrivediScherrer_2024,trivedi2024new}. The most general prescription is the \textit{Nojiri--Odintsov} (NO) cutoff \cite{Nojiri:2005pu,Nojiri:2017opc,nojiri2019holographic}, in which the cutoff is taken to be an arbitrary function of \(H\), its higher derivatives, the particle and future event horizons, and their derivatives:
\begin{equation} \label{nocutoff}
    L_{\text{NO}}= L\!\left(H,\dot{H},\ddot{H},\ldots,L_{p},L_{f},\dot{L}_{p},
    \dot{L}_{f},\ldots\right).
\end{equation}

A further complication common to all such cutoff choices is the classical instability of the corresponding HDE models under cosmological perturbations \cite{Myung:2007pn}. In general, any HDE density expressible as a function of \(H\), the particle horizon, and the future event horizon lies within the class defined by Eq. (\ref{nocutoff}). Indeed, all HDE models discussed in the literature, including Tsallis, Barrow, and the FHDE model introduced in this thesis, may in principle be embedded into this general structure. Although this observation is mathematically straightforward, the construction of a physically motivated and self-consistent form for the cutoff in Eq. (\ref{nocutoff}) remains highly non-trivial, even with complete generality in principle allowed. This difficulty is precisely what motivates the wide variety of HDE proposals, each based on its own physical and mathematical rationale \cite{TrivediScherrer_2024}.

To be more explicit, the standard HDE model is unable to realise dynamical DE within the Hubble cutoff, which motivates the search for extensions beyond the conventional HDE framework. This is the point at which we invoke FC effects at the fundamental level of quantum mechanics. Incorporating FC into quantum mechanics yields FQM, which preserves important quantum features such as memory effects and non-locality. The FHDE programme is motivated by a recent study in which fractional quantisation of the Schwarzschild BH was shown to reproduce the relevant BH observables, in particular the entropy \cite{Jalalzadeh_2021}. We place special emphasis on this construction in Chapter (\ref{Chapter3}), where these ideas are briefly introduced before applying FC as a new tool to the HDE problem with the Hubble cutoff.

To reiterate, our framework adopts the Hubble horizon as the IR cutoff, and this choice is motivated by two main considerations. First, we seek the simplest setting in which FC-inspired HDE, which we shall denote by FHDE, can generate viable cosmological evolution; in this respect, the Hubble horizon is the most natural and minimal option. Second, since the Hubble cutoff is known to be problematic in several existing constructions \cite{Myung:2007pn,Li:2004rb}, demonstrating that FHDE can remain observationally viable even in this setting would represent a meaningful advance in the broader reassessment of its viability.

\subsection{Cosmological Observations}\label{cosmoobs}

The DESI collaboration has released large-scale structure (LSS) data in DR1 and DR2, consisting of BAO and redshift-space distortion measurements from millions of galaxies and quasars up to \(z\lesssim 2.5\) \cite{DESI:2025fxa,Abdul_Karim_2025,DESI:2025wyn}\footnote{DESI has recently completed a three-dimensional map of the Universe containing nearly 47 million galaxies and quasars, which is expected to sharpen future observational constraints on dynamical DE. The next public release, DR3, is expected in 2028.}. These datasets, often combined with CMB measurements, the Pantheon+ supernova sample, and local \(H_0\) probes, have been used to test a broad class of dynamical DE models. Analyses based on the two-parameter Chevallier--Polarski--Linder (CPL) parametrisation,
\[
w(a)=w_0+w_a(1-a),
\]
indicate that DESI BAO data exhibit a preference for a time-evolving DE EoS over the constant CC value \(w_{\text{DE}}=-1\), with a tension of approximately \(2\) to \(3\sigma\) in some combinations with CMB and supernova data. In addition, fits of the \(w_0w_a\)CDM model indicate a mild but non-negligible deviation from the vacuum-energy hypothesis, often favouring a quintessence-like sector with \(w_0>-1\) and \(w_a>0\), or in some cases a recent phantom crossing. Model-independent reconstructions of the DE EoS also suggest a slowly evolving DE component at low redshift, driven mainly by the combination of DESI BAO and nearby supernova data.

When DESI data are combined with additional probes such as big bang nucleosynthesis (BBN), observational Hubble data (OHD), and Planck, the constraints on the \(w_0\) parameter of dynamical DE models become tighter, typically yielding \(w_0 \approx -1.10\) to \(-0.95\) at roughly \(1\) to \(2\sigma\), together with a non-zero \(w_a\) whose sign remains mildly degenerate with priors and with the choice of spatial curvature. Overall, these analyses suggest that a slowly varying DE EoS is statistically favoured over \(\Lambda\)CDM when the full DESI BAO sample is combined with CMB and supernova data, although the exact significance remains sensitive to systematics and to the treatment of low-redshift supernovae. Forthcoming DESI, DES, Euclid, and next-generation CMB observations will provide stronger tests of this possibility.

\subsection{Observational Constraints on the HDE Model}

Attempts to constrain the HDE model using cosmological observations are reviewed in detail in Section~(3.3) of \cite{Wang:2016och}, where the standard HDE model is studied systematically. In this framework, the only defining model parameter is the dimensionless constant \(C\), which enters the DE density relation in Eq. (\ref{HDEdensity}). The physical interpretation of the model is controlled entirely by this parameter: $C>1$ corresponds to quintessence-like behaviour, $C\to1$ approaches the CC limit, and $C<1$ drives the model into the phantom regime, leading to distinct cosmic futures. Determining $C$ from observations is therefore of central importance. To constrain $C$, one studies the evolution equation for the DE density parameter
\[
\Omega_{\rm de}= \rho_{\rm de}/3M_p^2H^2.
\]
In a spatially flat FLRW Universe, differentiating Eq. (\ref{HDEdensity}) and using the Friedmann equations gives
\begin{equation}
    \frac{d\Omega_{\text{DE}}}{d\ln(a)} 
    = \Omega_{\text{DE}}\!\left(1 - \Omega_{\text{DE}}\right)
      \!\left(1 + \frac{2\sqrt{\Omega_{\text{DE}}}}{C}\right),
    \label{eq:HDEevolution}
\end{equation}
which, together with the initial condition \(\Omega_{\text{DE}}(z_i)\approx0\) deep in the matter-dominated era, uniquely determines the expansion history once \(C\) and \(\Omega_{\text{DM}0}\) are specified. The corresponding EoS parameter follows algebraically as
\begin{equation}
    w_{\text{DE}} = -\frac{1}{3} - \frac{2\sqrt{\Omega_{\text{DE}}}}{3C},
    \label{eq:HDEeos}
\end{equation}
so that the sign of \(w_{\text{DE}}+1\) is fixed entirely by whether \(C\) is greater or less than unity. In the flat case, the model is therefore minimally parametrised by the two quantities \(C\) and \(\Omega_{\text{DM}0}\), which together determine the full late-time cosmological evolution.

As explained in \cite{wang2017holographic}, observational constraints are obtained by constructing the joint likelihood
\[
\mathcal{L}\propto e^{-\chi^2/2},
\]
with the total chi-squared defined by
\begin{equation}
    \chi^2_{\rm tot} = \chi^2_{\rm SNIa} + \chi^2_{\rm BAO} + \chi^2_{\rm CMB}.
\end{equation}
The CMB contribution is usually evaluated through the shift parameter
\[
\mathcal{R} \equiv \sqrt{\Omega_{m0}H_0^2}\,r(z_*),
\]
and the acoustic scale
\[
l_A \equiv \pi r(z_*)/r_s(z_*),
\]
which efficiently compress the full Planck likelihood into observables sensitive to the integrated expansion history from the decoupling redshift \(z_*\approx1090\) to the present epoch. Markov Chain Monte Carlo (MCMC) sampling then yields the marginalised posterior constraints on \(C\) and \(\Omega_{\text{DM}0}\).

The results obtained from progressively improved datasets present a coherent phenomenological picture. Early analyses based on WMAP-era CMB data combined with supernova and large-scale structure measurements found \(C\approx0.6\)--\(0.8\) \cite{Zhang:2005hs}, clearly below unity and therefore indicative of a phantom EoS. With the availability of higher-quality data, including the JLA supernova compilation, SDSS DR12 BAO measurements, and Planck 2015 distance priors, the allowed parameter range became tighter, but the preferred value of \(C\) remained below unity whenever CMB data were included. A representative fit from the combined CMB+BAO+SNIa dataset gives
\begin{equation}
    C = 0.495^{+0.039}_{-0.040}, \quad 
    \Omega_{\text{DM}0} = 0.289^{+0.016}_{-0.014}, \quad 
    H_0 = 74.2^{+2.1}_{-2.0}\ \mathrm{km\,s^{-1}\,Mpc^{-1}},
\end{equation}
at the \(68\%\) confidence level \cite{Wang_2017}. This result has a direct physical implication: substituting \(C<1\) into Eq. (\ref{eq:HDEeos}) yields \(w_{\text{DE}}<-1\) at the present epoch, which drives the Universe toward a Big Rip singularity within finite cosmic time.

There is, however, a noteworthy dataset dependence in the inferred value of \(C\). When supernova data are analysed without the CMB shift parameter, the fits often favour \(C\gtrsim1\), corresponding to quintessence-like behaviour and a future DE Sitter attractor. Once CMB information is included, the preferred value is shifted below unity with substantial statistical weight, since the shift parameter \(\mathcal{R}\) is sensitive to the integrated comoving distance to \(z_*\), which is modified by the HDE component relative to \(\Lambda\)CDM. This discrepancy between the supernova-preferred and CMB-preferred values of \(C\) is not merely numerical. Rather, it reflects a persistent internal tension in the HDE parameter space when mainstream cosmological datasets are combined. This issue reappears in more recent analyses that include DESI BAO data; see \cite{mnras}.

\chapter{Fractional \emph{swirl} to Holographic Dark Energy Model}\label{Chapter3}

The aim of this chapter is to formulate the FHDE framework and investigate several reconstructions through a correspondence between FHDE and a selected set of dynamical effective field configurations. To this end, we begin by outlining the central structural elements of the FHDE model, starting with the application of FQM to BHT, which yields a fractionally modified Bekenstein--Hawking entropy. We also briefly review the quantisation of a static, spherically symmetric BH within the framework of quantum geometrodynamics (QGD).

\section{Prelude to Fractional HDE Model}

In this section, we summarise the mathematical background and technical ingredients underlying the FHDE framework. The discussion is necessarily concise and is intended to provide sufficient context for the FC-based developments in quantum mechanics that motivate the FHDE construction. These preliminaries prepare the ground for the detailed formulation of FHDE that follows.

\subsection{Interplay Between FQM \& BHT}\label{FQM&BHT}

Extended versions of the SE arise by enlarging the set of allowed trajectories in the path-integral formulation. Once non-Brownian paths are admitted, one obtains space-fractional, time-fractional, and spacetime-fractional generalisations of the standard SE.

In the space-fractional case, the Gaussian measure of the Feynman path integral is replaced by a L\'evy distribution. The construction starts from the standard Hamiltonian
\begin{equation}
    H = \frac{\textbf{p}^{2}}{2m} + V(\textbf{r}, t),
\end{equation}
whose fractional generalisation is
\begin{equation}
    H_{\alpha}(\textbf{p}, \textbf{r}) := D_{\alpha}|\textbf{p}|^{\alpha} 
    + V(\textbf{r}), \quad \alpha \in (1,2].
\end{equation}
Here, \(D_{\alpha}\) is a coefficient with physical dimension
\[
[D_{\alpha}] = \text{erg}^{1-\alpha}\,\text{cm}^{\alpha}\,\text{sec}^{-\alpha},
\]
and \(\alpha\) is the L\'evy fractional parameter that determines the fractal dimension of the associated L\'evy paths. In the standard Feynman path integral, the measure is generated by Brownian motion, for which the fractal dimension is \(d^{(\text{Feynman})}_{(\text{fractal})}=2\). In the L\'evy generalisation, this becomes
\[
d^{(\text{L\'evy})}_{(\text{fractal})}=\alpha.
\]
Passing to the representation through \(\hat{\textbf{p}}\to-i\hbar\nabla\) and \(\hat{\textbf{r}}\to\textbf{r}\), one obtains the space-fractional SE:
\begin{equation}
    i\hbar\frac{\partial\psi(\textbf{r},t)}{\partial t} = 
    D_{\alpha}\left(-\hbar^{2}\Delta\right)^{\alpha/2}\psi(\textbf{r},t) 
    + V(\textbf{r},t)\psi(\textbf{r},t),
\end{equation}
where the fractional Riesz derivative \(\left(-\hbar^{2}\Delta\right)^{\alpha/2}\) is defined through the Fourier transform \(\mathcal{F}\) as
\begin{equation}
    \left(-\hbar^{2}\Delta\right)^{\alpha/2}\psi(\textbf{r},t) = 
    \mathcal{F}^{-1}|\textbf{p}|^{\alpha}\mathcal{F}\psi(\textbf{r},t) = 
    \frac{1}{(2\pi\hbar)^{3}}\int d^{3}p\; 
    e^{i\textbf{p}\cdot\textbf{r}/\hbar}|\textbf{p}|^{\alpha}
    \int e^{-i\textbf{p}\cdot\textbf{r}'/\hbar}
    \psi(\textbf{r}',t)\,d^{3}r'.
\end{equation}

The infinite square well was one of the earliest systems solved within the space-fractional SE framework, as studied by Laskin \cite{laskin2017time,laskin2000fractional,laskin2000fractals,laskin2000fractional1,laskin2002fractional}. Although simple, this system provides an instructive prototype for a quantum detector with internal degrees of freedom. In the time-fractional formulation, the temporal evolution is governed by the Caputo fractional derivative, and the resulting Hamiltonian is non-Hermitian and time-nonlocal. A unified spacetime-fractional SE was later introduced by Wang and Xu \cite{wang2007generalized}, combining the space- and time-fractional constructions into
\begin{equation}
    \hbar_{\beta}\,i^{\beta}\,\partial^{\beta}_{t}\psi(\textbf{r},t) = 
    \left[D_{\alpha,\beta}\left(-\hbar^{2}\Delta\right)^{\alpha/2} 
    + V(\textbf{r},t)\right]\psi(\textbf{r},t),
\end{equation}
where \(\alpha \in (1,2]\), \(\beta \in (0,1]\), and \(\hbar_{\beta}\), \(D_{\alpha,\beta}\) are scale coefficients with dimensions
\[
[\hbar_{\beta}] = \text{erg}\cdot\text{sec}^{\beta}, \qquad
[D_{\alpha,\beta}] = \text{erg}^{1-\alpha}\cdot\text{cm}^{\alpha}\cdot\text{sec}^{-\alpha\beta}.
\]
The operator \(\partial^{\beta}_{t}\) denotes the left Caputo fractional derivative of order \(\beta\):
\begin{equation}
    \partial^{\beta}_{t}f(t) = \frac{1}{\Gamma(1-\beta)}\int^{t}_{0} d\tau\;
    \frac{\dot{f}(\tau)}{(t-\tau)^{\beta}}, \qquad 
    \dot{f}(\tau) \equiv \frac{df(\tau)}{d\tau}.
\end{equation}

More broadly, Wheeler's proposal of a foam-like spacetime structure at the Planck scale makes the incorporation of fractal geometry into quantum gravity a natural possibility \cite{wheeler1955geons}. In recent years, FQM has therefore been applied actively in quantum gravity and quantum cosmology \cite{moniz2020fractional}. A concrete realisation of this programme is provided by the \textit{Fractional Wheeler--DeWitt} (FWDW) equation, whose study has revealed non-trivial connections between different areas of mathematics and physics within quantum gravity and quantum cosmology. For a recent survey of FQM in the canonical approach to quantum gravity and cosmology, see \cite{Varao:2024eig}; related developments may be found in \cite{moniz2020fractional,Jalalzadeh_2021,jalalzadeh2022sitter}.

We now briefly review the application of space-FQM to the canonical quantisation of the Schwarzschild BH, which leads to a fractionally modified Bekenstein--Hawking horizon entropy \cite{Jalalzadeh_2021}. The standard Bekenstein--Hawking entropy for a Schwarzschild BH of mass \(M\) is
\begin{equation}
    S_{\text{B-H}} = 4\pi G M^{2} = \frac{A}{4G},
\end{equation}
where \(A = 16\pi G^{2}M^{2}\) is the horizon area. A fractional extension of the WDW equation for the Schwarzschild BH yields modified expressions for the entropy and other thermodynamic observables. As a starting point, we recall the non-fractional canonical quantisation of the Schwarzschild BH. In 1994, Kuch\v{a}\v{r} used the ADM formalism to derive a WDW equation for a spherically symmetric spacetime line element \cite{Kuchar:1994zk}. The corresponding reduced action is
\begin{equation}
    \mathcal{S} = \int \left\{p\dot{x} - H\right\}dt,
\end{equation}
with reduced Hamiltonian
\begin{equation}
    H = \frac{M^{2}_{pl}}{2}\left(\frac{p^{2}}{M^{4}_{pl}x} + x\right).
\end{equation}
Applying the canonical quantisation prescription \(\hat{p}\to-i\,d/dx\), \(\hat{x}\to x\), one obtains the time-independent WDW equation for the one-dimensional minisuperspace of the Schwarzschild BH \cite{Kuchar:1994zk}:
\begin{equation}
    -\frac{1}{2M_{pl}}\frac{d^{2}}{dx^{2}}\Psi(x) + 
    \frac{1}{2}M_{pl}\omega^{2}_{pl}
    \left(x - \frac{M}{M_{pl}}\right)^{2}\Psi(x) = 
    \frac{M^{2}}{2M_{pl}}\Psi(x),
\end{equation}
where \(\omega_{pl}=t^{-1}_{pl}\) is the Planck angular frequency and \(t_{pl}=M^{-1}_{pl}\) is the Planck time. Imposing the DeWitt boundary condition \(\Psi(x)\big|_{x=0}=0\), the general solution is obtained as shown in \cite{DeWitt:1967yk}. Enforcing the boundary conditions on \(\Psi(x)\) yields the BH mass spectrum \(M = M_{pl}\sqrt{2n+1}\), where \(n\) is a non-negative integer with \(n\gg1\). From this spectrum, one can derive further thermodynamic observables, including the thermal radiation frequency \(\omega_{0}\), evaporation rate \(\dot{M}\), temperature \(T\), and horizon entropy \(S\).

Fractional generalisations of these observables are obtained by applying FQM to the Schwarzschild BH, thereby extending each quantity to depend explicitly on the L\'evy index \(\alpha\in(1,2]\). In particular, a FWDW equation is constructed in the corresponding minisuperspace. Following \cite{Jalalzadeh_2021}, the FWDW equation is
\begin{equation}
    \frac{M^{1-\alpha}_{pl}}{2}(-\Delta)^{\frac{\alpha}{2}}\Psi(z) + 
    \frac{M^{\alpha-1}_{pl}}{2}\omega^{2}_{pl}z^{\alpha}\Psi(z) = 
    \frac{M^{2}}{2M_{pl}}\Psi(z),
\end{equation}
where \(z = x - M/M^{2}_{pl}\) is the shifted coordinate in the one-dimensional minisuperspace, \(\Delta = d^{2}/dz^{2}\), and \((-\Delta)^{\alpha/2}\) is the Riesz fractional derivative \cite{Jalalzadeh_2021,bayin2016definition,cai2019riesz}. A solution \(\Psi(z)\) is obtained in \cite{Jalalzadeh_2021} through the Bohr--Sommerfeld quantisation rule. The resulting standard and fractionally modified thermodynamic observables are listed in Table (\ref{BHthermoobservables}) and Table (\ref{BHthermoobservables1}), respectively. It is important to note that the analysis in \cite{Jalalzadeh_2021} assumes that FC modifies only the QGD sector, while the thermodynamic laws themselves remain unchanged.

\begin{table}[h]
    \centering
    \caption{Schwarzschild BH thermodynamic observables in the 
    standard QGD framework.}
    \label{BHthermoobservables}
    \renewcommand{\arraystretch}{2.2}
    \begin{tabular}{|c|c|}
        \hline
         & \textbf{Standard (semi-classical) expressions} \\\hline
        $S^{(\alpha=2)}$ & $S_{\text{B-H}} - 2\pi\ln(S_{\text{B-H}}) 
        + \text{const.}$ \\\hline
        $T^{(\alpha=2)}$ & $\left(\dfrac{\beta}{16\pi\sigma_{\text{S}}}
        \right)^{\frac{1}{4}}\dfrac{M^{2}_{pl}}{M}
        \left[1+\dfrac{1}{4}\left(\dfrac{M_{pl}}{M}\right)^{2}\right]$ \\\hline
        $\dot{M}^{(\alpha=2)}$ & $\dfrac{\beta M^{4}_{pl}}{M^{2}}
        \left[1+\left(\dfrac{M_{pl}}{M}\right)^{2}\right]$ \\\hline
        $\omega^{(\alpha=2)}_{0}$ & $\dfrac{M^{2}_{pl}}{M}
        \left[1+\dfrac{1}{2}\left(\dfrac{M_{pl}}{M}\right)^{2}\right]$ \\
        \hline
    \end{tabular}
\end{table}

\begin{table}[h]
    \centering
    \caption{Schwarzschild BH thermodynamic observables in the fractional QGD framework.}
    \label{BHthermoobservables1}
    \renewcommand{\arraystretch}{2.2}
    \begin{tabular}{|c|c|}
        \hline
         & \textbf{Fractional (semi-classical) expressions} \\\hline
        $S^{(\alpha\neq2)}$ & 
        $S^{\frac{2+\alpha}{2\alpha}}_{\text{B-H}} + 
        \dfrac{\alpha(4-\alpha)(2+\alpha)4^{\frac{\alpha+2}{\alpha}}
        \pi^{\frac{3\alpha+2}{2\alpha}}\Gamma\!\left(\frac{2}{\alpha}\right)}
        {(2-\alpha)\,\Gamma\!\left(\frac{1}{\alpha}\right)^{2}}$ \\\hline
        $T^{(\alpha\neq2)}$ & 
        $\left(\dfrac{\beta\alpha^{2}B^{\frac{8}{\alpha}}}
        {16^{2}\pi\sigma_{\text{S}}}\right)^{\frac{1}{4}}
        \dfrac{M^{\frac{2}{\alpha}+1}_{pl}}{M^{\frac{2}{\alpha}}}
        \left[1+\dfrac{1}{4}\!\left(1-\dfrac{\alpha}{4}\right)
        \!\left(\dfrac{BM_{pl}}{M}\right)^{\frac{4}{\alpha}}\right]$ \\\hline
        $\dot{M}^{(\alpha\neq2)}$ & 
        $\dfrac{\alpha^{2}\beta 
        B^{\frac{8}{\alpha}}M^{\frac{8}{\alpha}}_{pl}}
        {16M^{\frac{8}{\alpha}-2}}
        \left[1+\!\left(1-\dfrac{\alpha}{4}\right)
        \!\left(\dfrac{BM_{pl}}{M}\right)^{\frac{4}{\alpha}}\right]$ \\\hline
        $\omega^{(\alpha\neq2)}_{0}$ & 
        $\dfrac{\alpha 
        B^{\frac{4}{\alpha}}M^{\frac{4}{\alpha}}_{pl}}
        {4M^{\frac{4}{\alpha}-1}}
        \left[1+\dfrac{1}{2}\!\left(1-\dfrac{\alpha}{4}\right)
        \!\left(\dfrac{BM_{pl}}{M}\right)^{\frac{4}{\alpha}}\right]$ \\
        \hline
    \end{tabular}
\end{table}

This discussion shows how non-trivial modifications can be introduced at the QGD level through the incorporation of the Riesz space-fractional derivative into the canonical quantisation of the Schwarzschild BH; for a complete treatment, see \cite{Jalalzadeh_2021}. The central quantity that will underlie the remainder of this thesis is the fractional Bekenstein--Hawking entropy
\[
\tilde{S}_{\text{B-H}} \propto A^{\frac{2+\alpha}{2\alpha}},
\]
where \(A\) is the horizon area and \(\alpha\) is the L\'evy index. The FHDE framework will be introduced in Section (\ref{FHDE}) with the Hubble horizon chosen as the IR cutoff. Before doing so, however, it is useful to examine the physical meaning of the parameter range \(1<\alpha\leq2\), since it is precisely through this fractional parameter that hidden structure in quantum gravity and quantum cosmology may become accessible.

\subsection{L\'evy's Distribution Theory}\label{LevyTheory}

As discussed above, the application of FC to quantum mechanics has been developed systematically within the framework of FQM. This construction is rooted in the Feynman--Hibbs path-integral formulation \cite{feynman2010quantum}, together with Nelson's clarification of the correspondence between Brownian motion and quantum behaviour \cite{schilling2014brownian}, and the result of Abbott and Wise that the fractal, or Hausdorff, dimension of one-dimensional quantum-mechanical paths is equal to \(2\). The central motivation for FQM is that restricting Feynman's path integral to Brownian trajectories alone leaves a wider class of quantum phenomena unexplored. This limitation is addressed by replacing the Gaussian probability distribution with a L\'evy distribution, so that the Hausdorff dimension of the resulting L\'evy paths is determined by the fractional parameter \(\alpha\).

Both Brownian and L\'evy paths arise from stochastic processes involving segment-like motion between spatial points and are characterised by a common set of mathematical assumptions. The Brownian process is continuous, whereas more general members of the L\'evy class may be discontinuous. The possibility of such jumps has proved particularly relevant in quantum physics \cite{laskin2000fractional1}. More precisely, Feynman's path integral is based on Brownian-type paths, but Brownian motion itself is only a special case within the broader family of \(\alpha\)-stable probability distributions.\footnote{In probability theory, a distribution is said to be stable if a linear combination of two independent random variables drawn from it has the same distribution, up to scaling and translation, as the original variables \cite{levy1925calcul}.}

The main mathematical points may be summarised as follows:
\begin{itemize}
    \item Consider the sum of \(N\) independent, identically distributed random variables \(X = X_{1} + X_{2} + \dots + X_{N}\). One may then ask whether \(X\) has the same probability distribution as each individual \(p_{i}(X_{i})\), where \(i=1,\dots,N\).

    \item If each \(p_{i}(X_{i})\) is Gaussian, this property follows from the central limit theorem.

    \item In particular, the sum of \(N\) Gaussian random variables is again Gaussian, so the Gaussian distribution is closed under addition.
\end{itemize}

The central limit theorem, however, admits a natural extension:
\begin{itemize}
    \item There exists a class of non-Gaussian, \(\alpha\)-stable probability distributions parametrised by the L\'evy index \(\alpha\), with \(0<\alpha\leq2\).

    \item The standard Brownian process is recovered at \(\alpha=2\). If the fractal dimension of a Brownian path is \(d^{(\text{Brownian})}_{(\text{fractal})}=2\), then a L\'evy path has fractal dimension \(d^{(\text{L\'evy})}_{(\text{fractal})}=\alpha\), with \(\alpha\) in the range \(1<\alpha\leq2\).
\end{itemize}

The L\'evy index \(\alpha\) therefore plays a fundamental role in FQM. The difference between the fractal dimensions of Brownian and L\'evy paths leads to qualitatively distinct physical behaviour. Several works have explored related developments in solid-state systems, particularly through the effective mass \(m(k)\) in certain Bose--Einstein condensate settings. However, essentially no direct observational or experimental confirmation has yet been reported. Although FQM remains experimentally untested, it is both internally consistent and, in principle, falsifiable, since ordinary quantum mechanics is recovered in the limit \(\alpha\to2\).

\section{An Introduction to FHDE Model}\label{FHDE}

In Section (\ref{HDEtheory}) of Chapter (\ref{Chapter2}), we emphasised the difficulties associated with choosing the Hubble cutoff \(L_{\text{IR}}=H^{-1}\), while also summarising several alternative prescriptions that are widely studied in the literature \cite{wang2017holographic}. In the present section, however, our aim is to overcome the absence of dynamical DE in conventional HDE with the Hubble cutoff by incorporating the distinctive features of FQM, specifically through the L\'evy index \(\alpha\).

Jalalzadeh \textit{et al.} \cite{Jalalzadeh_2021} recently investigated the effect of FQM on Schwarzschild BH thermodynamics. Using a second-order space-fractional Riesz derivative, they derived a fractionally modified WDW equation and thereby obtained the fractional Bekenstein--Hawking horizon entropy \(\tilde{S}_{\text{B-H}}\), which can be written as
\begin{equation} \label{fracentropy}
    \tilde{S}_{\text{B-H}} = \tilde{C} A^{\frac{2+\alpha}{2\alpha}},
\end{equation}
where \(\alpha\) is the L\'evy index in the range \(1<\alpha\leq2\), and the standard Bekenstein--Hawking entropy is recovered in the limit \(\alpha\to2\). The constant \(\tilde{C}\) is chosen so as to contain the appropriate Planck-unit combinations required to recover the standard expression at \(\alpha=2\). Equation (\ref{fracentropy}) therefore expresses the entropy as a power law in the horizon area, resembling the structure of Barrow and Tsallis entropies, although the underlying physical motivations are different. Combining Cohen's inequality, Eq. (\ref{Cohenineq}), with the fractional Bekenstein--Hawking entropy, Eq. (\ref{fracentropy}), one obtains
\begin{equation}
    \Lambda^{3}_{\text{UV}} L^{3}_{\text{IR}} \leq \left(\tilde{C} A^{\frac{2+\alpha}{2\alpha}}\right)^{\frac{3}{4}}.
\end{equation}
Using the identification \(\rho_{\text{DE}}\sim\Lambda^{4}_{\text{UV}}\), one is led to a DE density that encodes the fractional structure introduced at the QGD level \cite{Jalalzadeh_2021}:
\begin{equation}
    \rho^{(\text{fr})}_{\text{DE}} := \tilde{\gamma} L^{\frac{2 - 3\alpha}{\alpha}}_{\text{IR}},
\end{equation}
where \(\tilde{\gamma}\) is a constant parameter. In the limit \(\alpha\to2\), the standard HDE expression in Eq. \eqref{HDE} is correctly recovered, consistently with the corresponding limit of the fractional entropy in Eq. (\ref{fracentropy}). Setting \(\tilde{\gamma}=3c^2\), with \(c\) a constant of order \(\mathcal{O}(1)\), brings the expression into closer analogy with the conventional HDE model. The FHDE density may then be written as
\begin{equation} \label{fracrho}
    \rho^{(\text{fr})}_{\text{DE}} = 3c^{2}M^{2}_{pl} L^{\frac{2 - 3\alpha}{\alpha}}_{\text{IR}}.
\end{equation}

Equation (\ref{fracrho}) is the defining relation of the FHDE model introduced in this thesis. We now examine whether this model can provide a consistent description of late-time cosmological evolution compatible with the latest DESI DR1 and DR2 results, which point to a dynamically evolving DE sector \cite{desicollaboration2024desi2024vicosmological,DESI:2025zgx}. In this section, the reduced Planck mass is set equal to unity within GR; this will no longer be the case when dealing with modified gravity theories involving a time-dependent gravitational constant \(G_{\text{N}}(t)\), which will be addressed in Chapter (\ref{Chapter4}). Although FHDE exhibits a structural resemblance to the Barrow and Tsallis constructions, owing to the common power-law entropy--area dependence, the physical motivation is fundamentally different: FHDE originates from the application of space-FQM to BH thermodynamics, whereas the Tsallis and Barrow models arise from non-extensive thermodynamics and BH horizon deformations, respectively.

We restrict attention to a spatially flat FLRW Universe, \(k=0\), as supported by observations, and assume that the late-time energy budget is dominated by the dark sector, namely DE and DM. This is the standard approximation in late-time cosmology. The Friedmann equation for such a flat Universe is
\begin{equation} \label{fried}
    H^2 = \frac{\rho_{\text{DE}} + \rho_{\text{DM}}}{3},
\end{equation}
where \(H=\dot{a}/a\), and \(\rho_{\text{DE}}\) and \(\rho_{\text{DM}}\) denote the DE and DM energy densities, respectively. Conservation of energy leads to the interacting continuity equations
\begin{equation} \label{contm}
    \dot{\rho}_{\text{DM}} + 3H\rho_{\text{DM}}(1 + w_{\text{DM}}) = +Q,
\end{equation}
\begin{equation} \label{contd}
    \dot{\rho}_{\text{DE}} + 3H\rho_{\text{DE}}(1 + w_{\text{DE}}) = -Q,
\end{equation}
where \(w_{\text{DM}}\) and \(w_{\text{DE}}\) are the EoS parameters of DM and DE, respectively. The term \(Q\) represents the interaction between the two dark-sector components. Since a first-principles derivation of the interaction term would require a consistent theory of quantum gravity, which is not yet available, one generally adopts phenomenological forms. Two common choices are: \((i)\) a linear interaction, \(Q=\tilde{\beta}H(\rho_{\text{DE}}+\rho_{\text{DM}})\), and \((ii)\) a non-linear interaction, \(Q=\tilde{\beta}H\rho_{\text{DE}}/\rho_{\text{DM}}\), where \(\tilde{\beta}\) is the coupling parameter. In the present analysis, however, we neglect dark-sector interactions and set \(Q=0\). We also assume pressureless DM, as supported observationally \cite{wang2007generalized}, so that \(w_{\text{DM}}\simeq0\). Substituting the Hubble cutoff \(L_{\text{IR}}=H^{-1}\) into Eq. (\ref{fracrho}), the FHDE density becomes
\begin{equation} \label{h1frarho}
    \rho^{(\text{fr})}_{\text{DE}} = 3c^2 H^{\frac{3\alpha - 2}{\alpha}}.
\end{equation}
The corresponding fractional density parameters for DE and DM are defined as
\begin{equation} \label{omegas}
    \Omega^{(\text{fr})}_{\text{DE}} = \frac{\rho^{(\text{fr})}_{\text{DE}}}{3H^2} = 
    c^2 H^{\frac{\alpha - 2}{\alpha}}, \qquad 
    \Omega^{(\text{fr})}_{\text{DM}} = \frac{\rho^{\text{(fr)}}_{\text{DM}}}{3H^2}.
\end{equation}
In the limit \(\alpha\to2\), one has \(\Omega^{(\text{fr})}_{\text{DE}}\to\Omega_{\text{DE}}=c^2\) and \(\Omega^{(\text{fr})}_{\text{DM}}\to\Omega_{\text{DM}}=1-c^2\). Thus, as the fractional effects disappear, the dynamics becomes trivial and the Hubble-cutoff evolution reduces to a constant behaviour. Using Eq. (\ref{fried}), the Friedmann equation can be recast as
\begin{equation}
    \Omega^{(\text{fr})}_{\text{DE}} + \Omega^{(\text{fr})}_{\text{DM}} = \Omega^{(\text{fr})}_{\text{DE}}(1 + y) = 1,
\end{equation}
where \(y = \Omega^{(\text{fr})}_{\text{DM}}/\Omega^{(\text{fr})}_{\text{DE}}\). Combining this relation with Eqs. (\ref{fried}) and (\ref{contm}) gives
\begin{equation} \label{h11}
    \frac{\dot{H}}{H^2} = -\frac{3}{2}\left(1 + w^{(\text{fr})}_{\text{DE}} + y\right)\Omega^{(\text{fr})}_{\text{DE}}.
\end{equation}
On the other hand, using Eqs. (\ref{contd}) and (\ref{h1frarho}) yields
\begin{equation} \label{h12}
    \frac{\dot{H}}{H^2} = -\frac{3\alpha}{3\alpha - 2}\left(1 + w^{(\text{fr})}_{\text{DE}}\right).
\end{equation}
Equating Eqs. (\ref{h11}) and (\ref{h12}), and using \(y = (1 - \Omega^{(\text{fr})}_{\text{DE}})/\Omega^{(\text{fr})}_{\text{DE}}\), one obtains the FHDE EoS parameter:
\begin{equation} \label{FHDEEoS}
    w^{(\text{fr})}_{\text{DE}} = -1 + \frac{(3\alpha - 2)\left(1 - \Omega^{(\text{fr})}_{\text{DE}}\right)}
    {2\alpha - \Omega^{(\text{fr})}_{\text{DE}}(3\alpha - 2)}.
\end{equation}
In the limit \(\alpha\to2\), this expression tends to zero, \(w^{(\text{fr})}_{\text{DE}}\to0\). Substituting Eq. (\ref{FHDEEoS}) back into Eq. (\ref{h12}), the Hubble-rate equation becomes
\begin{equation} \label{h13}
    \frac{\dot{H}}{H^2} = -\frac{3\alpha\left(1 - \Omega^{(\text{fr})}_{\text{DE}}\right)}
    {2\alpha - \Omega^{(\text{fr})}_{\text{DE}}(3\alpha - 2)}.
\end{equation}
The deceleration parameter then follows directly as
\begin{equation} \label{qeq}
    q^{(\text{fr})}_{\text{DE}} = -1 - \frac{\dot{H}}{H^2} = -1 + 
    \frac{3\alpha\left(1 - \Omega^{(\text{fr})}_{\text{DE}}\right)}{2\alpha - \Omega^{(\text{fr})}_{\text{DE}}(3\alpha - 2)}.
\end{equation}
The evolution equation for \(\Omega^{(\text{fr})}_{\text{DE}}\), where a prime denotes differentiation with respect to \(\ln(a)\), follows from Eq. (\ref{omegas}):
\begin{equation}
    \Omega'^{(\text{fr})}_{\text{DE}} = \frac{\alpha - 2}{\alpha}
    \left(\frac{\Omega^{(\text{fr})}_{\text{DE}}\dot{H}}{H^2}\right).
\end{equation}
Using Eq. (\ref{h13}), this reduces to
\begin{equation} \label{diffom}
    \Omega'^{(\text{fr})}_{\text{DE}} = \frac{3(\alpha - 2)\left(1 - \Omega^{(\text{fr})}_{\text{DE}}\right)
    \Omega^{(\text{fr})}_{\text{DE}}}{2\alpha - \Omega^{(\text{fr})}_{\text{DE}}(3\alpha - 2)}.
\end{equation}
With the relation \(d\ln(a) = -dz/(1+z)\), Eq. (\ref{diffom}) may be rewritten in terms of redshift as
\begin{equation}
    (1+z)\frac{d\Omega^{(\text{fr})}_{\text{DE}}}{dz} = 
    \frac{3(\alpha - 2)\left(1 - \Omega^{(\text{fr})}_{\text{DE}}\right)\Omega^{(\text{fr})}_{\text{DE}}}
    {2\alpha - \Omega^{(\text{fr})}_{\text{DE}}(3\alpha - 2)}.
\end{equation}
This differential equation admits the analytic solution
\begin{equation} \label{omegeq}
    \left(\frac{\Omega^{(\text{fr})}_{\text{DE}}}{\Omega_{\text{DE}0}}\right)^{2\alpha}
    \left(\frac{1 - \Omega_{\text{DE}0}}{1 - \Omega^{(\text{fr})}_{\text{DE}}}\right)^{2-\alpha}
    = (1+z)^{3(\alpha - 2)},
\end{equation}
where \(\Omega_{\text{DE}0}\) denotes the present-day value at \(z=0\). We adopt \(\Omega_{\text{DE}0}\simeq0.69\) for all \(\alpha\in(1,2]\), in accordance with current observational constraints \cite{adame2024desi3,adame2024desi4,adame2024desi6}, and solve Eq. (\ref{omegeq}) numerically.

\begin{figure*}[t]
\centering
\begin{minipage}[b]{0.49\textwidth}
        \centering
        \includegraphics[width=\textwidth]{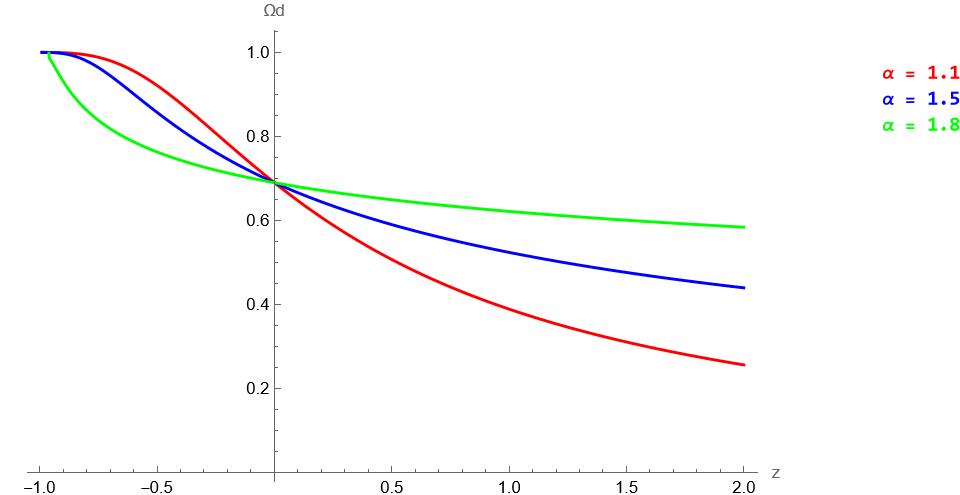}
  \caption{Redshift evolution of \(\Omega^{(\text{fr})}_{\text{DE}}\) for various values of \(\alpha\in(1,2]\).}
  \label{omegadfig}
\end{minipage}
\hfill
\begin{minipage}[b]{0.49\textwidth}
        \centering
        \includegraphics[width=\textwidth]{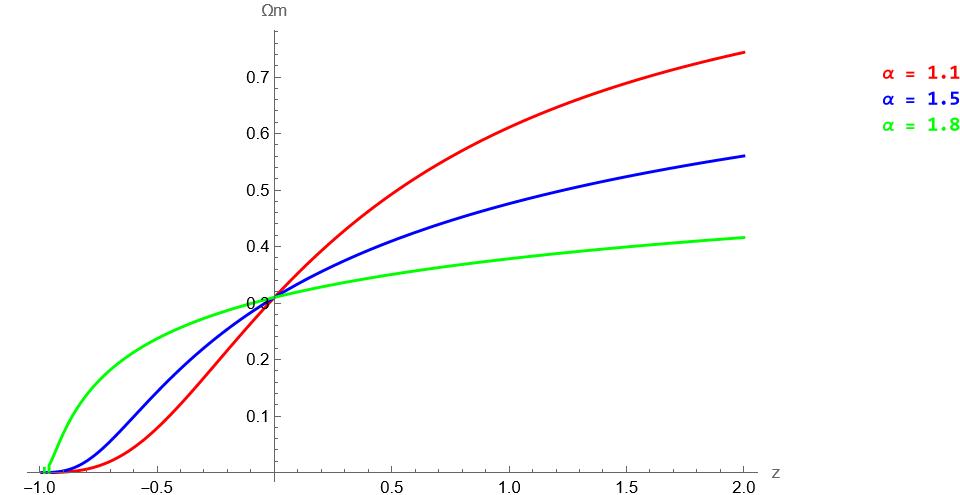}
  \caption{Redshift evolution of \(\Omega^{(\text{fr})}_{\text{DM}}\) for various values of \(\alpha\in(1,2]\).}
    \label{omegamfig}
\end{minipage}
\hfill
\begin{minipage}[b]{0.49\textwidth}
        \centering
        \includegraphics[width=\textwidth]{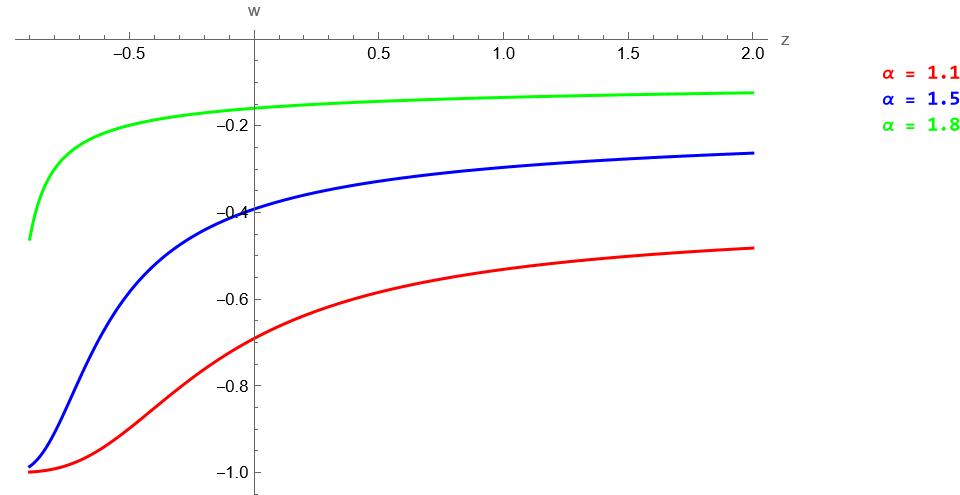}
  \caption{Redshift evolution of the EoS parameter \(w^{(\text{fr})}_{\text{DE}}\) for various values of \(\alpha\in(1,2]\).}
    \label{wfig}   
\end{minipage}
\hfill
\begin{minipage}[b]{0.49\textwidth}
        \centering
        \includegraphics[width=\textwidth]{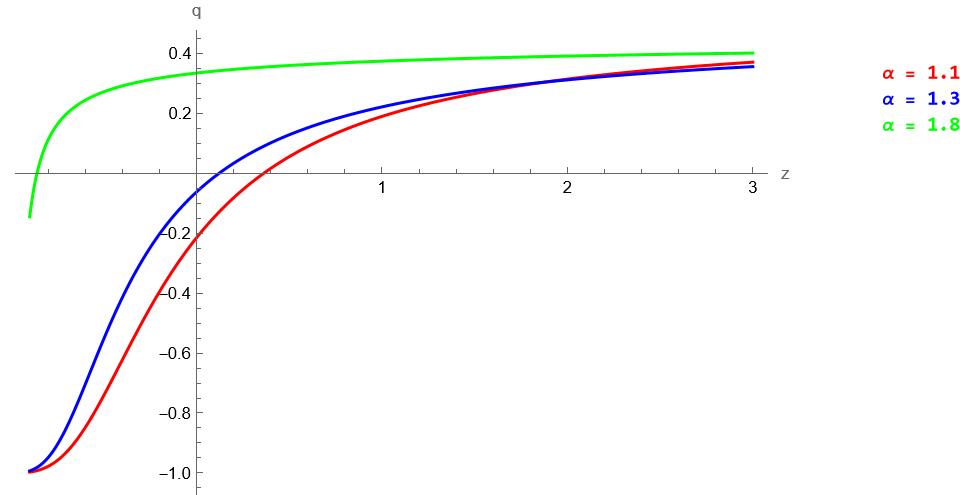}
  \caption{Redshift evolution of the deceleration parameter \(q^{(\text{fr})}_{\text{DE}}\) for various values of \(\alpha\in(1,2]\).}
    \label{qfig}
\end{minipage}
\end{figure*}

Figure (\ref{omegadfig}) shows the evolution of \(\Omega^{(\text{fr})}_{\text{DE}}\) as a function of redshift for several values of \(\alpha\). A viable and increasingly dynamical DE evolution is obtained as the L\'evy index decreases, that is, as fractional effects become stronger. For relatively larger values such as \(\alpha=1.8\), the growth of \(\Omega^{(\text{fr})}_{\text{DE}}\) with decreasing redshift remains comparatively mild. As \(\alpha\) decreases toward \(1.5\) and \(1.1\), the evolution becomes significantly more dynamical, with a more pronounced rise in the DE density, precisely as expected when fractional effects dominate. The behaviour of \(\Omega^{(\text{fr})}_{\text{DM}}\), displayed in Figure (\ref{omegamfig}), follows the complementary trend: the DM density parameter also evolves more dynamically for smaller \(\alpha\). Taken together, Figures (\ref{omegadfig}) and (\ref{omegamfig}) indicate DE domination in the far-future limit \(z\to-1\), in agreement with expectation and with current observations.

The FHDE EoS parameter in Eq. (\ref{FHDEEoS}) is plotted in Figure (\ref{wfig}). For smaller values of \(\alpha\), the EoS exhibits behaviour that is both physically reasonable and compatible with observational constraints. In particular, for \(\alpha=1.1\), the present-day value of \(w^{(\text{fr})}_{\text{DE}}\) at \(z=0\) lies close to the recent DESI bounds \cite{adame2024desi3,adame2024desi4,adame2024desi6}. As \(\alpha\) increases, the EoS progressively departs from observationally acceptable DE behaviour. For both \(\alpha=1.1\) and \(\alpha=1.5\), the model remains in the \textit{quintessence} regime throughout the interval \(-1\leq z\leq2\), with \(w^{(\text{fr})}_{\text{DE}}\to-1\) in the asymptotic future \(z\to-1\).

The deceleration parameter given in Eq. (\ref{qeq}) is displayed in Figure (\ref{qfig}). For smaller values of \(\alpha\), the transition from decelerated to accelerated expansion is clearly visible. In the case \(\alpha=1.1\), the transition redshift is \(z_{T}\simeq0.38\), which lies well within current observational bounds \cite{Kumar:2022mtx}. For \(\alpha=1.3\), one finds \(z_{T}\simeq0.17\), which is considerably less compatible with observations. For larger values such as \(\alpha=1.8\), the onset of acceleration is shifted into the future, \(z<0\), and is therefore unphysical. These results indicate that the observationally viable FHDE regime corresponds to smaller values of \(\alpha\), where the fractional effects are strongest and the Hubble cutoff remains compatible with late-time cosmology.

\begin{figure*}[t]
\centering
\begin{minipage}[b]{0.49\textwidth}
        \centering
        \includegraphics[width=\textwidth]{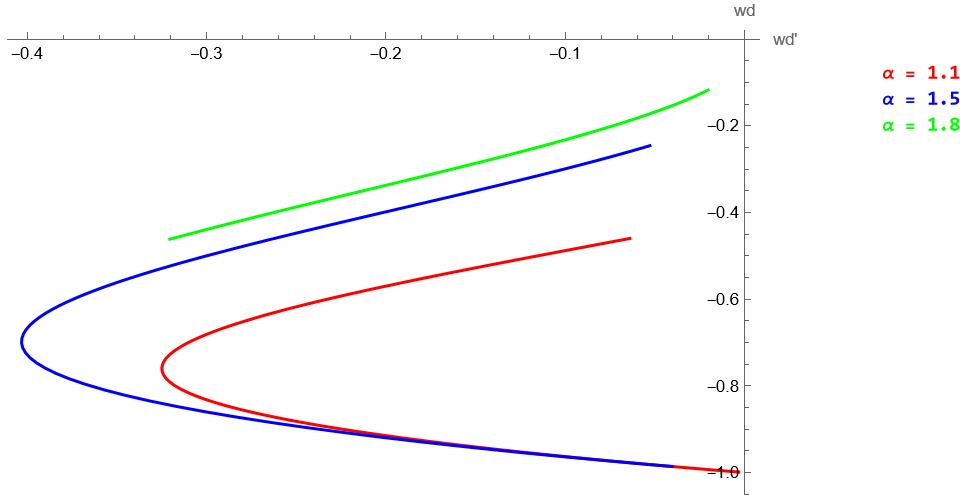}
  \caption{The \(w^{(\text{fr})}_{\text{DE}}\)--\(w'^{(\text{fr})}_{\text{DE}}\) plane over the redshift range 
  \(-1 \leq z \leq 2\), for various values of \(\alpha\).}
    \label{wdwd'} 
\end{minipage}
\hfill
\begin{minipage}[b]{0.49\textwidth}
        \centering
        \includegraphics[width=\textwidth]{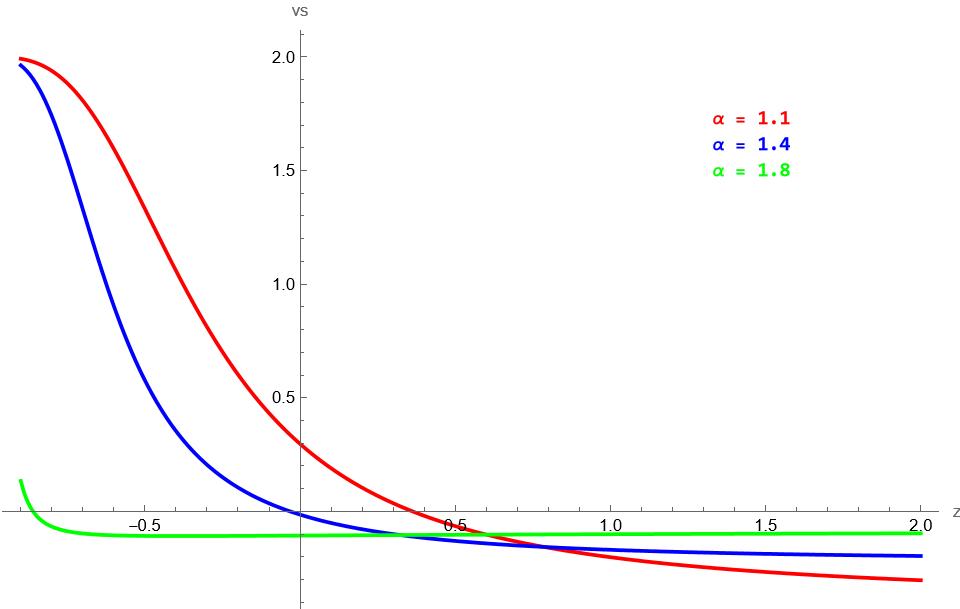}
  \caption{Redshift evolution of the squared sound speed \(v_{s}^2\) for various values of \(\alpha\).}
    \label{vsfig} 
\end{minipage}
\end{figure*}

The \(w^{(\text{fr})}_{\text{DE}}\)--\(w'^{(\text{fr})}_{\text{DE}}\) plane, introduced in \cite{Caldwell:2005tm}, provides a useful diagnostic for quintessence models. The region \(w^{(\text{fr})}_{\text{DE}}<0\) and \(w'^{(\text{fr})}_{\text{DE}}>0\) corresponds to thawing quintessence, while \(w^{(\text{fr})}_{\text{DE}}<0\) and \(w'^{(\text{fr})}_{\text{DE}}<0\) characterises freezing quintessence. As shown in Figure (\ref{wdwd'}), the FHDE model lies in the freezing region throughout the full range of \(\alpha\) considered. This classification is physically most relevant for \(\alpha=1.1\) and \(\alpha=1.5\), for which the EoS remains in the quintessence regime, as already seen in Figure (\ref{wfig}).

Finally, we consider the squared sound speed,
\begin{equation}
    v_{s}^2 = \frac{\dot{p}}{\dot{\rho}} = w + \frac{\dot{w}\rho}{\dot{\rho}}.
\end{equation}
Using Eq. (\ref{h13}) together with \(d\ln a = -dz/(1+z)\), this may be written as
\begin{equation}
    (v^{(\text{fr})}_{s})^{2} = w^{(\text{fr})}_{\text{DE}} + \frac{(1+z)\left(2\alpha - \Omega^{(\text{fr})}_{\text{DE}}(3\alpha - 2)\right)}
    {(3\alpha - 2)\left(1 - \Omega^{(\text{fr})}_{\text{DE}}\right)} \frac{dw^{(\text{fr})}_{\text{DE}}}{dz}.
\end{equation}
Substituting Eqs. (\ref{FHDEEoS}) and (\ref{omegeq}) gives \(v_{s}^2\) as a function of \(\Omega^{(\text{fr})}_{\text{DE}}(z)\), and the result is plotted for several values of \(\alpha\) in Figure (\ref{vsfig}). Classical stability requires \(0\leq v_{s}^2\leq1\). Negative values indicate instability in the energy-density perturbations, while \(v_{s}^2>1\) implies superluminal propagation of perturbations, which is generally regarded as another sign of classical instability. Although superluminal propagation has occasionally been invoked in both early- and late-Universe scenarios \cite{Bessada:2009ns,Afshordi:2006ad}, its physical justification remains non-trivial even in non-canonical models, and it is usually viewed as problematic. Classical instability has, in fact, been a persistent difficulty in HDE models more generally \cite{Myung:2007pn}, affecting both the standard HDE scenario and several of its extensions with alternative cutoffs. In the present FHDE model, classical stability is preserved at \(z=0\) for smaller values of \(\alpha\), where fractional effects are strongest. However, instability develops in the far-future regime. We expect that this issue may be alleviated by considering IR cutoffs beyond the Hubble horizon, although a systematic analysis of this possibility is left for future work. We also note that, since no specific value of \(c\) has been fixed in Eq. (\ref{fracrho}), the results presented here remain valid for any choice of this constant.

\section{Reconstructing the Fractional HDE Model}\label{ReconFHDE}

In the previous section (\ref{FHDE}), we developed the FHDE framework in detail and derived the principal cosmological quantities, together with their numerical evolution as functions of the redshift parameter \(z\), for several values of the L\'evy index \(\alpha\in(1,2]\). In the present section, we extend this programme by further exploring the bridge between the HP and FC within the FHDE setting, following the general framework developed in~\cite{Trivedi:2024inb}. Our main goal is to determine whether FHDE can serve as a viable and physically well-motivated explanation of late-time cosmic acceleration through an effective-field-theory realisation. To this end, we reconstruct the FHDE scenario using the following phenomenological models: \((i)\) Quintessence, \((ii)\) K-essence, \((iii)\) Dilaton, \((iv)\) Yang--Mills condensate (YMC), \((v)\) Dirac--Born--Infeld (DBI) essence, and \((vi)\) the Tachyon scalar field. In the appropriate regime, each of these models is governed by a rolling scalar degree of freedom that generalises the cosmological constant through dynamical evolution; when the field becomes stationary, its potential \(V(\varphi)\) reduces effectively to \(\Lambda\)~\cite{COPELAND_2006}. Although present observations do not conclusively establish dynamics beyond \(\Lambda\)CDM, recent data~\cite{desicollaboration2024desi2024vicosmological,
lodha2024desi2024constraintsphysicsfocused,Calderon_2024,DESI:2025zgx} suggest possible deviations that motivate a systematic reconstruction programme.

We consider homogeneous effective-field configurations --- namely, models \((i)\)--\((vi)\) above --- as candidates for dynamical DE, and study their redshift evolution within the FHDE framework. This provides a flexible arena for probing late-time expansion beyond the \(\Lambda\)CDM paradigm, particularly through the time-dependent EoS parameter \(w_{\text{DE}}(z)\), in contrast with the constant value \(w_{\Lambda\text{CDM}}=-1\). The reconstruction is carried out by establishing a \textit{correspondence} between FHDE and each effective field model, and then examining the evolution of the associated field potential \(V(\varphi)\) and kinetic term \(X\) over the redshift interval \(-1<z\leq2\). Holographic reconstructions of this type have been widely studied in the literature for different infrared cutoffs and entropy corrections, including Tsallis and Barrow entropies in modified-gravity settings~\cite{Sheykhi_2011,ZHANG20071,WU2008152,SHEYKHI2010329}. The present analysis extends this line of work to the FHDE framework~\cite{Trivedi:2024inb}, using the Hubble horizon as the IR cutoff together with an entropy correction motivated by fractional effects~\cite{Jalalzadeh_2021}, and builds directly on~\cite{Sheykhi_2011,WU2008152,Sheykhi_2010}.

\subsection{Bosonic \& String-Inspired Field Candidates}

\subsubsection{$(i)$ Quintessence}

Among alternatives to a bare cosmological constant, quintessence is one of the most extensively studied, with late-time acceleration driven by a dynamical scalar field. Several works, including~\cite{ZHANG20071} and~\cite{WU2008152}, have developed methods for reconstructing dynamical DE through a quintessence field associated with holographic vacuum energy. The quintessence field is described by the real scalar Lagrangian
\begin{equation}
    \mathcal{L}_{\text{q}} = X_{\text{q}} - V_{\text{q}}(\varphi),
\end{equation}
where \(X_{\text{q}}\equiv \frac{1}{2}\partial_{\mu}\varphi\partial^{\mu}\varphi\) and \(V_{\text{q}}(\varphi)\) denotes the scalar potential. In this framework, the potential acts as an effective \(\Lambda\) and must be sufficiently flat without excessive fine-tuning. From a high-energy perspective, however, this flatness is generically destabilised by quantum loop corrections, thereby compromising technical naturalness~\cite{de_Putter_2007}.\footnote{A natural alternative is to consider a non-canonical kinetic term in the scalar-field Lagrangian, leading to K-essence, which mitigates this difficulty.} The action for the quintessence field is
\begin{equation}
    \mathcal{S}_{\text{q}} = \int d^{4}x\sqrt{-g}\left[-\frac{1}{2}g^{\mu\nu}
    \partial_{\mu}\varphi\partial_{\nu}\varphi - V_{\text{q}}(\varphi)\right],
\end{equation}
where \(g^{\mu\nu}\) is the inverse metric with signature
\(g_{\mu\nu}=\text{diag}(-,+,+,+)\), so the field carries a canonical kinetic term. The corresponding energy-momentum tensor is
\begin{equation}\label{E-M Tensor}
    T_{\text{q}}^{\mu\nu} = \frac{\partial\mathcal{L}_{\text{q}}}
    {\partial(\partial_{\mu}\varphi)}\partial^{\nu}\varphi
    - g^{\mu\nu}\mathcal{L}_{\text{q}},
\end{equation}
from which the energy density and pressure follow as
\begin{equation}
    \rho_{\text{q}}(t) = \frac{1}{2}\dot{\varphi}_{\text{q}}^{2}
    + V_{\text{q}}(\varphi), \qquad
    p_{\text{q}}(t) = \frac{1}{2}\dot{\varphi}_{\text{q}}^{2}
    - V_{\text{q}}(\varphi).
\end{equation}
The corresponding EoS parameter is
\begin{equation}\label{eos-q}
    w_{\text{q}} = \frac{p_{\text{q}}}{\rho_{\text{q}}} =
    \frac{X_{\text{q}} - V_{\text{q}}(\varphi)}{X_{\text{q}}
    + V_{\text{q}}(\varphi)},
\end{equation}
where \(X_{\text{q}}=\dot{\varphi}_{\text{q}}^{2}/2\). Imposing the correspondence \(w^{(\text{fr})}_{\text{DE}}\leftrightarrow w_{\text{q}}\) between the FHDE EoS parameter in Eq.~(\ref{FHDEEoS}) and Eq.~(\ref{eos-q}) yields
\begin{equation}
    \frac{X_{\text{q}} - V_{\text{q}}(\varphi)}{X_{\text{q}}
    + V_{\text{q}}(\varphi)} = -1 + \frac{(3\alpha-2)(1-\Omega^{(\text{fr})}_{\text{DE}})}
    {2\alpha - \Omega^{(\text{fr})}_{\text{DE}}(3\alpha-2)}.
\end{equation}
Solving algebraically for the kinetic and potential terms gives the reconstructed FHDE expressions
\begin{equation}\label{q-field}
    X_{\text{q}} = \frac{\dot{\varphi}_{\text{q}}^{2}}{2} =
    \frac{3\Omega^{(\text{fr})}_{\text{DE}}H^{2}(3\alpha-2)(1-\Omega^{(\text{fr})}_{\text{DE}})}
    {4\alpha - \Omega^{(\text{fr})}_{\text{DE}}(6\alpha-4)},
\end{equation}
\begin{equation}
    V_{\text{q}}(\varphi) = 3H^{2}\Omega^{(\text{fr})}_{\text{DE}}\left[1 -
    \frac{(3\alpha-2)(1-\Omega^{(\text{fr})}_{\text{DE}})}
    {4\alpha - 2\Omega^{(\text{fr})}_{\text{DE}}(3\alpha-2)}\right].
\end{equation}
These expressions determine the redshift evolution of the quintessence kinetic term \(X_{\text{q}}\) and potential \(V_{\text{q}}(\varphi)\), shown in Figures (\ref{Figure 1}) and (\ref{Figure 2}).

\begin{figure*}[t]
\centering
\begin{minipage}[b]{0.47\textwidth}
        \centering
        \includegraphics[width=\textwidth]{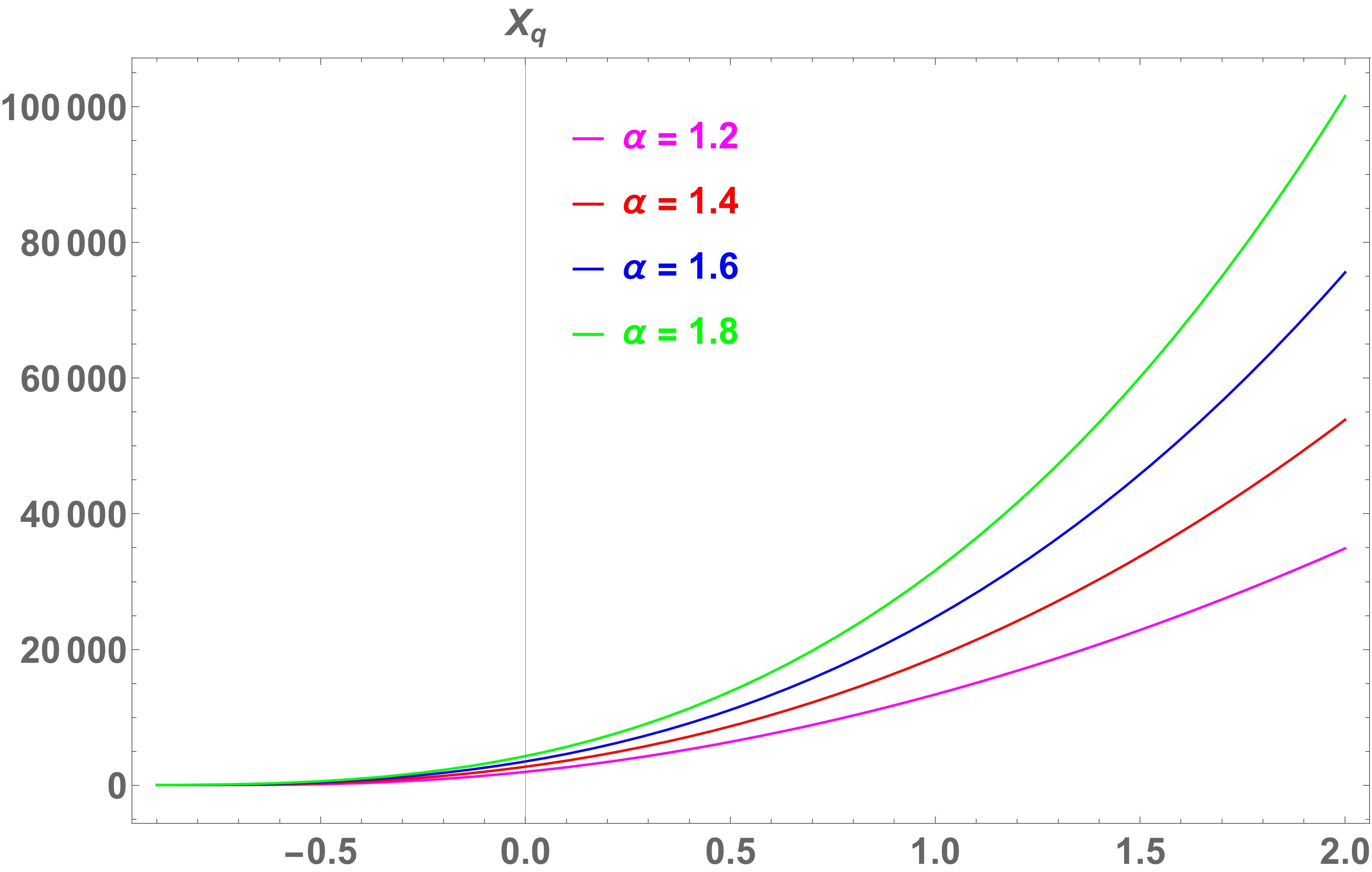}
  \caption{Redshift evolution of the quintessence kinetic energy \(X_{\text{q}}\).}
  \label{quintkinetic}
\end{minipage}
\hfill
\begin{minipage}[b]{0.47\textwidth}
        \centering
        \includegraphics[width=\textwidth]{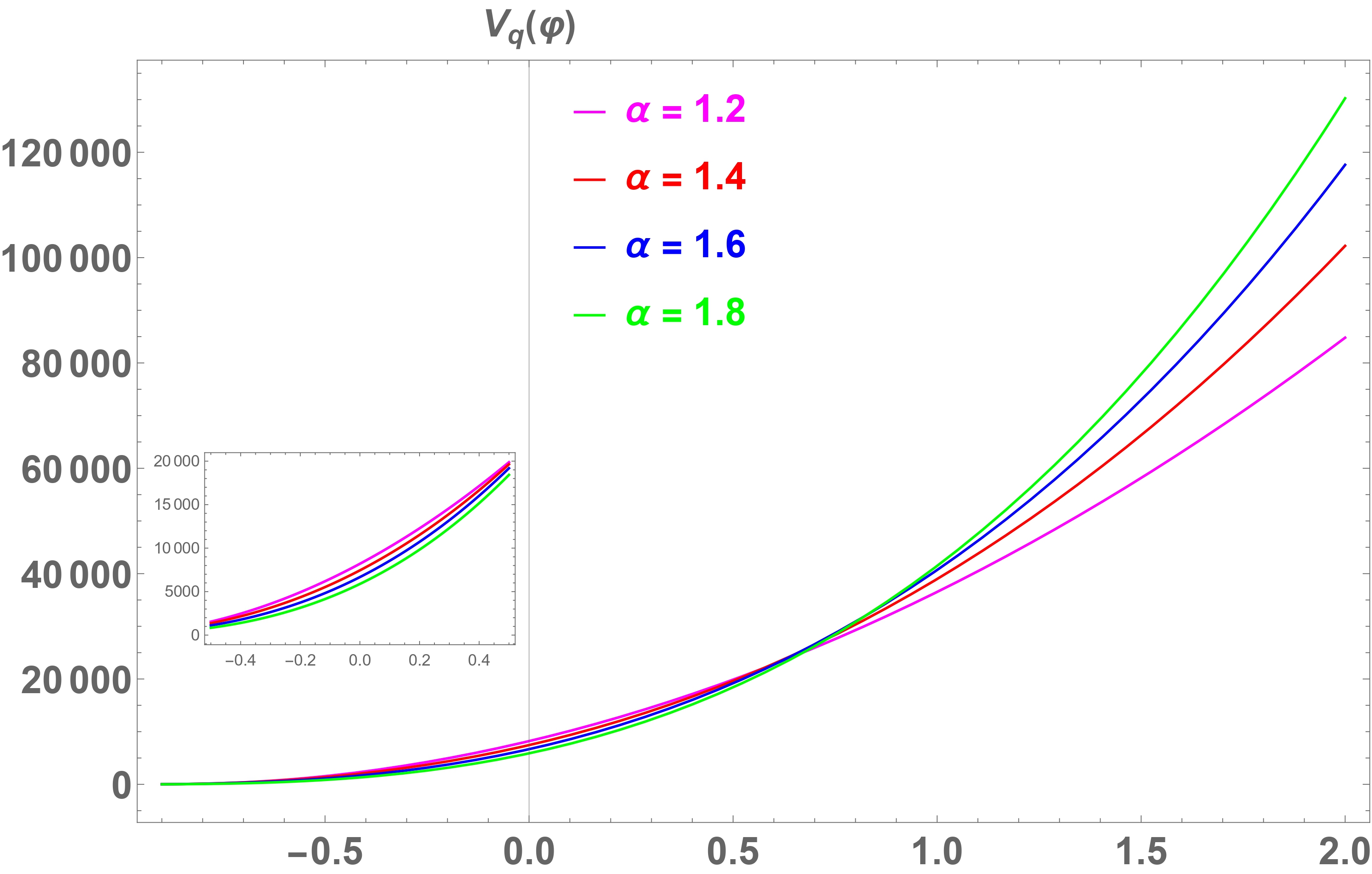}
  \caption{Redshift evolution of the quintessence potential \(V_{\text{q}}(\varphi)\).}
  \label{quintpotential}
\end{minipage}
\hfill
\begin{minipage}[b]{0.47\textwidth}
        \centering
        \includegraphics[width=\textwidth]{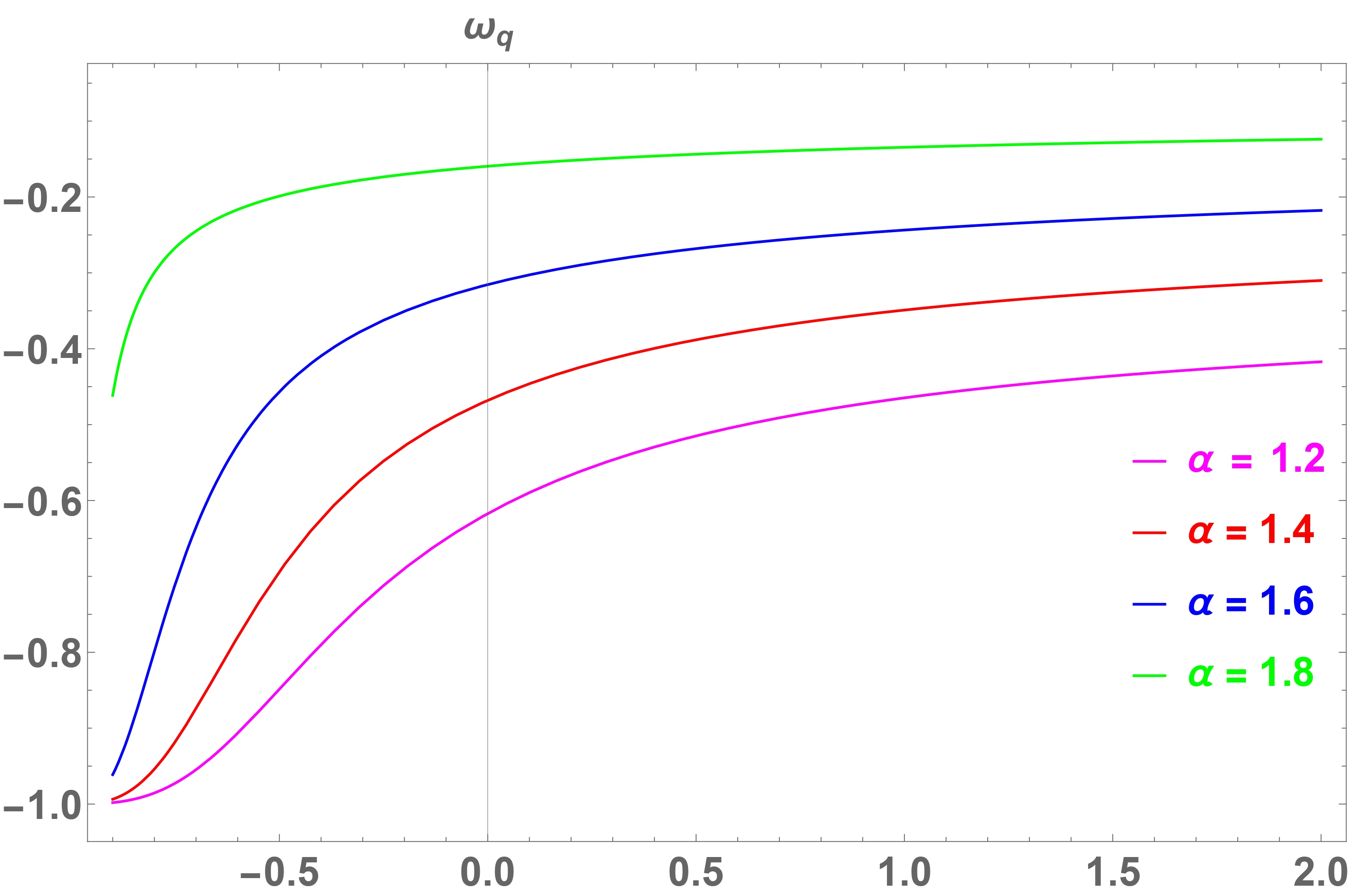}
\caption{Redshift evolution of the quintessence EoS parameter \(w_{\text{q}}\).}
\label{quintEoS}
\end{minipage}
\end{figure*}

Figures (\ref{quintkinetic}) and (\ref{quintpotential}) show that both \(X_{\text{q}}\) and \(V_{\text{q}}(\varphi)\) decay asymptotically and approach zero in the far-future limit \(z\to-1\), for all values of \(\alpha\). Several features of this behaviour are worth emphasising:
\begin{itemize}
    \item For larger values of \(\alpha\) (for example, \(\alpha=1.8\) and \(1.6\)), both \(X_{\text{q}}\) and \(V_{\text{q}}(\varphi)\) evolve more rapidly as redshift decreases.\footnote{By ``more rapid evolution'' we mean that the field quantity varies more steeply as a function of \(z\); the opposite interpretation applies to slower evolution.}

    \item For smaller values of \(\alpha\) (for example, \(\alpha=1.4\) and \(1.2\)), the evolution is comparatively more gradual.

    \item For \(z\lesssim0.55\) in Figure~\ref{quintpotential}, the potential \(V_{\text{q}}(\varphi)\) for \(\alpha=1.8\) decays more slowly than for smaller values of \(\alpha\), thereby preserving the condition \(V_{\text{q}}\gg X_{\text{q}}\) for a longer interval and keeping \(w_{\text{q}}\) closer to \(-1\).\footnote{A viable quintessence model requires \(V_{\text{q}}\gg X_{\text{q}}\) so that the EoS remains close to \(-1\) at late times; in this sense, a slower decay of the potential is preferable in the DE-dominated era.} As \(z\to-1\), both \(X_{\text{q}}\) and \(V_{\text{q}}\) vanish, yielding an asymptotic \(\Lambda\)CDM-like limit.
\end{itemize}

The EoS parameter \(w_{\text{q}}\), plotted in Figure (\ref{quintEoS}), approaches \(-1\) asymptotically in the far-future limit for all values of \(\alpha\). For smaller values, such as \(\alpha=1.2\) and \(1.4\), where fractional effects are strongest, the convergence toward the \(\Lambda\)CDM behaviour \(w_{\Lambda\text{CDM}}=-1\) is particularly pronounced during the DE-dominated phase. Larger values of \(\alpha\) instead lead to progressively larger deviations from this limit. Nevertheless, throughout the range \(\alpha\in[1.2,1.8]\), the EoS remains within the quintessence regime,
\(-1<w_{\text{q}}<-1/3\), over the redshift interval approaching \(z\to-1\).

Overall, Figure (\ref{quintkinetic}) suggests that larger values of \(\alpha\) are less favourable for quintessence reconstruction, since they induce a kinetic term that evolves too rapidly to resemble a slowly varying DE component. By contrast, Figure (\ref{quintpotential}) indicates that larger \(\alpha\) is more advantageous in the interval \(-1<z<0.55\), where the slower decay of the potential better supports a DE-dominated phase.

\subsubsection{$(ii)$ K-essence}

\begin{figure*}[t]
\centering
\begin{minipage}[b]{0.47\textwidth}
        \centering
        \includegraphics[width=\textwidth]{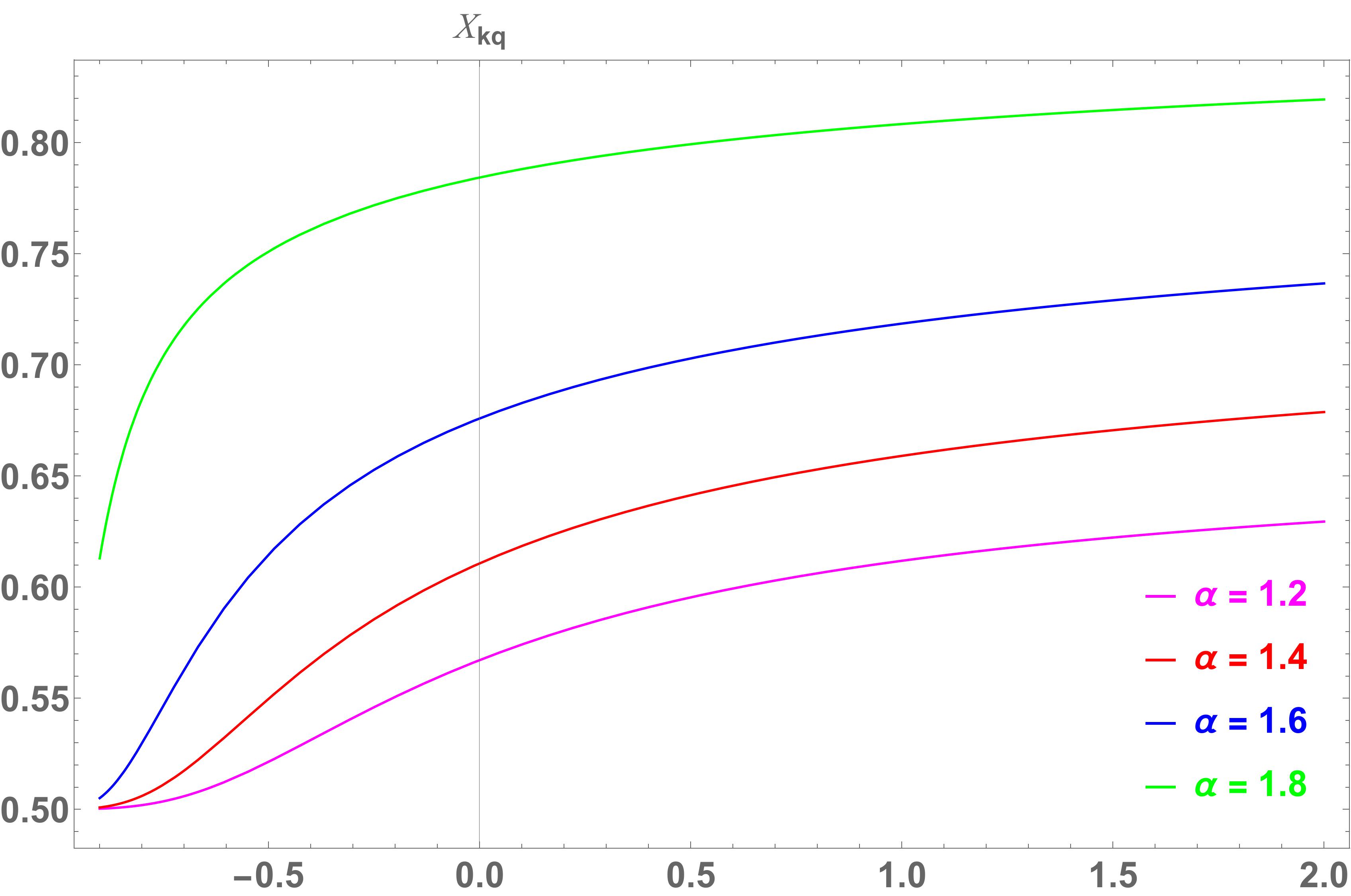}
  \caption{Redshift evolution of the K-essence kinetic energy \(X_{\text{kq}}\).}
  \label{kessencekinetic}
\end{minipage}
\hfill
\begin{minipage}[b]{0.47\textwidth}
        \centering
        \includegraphics[width=\textwidth]{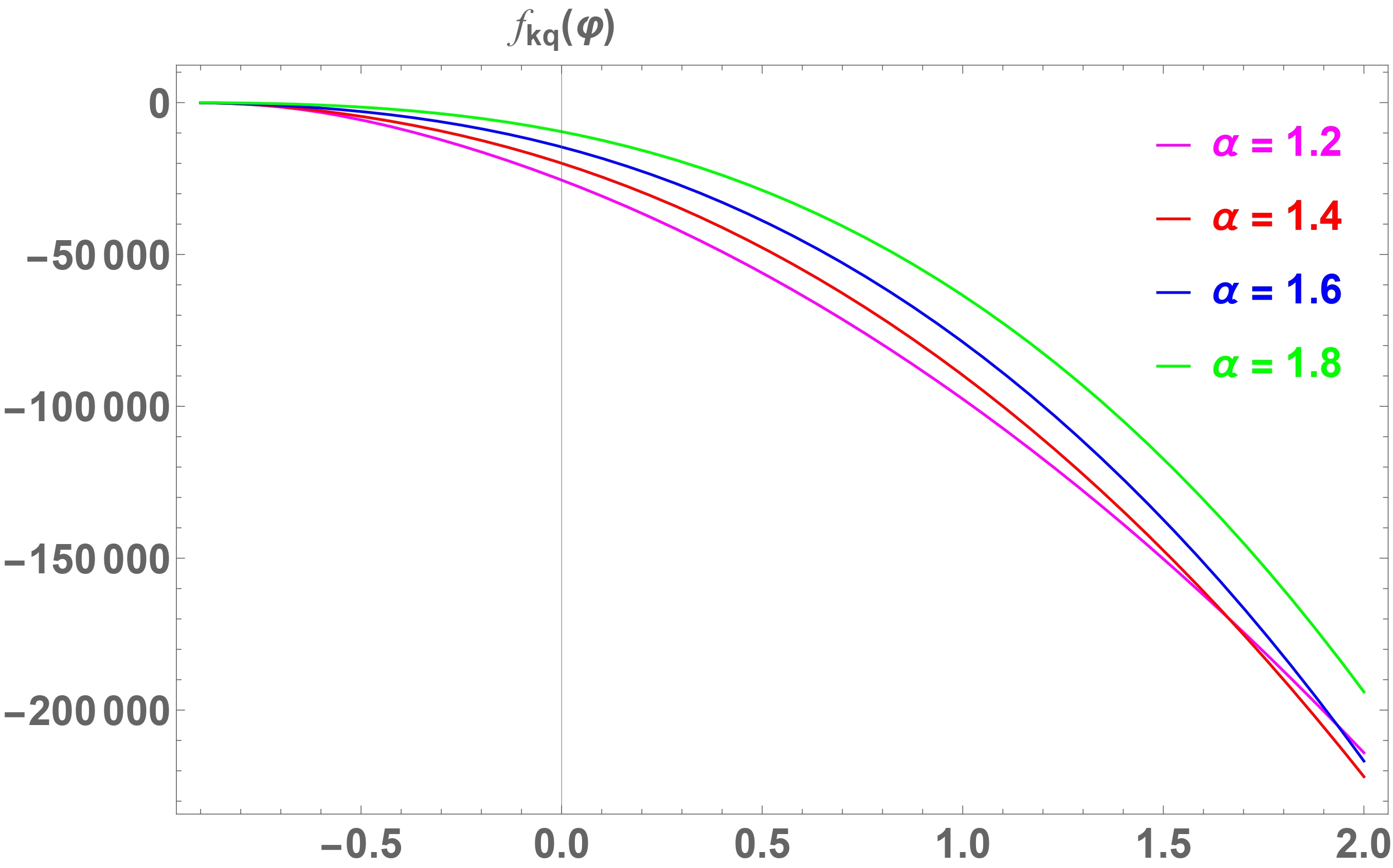}
  \caption{Redshift evolution of the K-essence coupling function
  \(f_{\text{kq}}(\varphi)\).}
  \label{kessencepotential}
\end{minipage}
\end{figure*}

Kinetic quintessence, or K-essence, generalises the standard quintessence scenario by allowing for a non-canonical kinetic term in the scalar-field Lagrangian. In contrast to quintessence, K-essence does not require an explicit scalar potential and is therefore less vulnerable to the quantum loop corrections that affect canonical models. The general K-essence Lagrangian may be written as
\begin{equation}\label{Lagrangian-kinetic}
    \mathcal{L}_{\text{kq}} = f_{\text{kq}}(\varphi)F(X) - V(\varphi).
\end{equation}
Standard quintessence is recovered by setting
\(f_{\text{kq}}(\varphi)=1\) and \(F(X)=X\). The principal motivation for K-essence is that late-time accelerated expansion --- and, in other contexts, inflation --- can be driven through non-standard kinetic structure alone, without the need for a potential. Following~\cite{Armend_riz_Pic_n_1999}, we adopt a general kinetic Lagrangian \(P(\varphi,X_{\text{kq}})\), depending on the scalar field \(\varphi\) and the kinetic variable
\(X_{\text{kq}}=-\dot{\varphi}_{\text{kq}}^{2}/2\). The corresponding action is
\begin{equation}\label{action-K}
    \mathcal{S}_{\text{kq}} = \int d^{4}x\sqrt{-g}\,
    P\!\left(\varphi, X_{\text{kq}}\right).
\end{equation}
The associated energy density and pressure are
\begin{equation}
    \rho_{\text{kq}}(\varphi, X_{\text{kq}}) =
    f_{\text{kq}}(\varphi)\left(-X_{\text{kq}} + 3X_{\text{kq}}^{2}\right),
    \qquad
    p_{\text{kq}}(\varphi, X_{\text{kq}}) =
    f_{\text{kq}}(\varphi)\left(-X_{\text{kq}} + X_{\text{kq}}^{2}\right),
\end{equation}
where \(f_{\text{kq}}(\varphi)\) specifies the scalar coupling. The EoS parameter is
\begin{equation}\label{EoS-K}
    w_{\text{kq}} = \frac{p_{\text{kq}}}{\rho_{\text{kq}}} =
    \frac{X_{\text{kq}} - 1}{3X_{\text{kq}} - 1}.
\end{equation}
Imposing the correspondence \(w^{(\text{fr})}_{\text{DE}}=w_{\text{kq}}\) together with Eq.~(\ref{EoS-K}) gives
\begin{equation}
    \frac{X_{\text{kq}} - 1}{3X_{\text{kq}} - 1} = -1 +
    \frac{(3\alpha-2)(1-\Omega^{(\text{fr})}_{\text{DE}})}
    {2\alpha - \Omega^{(\text{fr})}_{\text{DE}}(3\alpha-2)}.
\end{equation}
Solving algebraically yields the reconstructed FHDE expressions for the kinetic term and coupling function:
\begin{equation}\label{kq-field}
    X_{\text{kq}} = -\frac{\dot{\varphi}_{\text{kq}}^{2}}{2} =
    1 - \frac{2(\alpha-2)}{\alpha(1+3\Omega^{(\text{fr})}_{\text{DE}})
    - 2(\Omega^{(\text{fr})}_{\text{DE}}+3)},
\end{equation}
\begin{equation}
    f_{\text{kq}}(\varphi) = \frac{3H^{2}\Omega^{(\text{fr})}_{\text{DE}}
    \bigl(\alpha + 3\alpha\Omega^{(\text{fr})}_{\text{DE}} - 2(3+\Omega^{(\text{fr})}_{\text{DE}})\bigr)^{2}}
    {2\bigl(2+\alpha+2\Omega^{(\text{fr})}_{\text{DE}}-3\alpha\Omega^{(\text{fr})}_{\text{DE}}\bigr)
    \bigl(\alpha(3\Omega^{(\text{fr})}_{\text{DE}}-2)-2\Omega^{(\text{fr})}_{\text{DE}}\bigr)}.
\end{equation}
These determine the late-time behaviour of \(X_{\text{kq}}\) and \(f_{\text{kq}}(\varphi)\) within FHDE, illustrated in Figures (\ref{kessencekinetic}) and (\ref{kessencepotential}).

Figures (\ref{kessencekinetic}) and (\ref{kessencepotential}) show that \(X_{\text{kq}}\to1/2\) and \(f_{\text{kq}}(\varphi)\to0\) asymptotically as \(z\to-1\), for all values of \(\alpha\). The main features are as follows:
\begin{itemize}
    \item For larger values of \(\alpha\) (for example, \(\alpha=1.8\) and \(1.6\)), the kinetic term \(X_{\text{kq}}\) evolves more rapidly as redshift decreases, whereas smaller values such as \(\alpha=1.4\) and \(1.2\) yield a slower evolution.

    \item The coupling function \(f_{\text{kq}}(\varphi)\) rises from smaller values at high redshift and tends asymptotically toward zero. This evolution is slower for larger \(\alpha\), and significantly more rapid for smaller \(\alpha\) across the interval \(-1<z\leq2\).
\end{itemize}

The EoS parameter \(w_{\text{kq}}\) shows a behaviour qualitatively similar to that found in the quintessence case, as seen in Figure (\ref{quintEoS}). For smaller values of \(\alpha\), especially \(\alpha=1.2\) and \(1.4\), where fractional contributions are strongest, the EoS asymptotically approaches the \(\Lambda\)CDM value \(w_{\Lambda\text{CDM}}=-1\) in the DE-dominated regime.

As in the quintessence reconstruction, Figure (\ref{kessencekinetic}) indicates that larger values of \(\alpha\) are less suitable for the K-essence scenario, since they generate a kinetic term that evolves too rapidly to sustain an approximately constant DE component.

\subsubsection{$(iii)$ Dilaton}

\begin{figure*}[t]
\centering
\begin{minipage}[b]{0.47\textwidth}
        \centering
        \includegraphics[width=\textwidth]{ 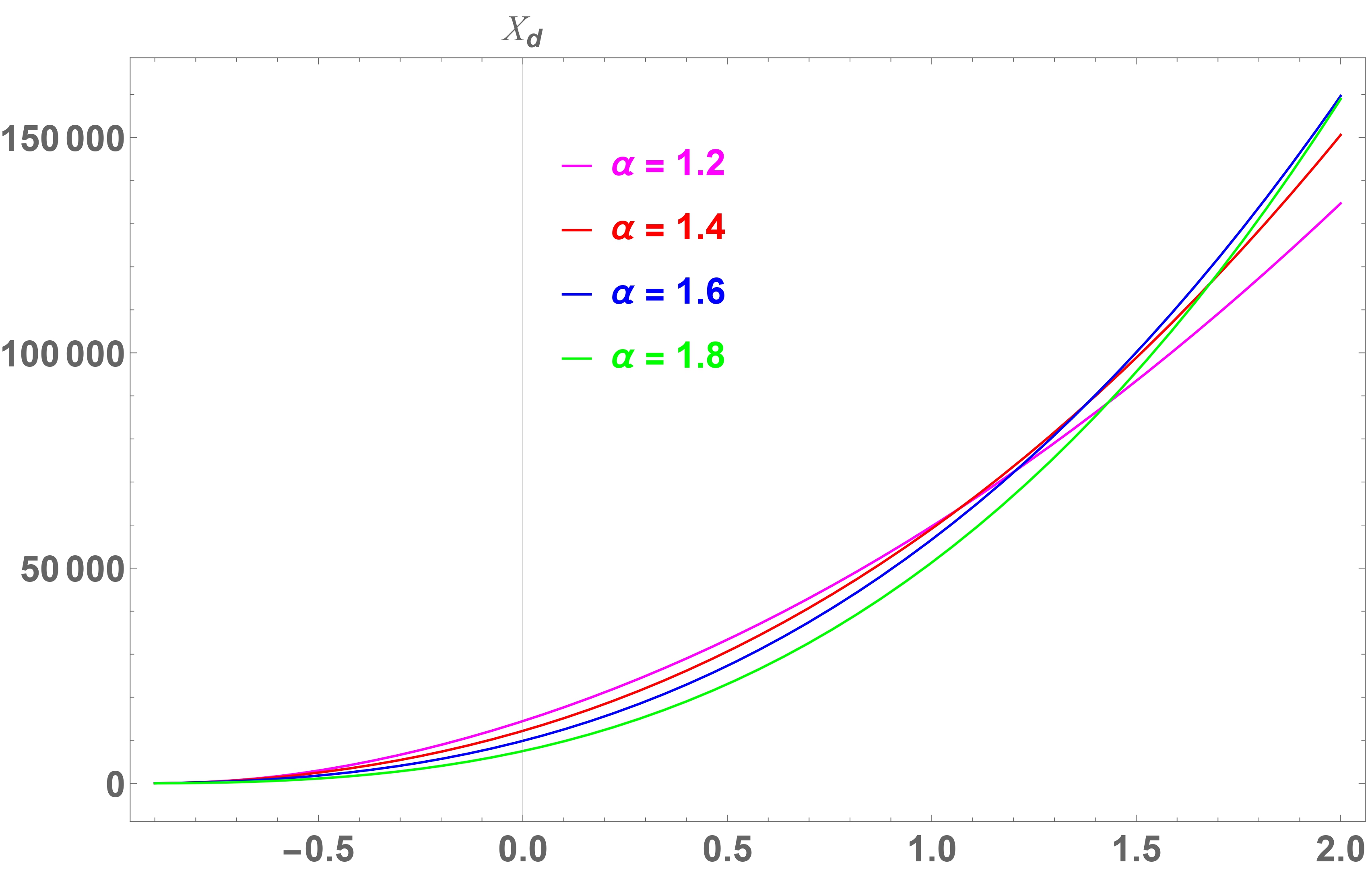}
  \caption{Redshift evolution of the dilaton kinetic energy \(X_{\text{d}}\).}
  \label{Figure 7}
\end{minipage}
\hfill
\begin{minipage}[b]{0.47\textwidth}
        \centering
        \includegraphics[width=\textwidth]{ 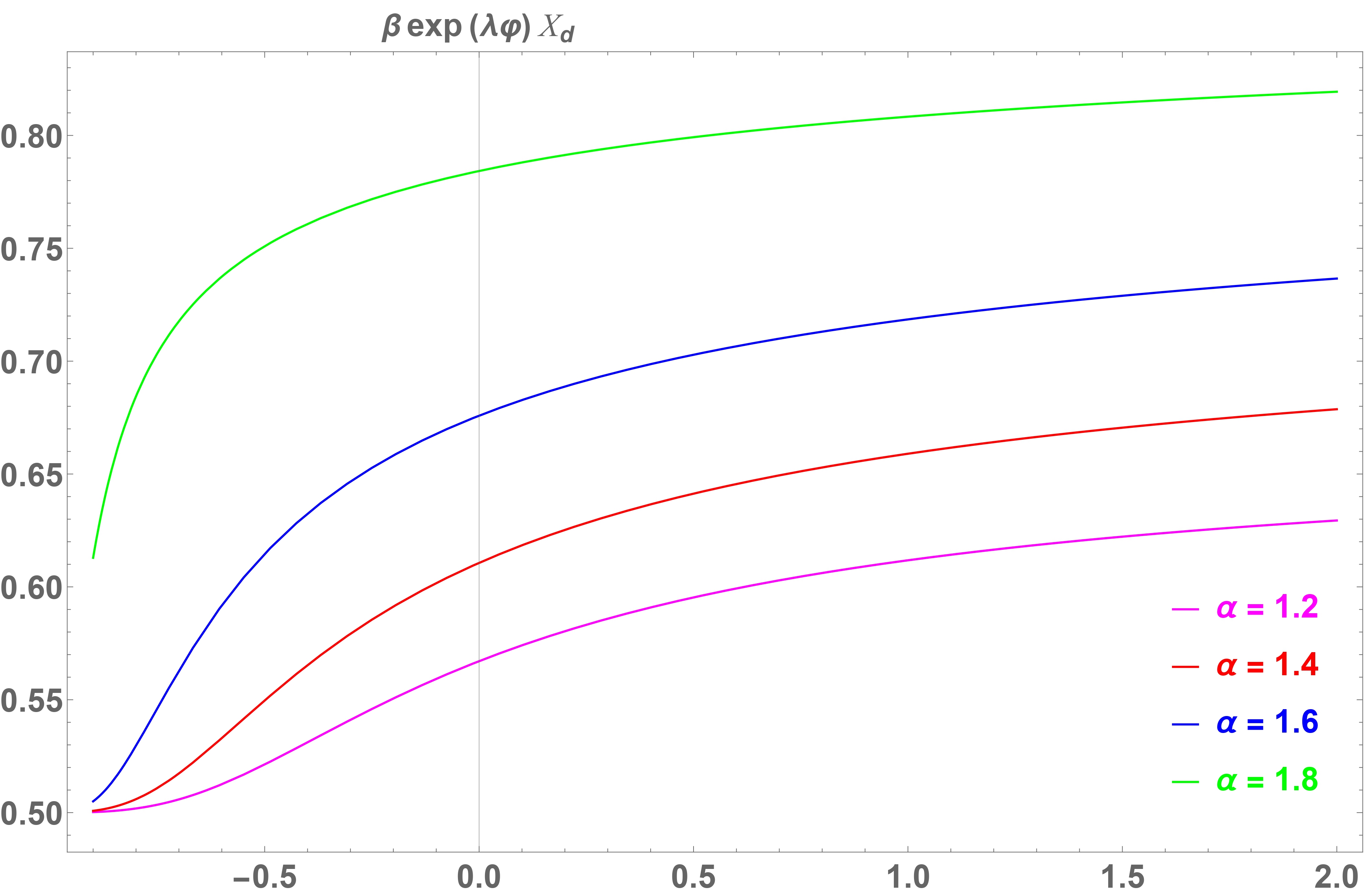}
  \caption{Redshift evolution of the dilaton exponential potential
  \(\beta\,e^{\lambda\varphi}X_{\text{d}}\).}
  \label{Figure 8}
\end{minipage}
\end{figure*}

The dilatonic ghost condensate is a string-theory-motivated DE model. In~\cite{Piazza_2004}, the dynamics of a scalar field with a negative kinetic term \(-X\) and a bell-shaped potential were analysed, showing that the dilaton evolves toward the potential maximum with an EoS satisfying \(w_{\text{d}}\leq-1\). Because this configuration is unstable under quantum fluctuations, it is stabilised by adding a term of the form \(e^{\lambda\varphi}X^{2}\). When the stability conditions are satisfied, the EoS is instead constrained by \(w_{\text{d}}\geq-1\), in contrast with phantom-field models~\cite{Carroll_2003}. The corresponding scalar-field Lagrangian is
\begin{equation}
    \mathcal{L}_{\text{d}} = \frac{1}{2}(\partial\varphi)^{2} +
    \frac{A}{m^{4}}(\partial\varphi)^{4}\exp\!\left(\frac{\lambda\varphi}{M_P}
    \right) + \text{higher-order terms}.
\end{equation}
Setting \(M_{pl}=1\) henceforth, the pressure is
\begin{equation}\label{pressure-D}
    p_{\text{d}}(X_{\text{d}}, \varphi) = -X_{\text{d}} +
    \beta\,e^{\lambda\varphi}X_{\text{d}}^{2},
\end{equation}
where \(\beta=A/m^{4}\). The energy-momentum tensor is
\begin{equation}
    T_{\mu\nu}^{\varphi} = g_{\mu\nu}p_{\text{d}} +
    \frac{\partial p_{\text{d}}}{\partial X}\partial_{\mu}\varphi
    \partial_{\nu}\varphi,
\end{equation}
which yields the energy density
\begin{equation}\label{dilaton-density}
    \rho_{\text{d}}(X_{\text{d}}, \varphi) = -X_{\text{d}} +
    3\beta\,e^{\lambda\varphi}X_{\text{d}}^{2},
\end{equation}
with \(X_{\text{d}}=\dot{\varphi}^{2}/2\). The dilaton EoS parameter then becomes
\begin{equation}\label{EoS-D}
    w_{\text{d}} = \frac{-1 + \beta\,e^{\lambda\varphi}X_{\text{d}}}
    {-1 + 3\beta\,e^{\lambda\varphi}X_{\text{d}}}.
\end{equation}
Imposing the FHDE correspondence \(w^{(\text{fr})}_{\text{DE}}\leftrightarrow w_{\text{d}}\) gives
\begin{equation}
    \frac{-1 + \beta\,e^{\lambda\varphi}X_{\text{d}}}
    {-1 + 3\beta\,e^{\lambda\varphi}X_{\text{d}}} = -1 +
    \frac{(3\alpha-2)(1-\Omega^{(\text{fr})}_{\text{DE}})}
    {2\alpha - \Omega^{(\text{fr})}_{\text{DE}}(3\alpha-2)}.
\end{equation}
Solving for the relevant quantities gives the reconstructed FHDE expressions
\begin{equation}\label{dilaton-kinetic}
    X_{\text{d}} = \frac{\dot{\varphi}^{2}}{2} =
    \frac{3H^{2}\Omega^{(\text{fr})}_{\text{DE}}\bigl(\alpha + 3\alpha\Omega^{(\text{fr})}_{\text{DE}}
    - 2(3+\Omega^{(\text{fr})}_{\text{DE}})\bigr)}{\alpha(6\Omega^{(\text{fr})}_{\text{DE}}-4)
    - 4\Omega^{(\text{fr})}_{\text{DE}}},
\end{equation}
\begin{equation}
    \beta\,e^{\lambda\varphi}X_{\text{d}} = 1 -
    \frac{2(\alpha-2)}{\alpha + 3\alpha\Omega^{(\text{fr})}_{\text{DE}}
    - 2(3+\Omega^{(\text{fr})}_{\text{DE}})}.
\end{equation}
These determine the redshift evolution of the kinetic term \(X_{\text{d}}\) and the exponential potential \(\beta\,e^{\lambda\varphi}X_{\text{d}}\), shown in Figures (\ref{Figure 7}) and (\ref{Figure 8}).

Figures (\ref{Figure 7}) and (\ref{Figure 8}) show that both \(X_{\text{d}}\) and \(\beta\,e^{\lambda\varphi}X_{\text{d}}\) evolve asymptotically, with
\(X_{\text{d}}\to0\) and \(\beta\,e^{\lambda\varphi}X_{\text{d}}\to1/2\) as \(z\to-1\). More specifically:
\begin{itemize}
    \item In the higher-redshift region \(1.5\leq z\leq2\), larger values of \(\alpha\) such as \(1.8\) and \(1.6\) produce a faster evolution of \(X_{\text{d}}\), whereas smaller values such as \(1.4\) and \(1.2\) yield a slower one. This trend reverses as redshift decreases toward \(z\to-1\), where larger \(\alpha\) leads to slower evolution.

    \item For the exponential potential shown in Figure~\ref{Figure 8}, larger values of \(\alpha\) produce a more rapid evolution over the entire interval \(-1<z\leq2\), while smaller values lead to a more gradual variation.
\end{itemize}

The EoS parameter \(w_{\text{d}}\) also approaches \(-1\) asymptotically in the far-future limit, in qualitative agreement with the quintessence and K-essence reconstructions. This convergence is again most pronounced for smaller \(\alpha\), where the fractional effects are strongest and the dilaton behaves most like an effective cosmological constant in the DE-dominated regime.

In summary, Figure (\ref{Figure 7}) suggests that larger values of \(\alpha\) are preferable for the dilaton kinetic term, since they lead to a sufficiently slow evolution compatible with nearly constant DE. By contrast, Figure (\ref{Figure 8}) favours smaller values of \(\alpha\), which yield a more suitable evolution of the exponential potential\footnote{The exponential potential \(\beta\,e^{\lambda\varphi}X_{\text{d}}\) is structurally similar to \(X_{\text{kq}}\), approaching \(1/2\) asymptotically as \(z\to-1\).} across the interval \(-1<z\leq2\).

\subsubsection{$(iv)$ DBI-essence}

\begin{figure*}[t]
\centering
\begin{minipage}[b]{0.47\textwidth}
        \centering
        \includegraphics[width=\textwidth]{ 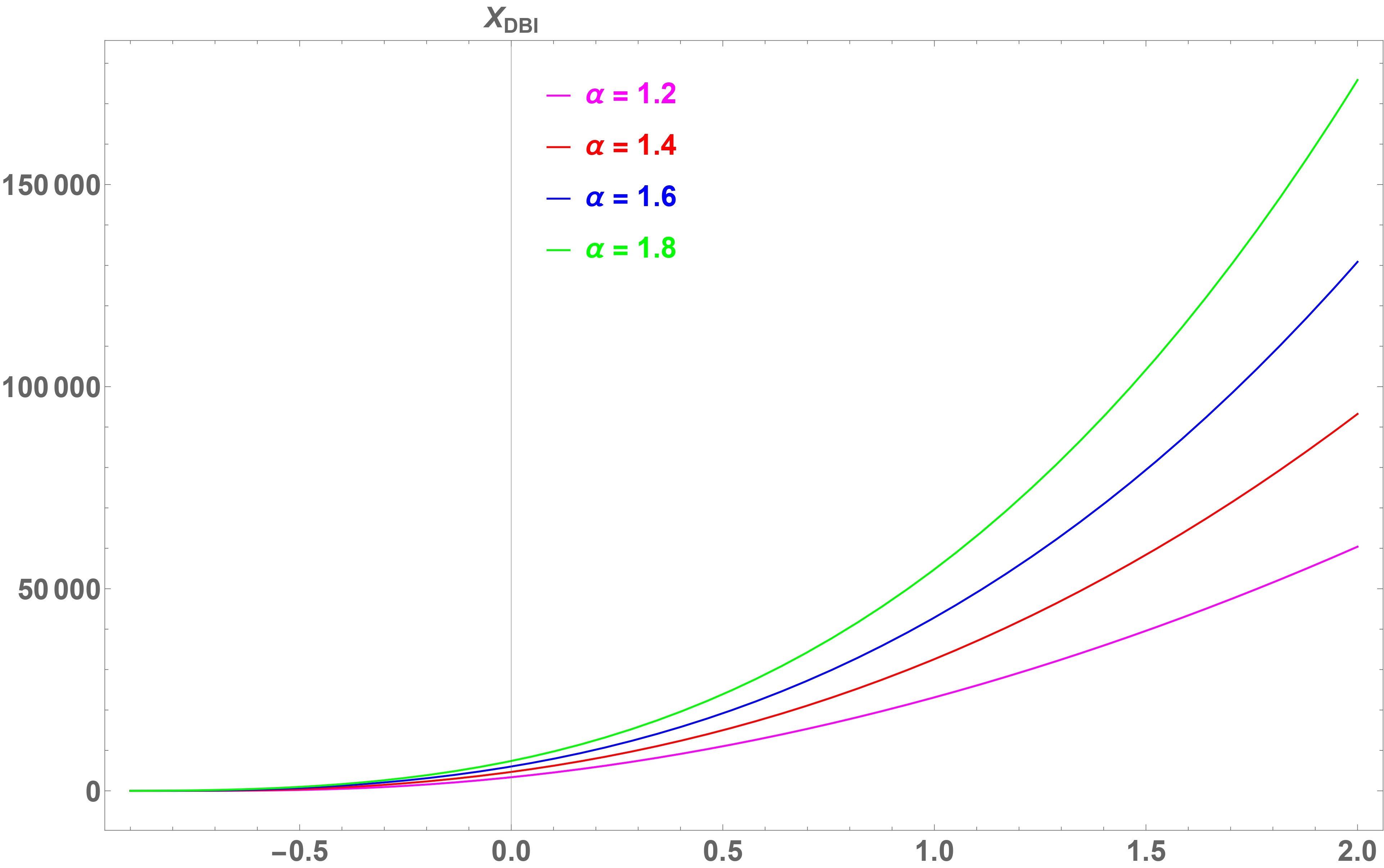}
  \caption{Redshift evolution of the DBI-essence kinetic energy \(X_{\text{DBI}}\).}
  \label{Figure 12}
\end{minipage}
\hfill
\begin{minipage}[b]{0.47\textwidth}
        \centering
        \includegraphics[width=\textwidth]{ 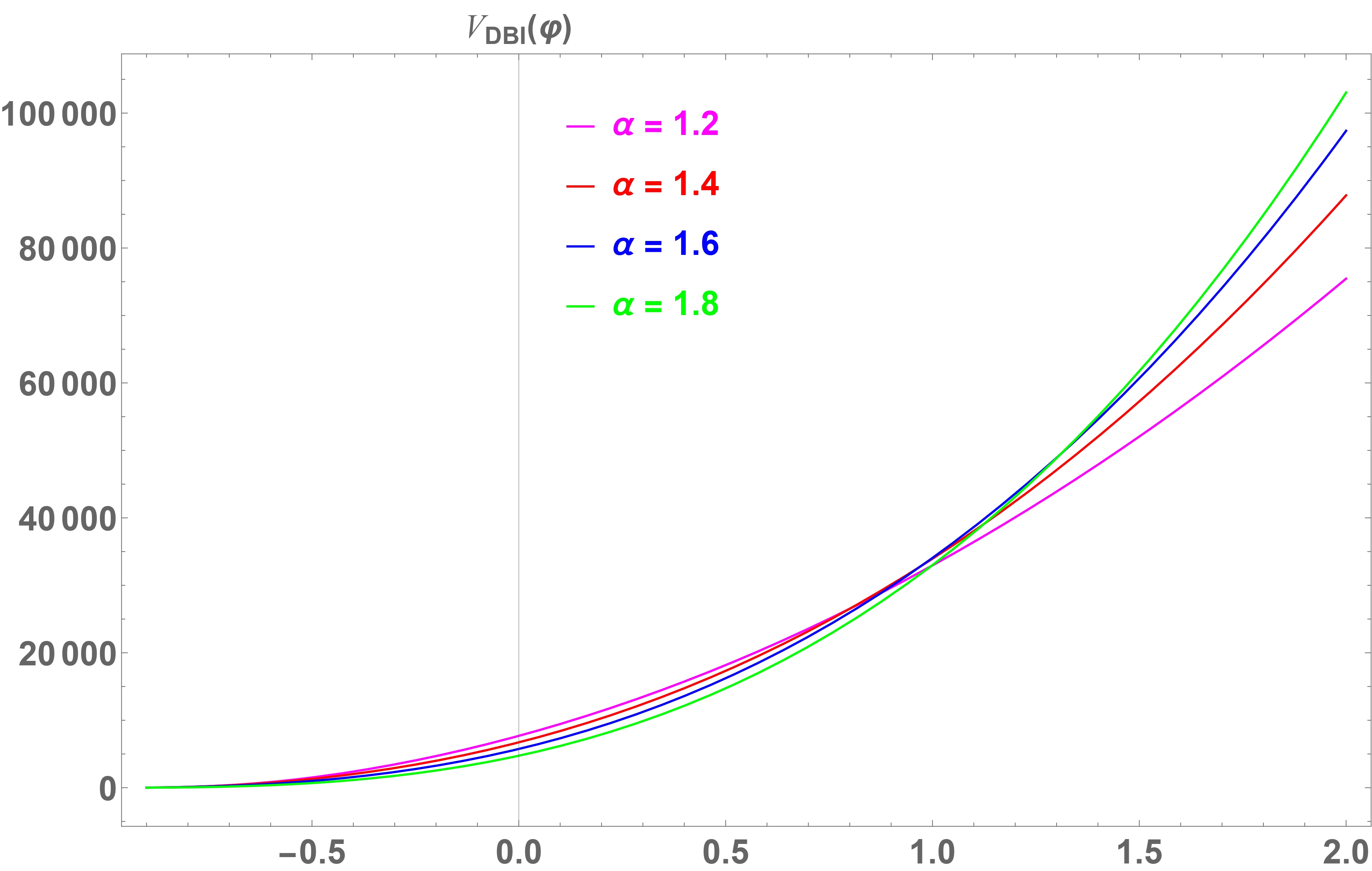}
  \caption{Redshift evolution of the DBI-essence potential \(V_{\text{DBI}}(\varphi)\).}
\label{Figure 13}
\end{minipage}
\hfill
\begin{minipage}[b]{0.47\textwidth}
        \centering
        \includegraphics[width=\textwidth]{ 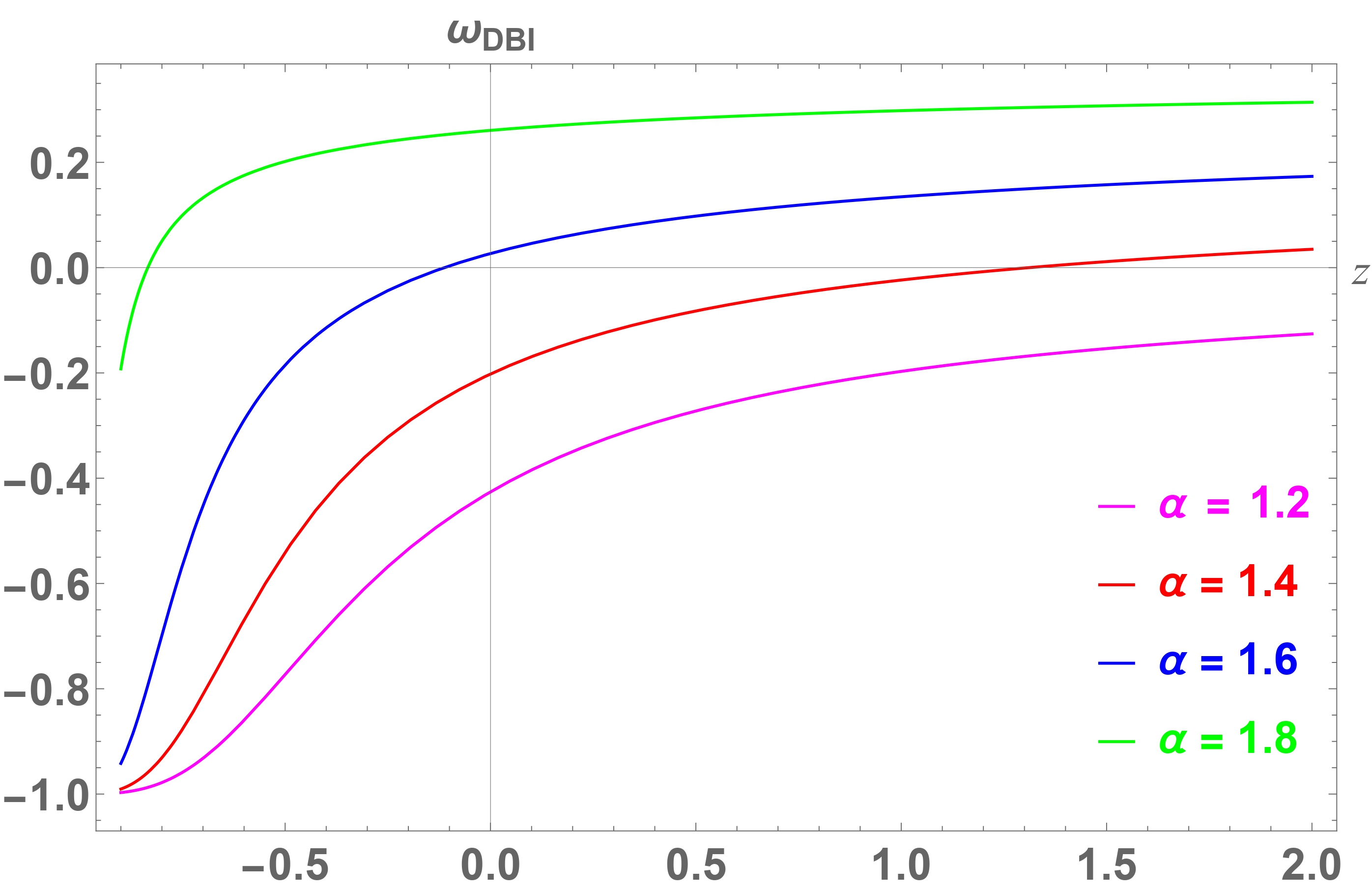}
  \caption{Redshift evolution of the DBI-essence EoS parameter
  \(w_{\text{DBI}}\).}
\label{Figure 14}
\end{minipage}
\end{figure*}

Recent developments in string theory have suggested scenarios in which the inflaton is identified with the separation between branes moving through extra dimensions within a warped throat geometry. This identification arises naturally because the corresponding action is proportional to the world-volume swept out by the moving brane. Since that volume is determined by the square root of the induced metric, a Dirac--Born--Infeld (DBI) kinetic term appears naturally, linking the construction to accelerated expansion in the early Universe; see~\cite{Ahn_2009} for details. Here we consider the DBI scalar field as a DE candidate, with action
\begin{equation}
    \mathcal{S}_{\text{DBI}} = -\int d^{4}x\; a^{3}(t)\left[T(\varphi)
    \sqrt{1 - \frac{\dot{\varphi}^{2}}{T(\varphi)}} + V(\varphi)
    - T(\varphi)\right],
\end{equation}
where \(T(\varphi)=n\dot{\varphi}^{2}\) denotes the warped brane tension, identified here with the kinetic energy \(X_{\text{DBI}}\), and \(V(\varphi)\) is the potential induced by interactions with Ramond--Ramond fluxes~\cite{Tong}. The energy density and pressure are
\begin{equation}
    \rho_{\text{DBI}} = (\eta - 1)T(\varphi) + V(\varphi), \qquad
    p_{\text{DBI}} = \left(\frac{\eta-1}{\eta}\right)T(\varphi) - V(\varphi),
\end{equation}
where \(\eta\) is the Lorentz boost factor,
\begin{equation}
    \eta = \sqrt{\frac{1}{1 - \dot{\varphi}^{2}/T(\varphi)}}.
\end{equation}
The corresponding DBI EoS parameter is
\begin{equation}
    w_{\text{DBI}} = \frac{(\eta-1)T(\varphi) - \eta V(\varphi)}
    {\eta(\eta-1)T(\varphi) + \eta V(\varphi)}.
\end{equation}
Imposing the FHDE correspondence \(w^{(\text{fr})}_{\text{DE}}\leftrightarrow w_{\text{DBI}}\) gives
\begin{equation}
    \frac{(\eta-1)T(\varphi) - \eta V(\varphi)}
    {\eta(\eta-1)T(\varphi) + \eta V(\varphi)} = -1 +
    \frac{(3\alpha-2)(1-\Omega^{(\text{fr})}_{\text{DE}})}
    {2\alpha - \Omega^{(\text{fr})}_{\text{DE}}(3\alpha-2)}.
\end{equation}
Solving this relation yields the reconstructed FHDE expressions for the kinetic term and potential:
\begin{equation}\label{DBI-field}
    X_{\text{DBI}} = n\dot{\varphi}_{\text{DBI}}^{2} =
    \frac{3(3\alpha-2)H^{2}(n-1)\sqrt{\dfrac{n}{n-1}}
    (1-\Omega^{(\text{fr})}_{\text{DE}})\Omega^{(\text{fr})}_{\text{DE}}}
    {2\alpha - (3\alpha-2)\Omega^{(\text{fr})}_{\text{DE}}},
\end{equation}
\begin{equation}\label{DBI-potential}
    V_{\text{DBI}}(\varphi) = 3H^{2}\Omega^{(\text{fr})}_{\text{DE}}\left[1 -
    \frac{(3\alpha-2)n\left(1-\sqrt{n-1}\right)(1-\Omega^{(\text{fr})}_{\text{DE}})}
    {2\alpha - (3\alpha-2)\Omega^{(\text{fr})}_{\text{DE}}}\right].
\end{equation}
Equations~(\ref{DBI-field}) and~(\ref{DBI-potential}) determine the late-time evolution of \(X_{\text{DBI}}\) and \(V_{\text{DBI}}(\varphi)\) within the FHDE framework. We set \(n=1.5\) throughout. The resulting evolution is shown in Figures (\ref{Figure 12}) and (\ref{Figure 13}).

Figures (\ref{Figure 12}) and (\ref{Figure 13}) show that both \(X_{\text{DBI}}\) and \(V_{\text{DBI}}(\varphi)\) decay asymptotically to zero as \(z\to-1\), for all values of \(\alpha\). The main features are:
\begin{itemize}
    \item For larger values of \(\alpha\), such as \(1.8\) and \(1.6\), the kinetic term \(X_{\text{DBI}}\) evolves more rapidly with decreasing redshift, whereas smaller values such as \(1.4\) and \(1.2\) yield a slower evolution.

    \item The potential \(V_{\text{DBI}}(\varphi)\) behaves qualitatively like the quintessence potential in Figure~\ref{quintpotential}, though here the behaviour is particularly clear over the interval \(-1<z\leq0.99\): larger \(\alpha\) leads to a slower decay of the potential and thereby favours a more sustained DE-dominated phase.
\end{itemize}

The EoS parameter \(w_{\text{DBI}}\), shown in Figure (\ref{Figure 14}), approaches \(-1\) asymptotically as fractional effects become more important, indicating that the DBI field increasingly mimics \(\Lambda\)CDM behaviour in the late-time DE regime. It is noteworthy that the present-day value of \(w_{\text{DBI}}\) at \(z=0\) differs from the corresponding values obtained in the quintessence, K-essence, dilaton, and YMC reconstructions.

Overall, Figure (\ref{Figure 12}) indicates that larger values of \(\alpha\) are disfavoured for the DBI kinetic sector, since they produce an overly rapid evolution that is difficult to reconcile with nearly constant DE. By contrast, Figure (\ref{Figure 13}) favours larger values of \(\alpha\) over the interval \(-1<z\leq0.99\), where the slower decay of the potential more effectively sustains late-time acceleration.

\subsubsection{$(v)$ Tachyon Field}

\begin{figure*}[t]
\centering
\begin{minipage}[b]{0.47\textwidth}
        \centering
        \includegraphics[width=\textwidth]{ 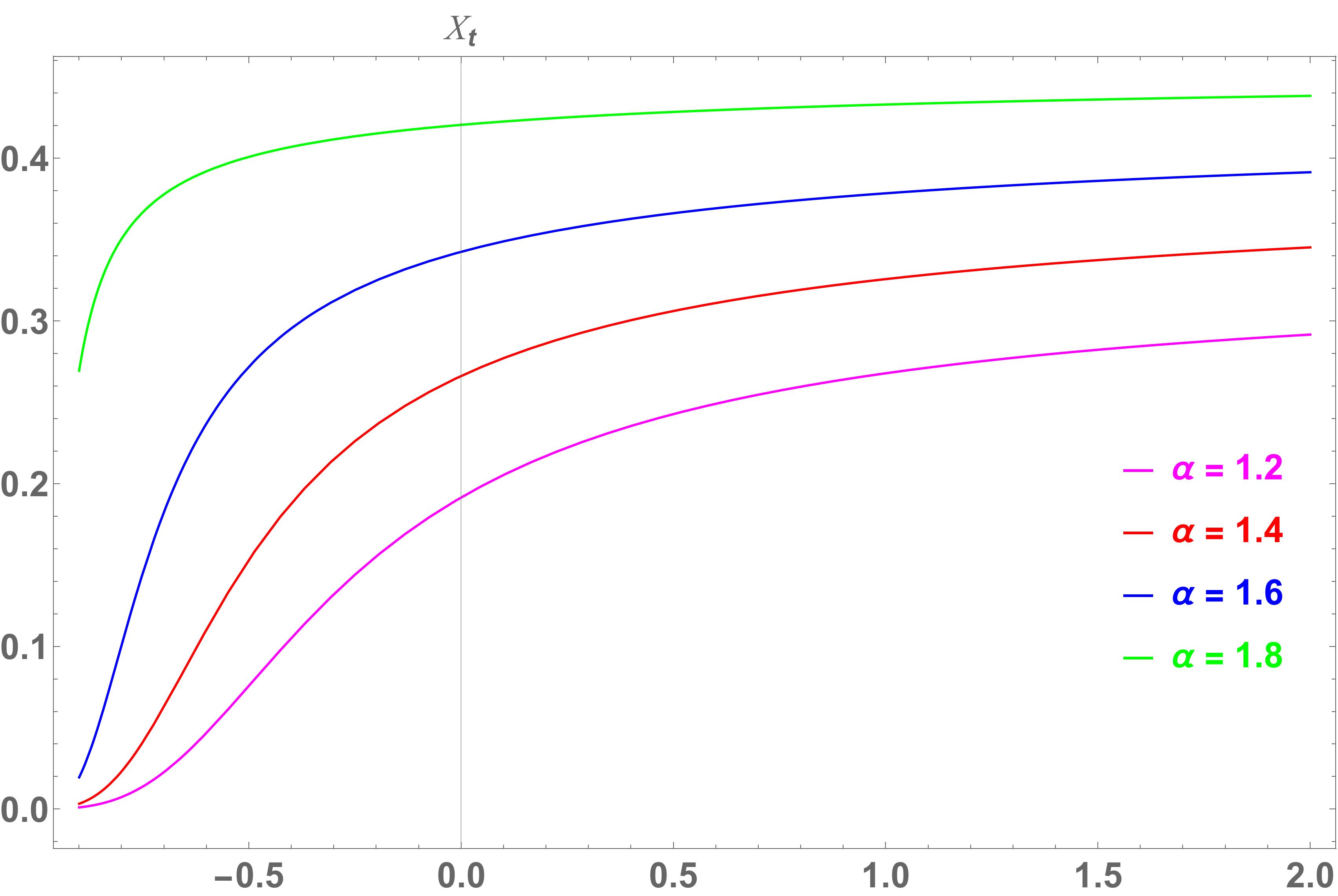}
  \caption{Redshift evolution of the tachyon kinetic energy \(X_{\text{t}}\).}
  \label{Figure 15}
\end{minipage}
\hfill
\begin{minipage}[b]{0.47\textwidth}
        \centering
        \includegraphics[width=\textwidth]{ 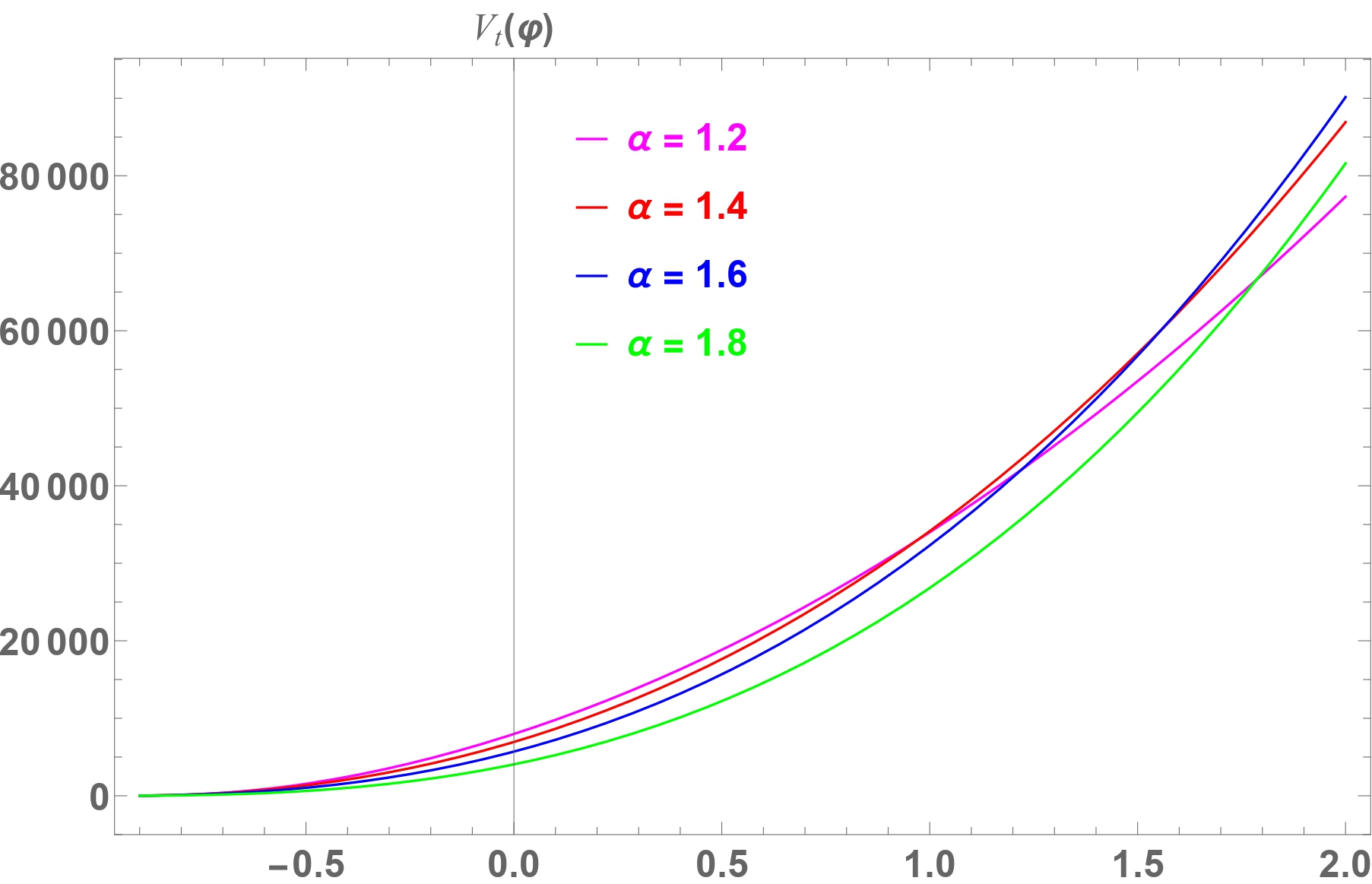}
  \caption{Redshift evolution of the tachyon potential \(V_{\text{t}}(\varphi)\).}
  \label{Figure 16}
\end{minipage}
\end{figure*}

The tachyon field is a well-motivated candidate for both inflation and DE at high energies~\cite{Mazumdar_2001,Feinstein_2002,Piao_2002,
gibbons2002cosmological}. The form of its self-interaction potential \(V(\varphi)\) plays a decisive role in determining whether the desired DE behaviour can be realised. By imposing a correspondence with FHDE, one obtains an explicit expression for the potential and can then study the resulting cosmological dynamics. The effective tachyon-field Lagrangian is
\begin{equation}\label{T-Lag}
    \mathcal{L}_{\text{t}} = -V_{\text{t}}(\varphi)
    \sqrt{1 - g^{\mu\nu}\partial_{\mu}\varphi\partial_{\nu}\varphi}.
\end{equation}
The corresponding energy-momentum tensor takes the perfect-fluid form
\begin{equation}
    T_{\mu\nu}^{\varphi} = (p_{\text{t}} + \rho_{\text{t}})u_{\mu}u_{\nu}
    + p_{\text{t}}g_{\mu\nu},
\end{equation}
with energy density and pressure
\begin{equation}\label{pressureanddensity}
    p_{\text{t}}(t) = -V_{\text{t}}(\varphi)
    \sqrt{1 - \dot{\varphi}_{\text{t}}^{2}}, \qquad
    \rho_{\text{t}}(t) = \frac{V_{\text{t}}(\varphi)}
    {\sqrt{1 - \dot{\varphi}_{\text{t}}^{2}}}.
\end{equation}
The tachyon EoS parameter is
\begin{equation}\label{EoS-T}
    w_{\text{t}} = \dot{\varphi}_{\text{t}}^{2} - 1.
\end{equation}
The FHDE correspondence \(w^{(\text{fr})}_{\text{DE}}\leftrightarrow w_{\text{t}}\) therefore gives
\begin{equation}
    \dot{\varphi}_{\text{t}}^{2} - 1 = -1 +
    \frac{(3\alpha-2)(1-\Omega^{(\text{fr})}_{\text{DE}})}
    {2\alpha - \Omega^{(\text{fr})}_{\text{DE}}(3\alpha-2)},
\end{equation}
which yields the reconstructed FHDE expressions
\begin{equation}\label{t-field}
    X_{\text{t}} = \frac{\dot{\varphi}_{\text{t}}^{2}}{2} =
    \frac{(3\alpha-2)(1-\Omega^{(\text{fr})}_{\text{DE}})}
    {4\alpha - \Omega^{(\text{fr})}_{\text{DE}}(6\alpha-4)},
\end{equation}
\begin{equation}
    V_{\text{t}}(\varphi) = 3H^{2}\Omega^{(\text{fr})}_{\text{DE}}\left[1 -
    \frac{(3\alpha-2)(1-\Omega^{(\text{fr})}_{\text{DE}})}
    {2\alpha - \Omega^{(\text{fr})}_{\text{DE}}(3\alpha-2)}\right]^{\frac{1}{2}}.
\end{equation}
The corresponding cosmological evolution is displayed in Figures (\ref{Figure 15}) and (\ref{Figure 16}).

Both \(X_{\text{t}}\) and \(V_{\text{t}}(\varphi)\) approach zero asymptotically as \(z\to-1\), for all values of \(\alpha\). More specifically:
\begin{itemize}
    \item For larger values of \(\alpha\), such as \(1.8\) and \(1.6\), the kinetic term \(X_{\text{t}}\) evolves more rapidly as redshift decreases, whereas smaller values such as \(1.4\) and \(1.2\) give a correspondingly slower evolution across \(-1\leq z\leq2\).

    \item The potential \(V_{\text{t}}(\varphi)\) shows the opposite trend, as seen in Figure~\ref{Figure 16}: larger \(\alpha\) produces a slower and more gradual decay, while smaller \(\alpha\) leads to more rapid variation over the interval \(-1<z\leq1.8\).
\end{itemize}

The EoS parameter \(w_{\text{t}}\) also approaches \(-1\) asymptotically as fractional effects become stronger, indicating that the FHDE-reconstructed tachyon field tends toward \(\Lambda\)CDM-like behaviour during the DE-dominated phase. This is qualitatively consistent with the quintessence reconstruction shown in Figure~\ref{quintEoS}.

In summary, Figure (\ref{Figure 15}) disfavors larger values of \(\alpha\) for the tachyon kinetic term, since they produce an evolution that is too dynamical to support nearly constant DE. Conversely, Figure (\ref{Figure 16}) favours larger values of \(\alpha\) over the interval \(-1<z\leq1.8\), where the potential decays more slowly and better maintains a DE-dominated phase.

This completes the analysis of bosonic and string-inspired effective field configurations as candidates for dynamical DE within FHDE. We now turn to a gauge-field candidate.

\subsection{Gauge Bosonic Field Candidate}
\subsubsection{$(i)$ Yang--Mills Condensate}

The Yang--Mills condensate (YMC) provides a geometrically distinct approach to DE based on gauge-boson fields in a cosmological background. There are two main reasons for considering YMC alongside the scalar-field models. First, the effective Yang--Mills Lagrangian is fixed by quantum field theory, so the only free parameter is the characteristic energy scale; ad hoc modifications of the effective Lagrangian are therefore not allowed. Second, in scalar-field models such as quintessence, the EoS is typically restricted to the range \(-1<w_{\text{q}}<1\), and achieving \(w_{\text{q}}<-1\) requires the introduction of a phantom field, which is generically unstable at the quantum level. The YMC framework avoids this problem; see~\cite{Zhao_2006,
zhao2009quantumyangmillscondensatedark} for detailed discussions.

\begin{figure*}[t]
\centering
\begin{minipage}[b]{0.47\textwidth}
        \centering
        \includegraphics[width=\textwidth]{ 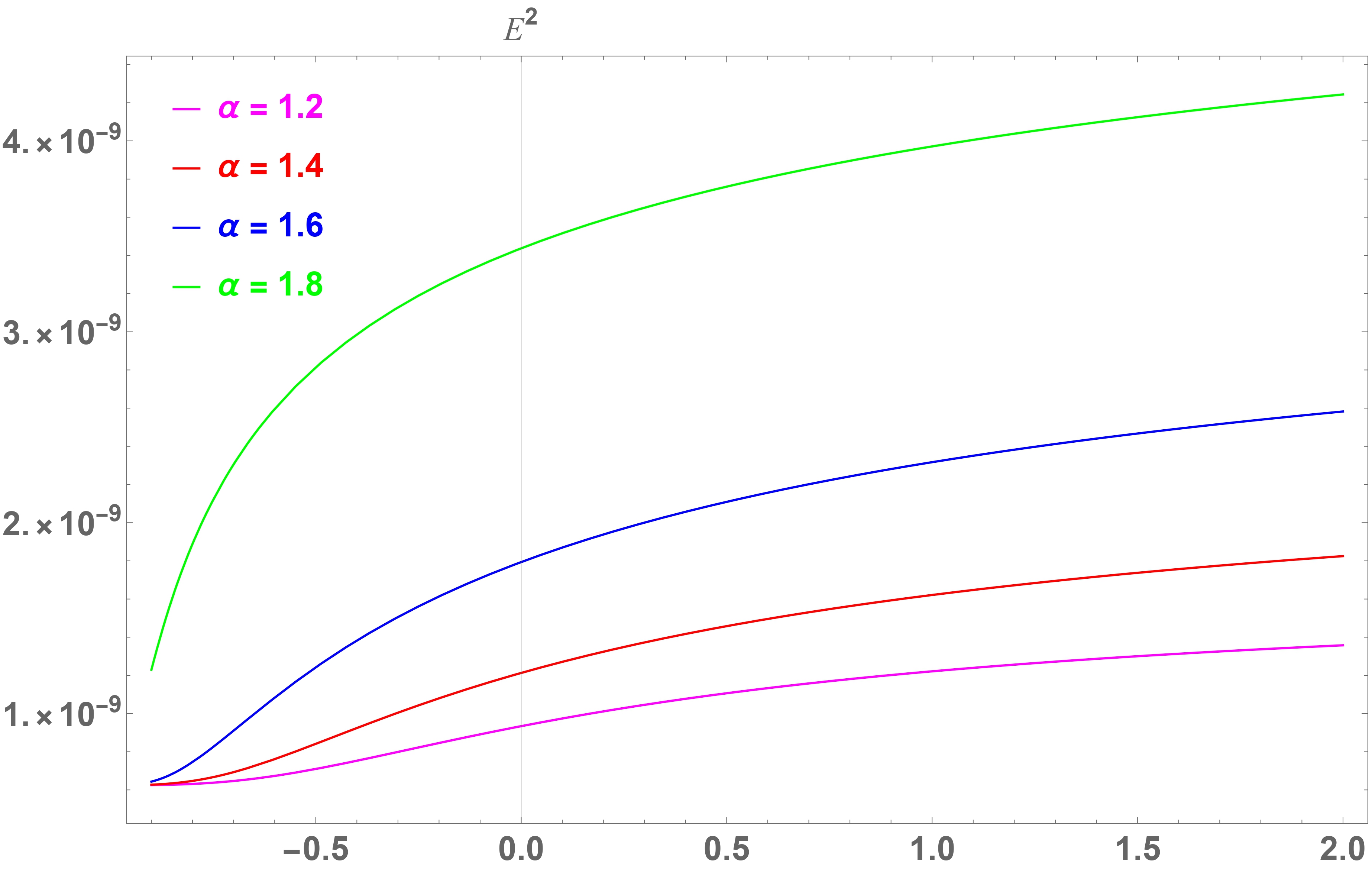}
  \caption{Redshift evolution of the YMC electric field squared \(E^{2}\).}
  \label{Figure 10}
\end{minipage}
\hfill
\begin{minipage}[b]{0.47\textwidth}
        \centering
        \includegraphics[width=\textwidth]{ 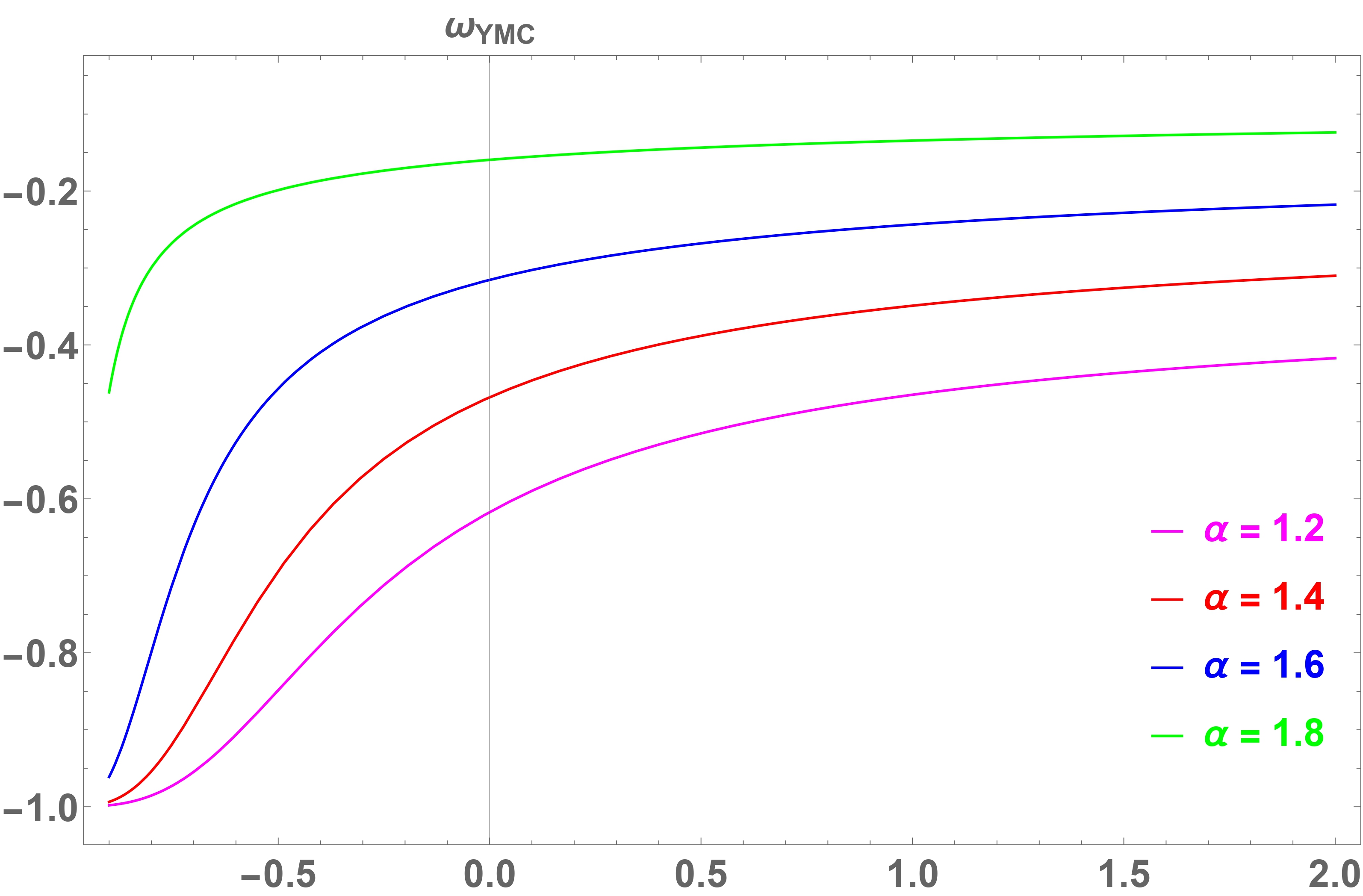}
  \caption{Redshift evolution of the YMC EoS parameter \(w_{\text{YMC}}\).}
  \label{Figure 11}
\end{minipage}
\end{figure*}

In the effective YMC DE model, the Lagrangian is~\cite{Zhao_2006,PhysRevD.23.2905,ZHao1_2006}
\begin{equation}\label{Lag-YMC}
    \mathcal{L}_{\text{YMC}} = \frac{b}{2}F\left(\ln\left|\frac{F}{\kappa^{2}}
    \right| - 1\right),
\end{equation}
where \(\kappa\) is the renormalisation scale with dimensions of squared mass,
\(b\) is the Callan--Symanzik coefficient \(b=(11N-2N_{f})/24\pi^{2}\) for
\(SU(N)\) with \(N_{f}\) quark flavours, and \(F\equiv-\frac{1}{2}F^{a}_{\mu\nu}
F^{a\mu\nu}\) is the order parameter of the condensate. The index \(a\) runs over the gauge group, with \(a=1,2,3\) for \(SU(2)\) and \(a=1,\ldots,8\) for \(SU(3)\). Assuming that the Universe contains only YMC minimally coupled to gravity, the effective action is
\begin{equation}
    \mathcal{S}_{\text{YMC}} = \int d^{4}x\sqrt{-g}\left[\frac{\mathcal{R}}
    {16\pi G} + \frac{b}{2}F\left(\ln\left|\frac{F}{\kappa^{2}}\right| - 1
    \right)\right].
\end{equation}
Variation with respect to \(g^{\mu\nu}\) yields the Einstein equation
\(G_{\mu\nu}=8\pi G T_{\mu\nu}\), with stress tensor
\begin{equation}\label{Stress-Energy-YM}
    T_{\mu\nu} = \sum_{a=1}^{3}\left[\frac{g_{\mu\nu}}{4g^{2}}
    F^{a}_{\sigma\delta}F^{a\sigma\delta} + \epsilon
    F^{a}_{\mu\sigma}F^{a\sigma}{}_{\nu}\right].
\end{equation}
We restrict the gauge fields to depend on time only,\footnote{Suitable ans\"atze for the gauge fields are discussed in detail in~\cite{PVMoniz_1991,PVMoniz_1993}.} writing
\(A_{\mu}=\frac{i}{2}\sigma_{a}A^{a}_{\mu}(t)\), where \(\sigma_{a}\) are the Pauli matrices, with \(A_{0}=0\) and \(A^{a}_{i}=\delta^{a}_{i}(t)\). The Yang--Mills field-strength tensor is
\begin{equation}\label{Field Strength Tensor}
    F^{a}_{\mu\nu} = \partial_{\mu}A^{a}_{\nu} - \partial_{\nu}A^{a}_{\mu}
    + gf^{abc}A^{b}_{\mu}A^{c}_{\nu},
\end{equation}
where \(f^{abc}\) are the structure constants of the gauge group. The dielectric constant is defined by \(\epsilon = 2\partial\mathcal{L}_{\text{YMC}}/\partial F\), and it is convenient to introduce the dimensionless variable
\begin{equation}\label{dimensionless}
    y = \frac{\epsilon}{b} = \ln\left|\frac{F}{\kappa^{2}}\right|.
\end{equation}
Using the stress tensor in Eq.~(\ref{Stress-Energy-YM}), the energy density and pressure are
\begin{equation}
    \rho_{\text{YMC}} = \frac{1}{2}\epsilon(E^{2}+B^{2}) +
    \frac{1}{2}b(E^{2}-B^{2}), \qquad
    p_{\text{YMC}} = \frac{1}{6}\epsilon(E^{2}+B^{2}) -
    \frac{1}{2}b(E^{2}-B^{2}).
\end{equation}
In an expanding Universe, the magnetic part of the Yang--Mills field decays much more rapidly, so that the late-time field becomes effectively electric~\cite{ZHAO_2007}. Under this approximation, the energy density and pressure reduce to
\begin{equation}
    \rho_{\text{YMC}} = \frac{E^{2}}{2}(\epsilon + b), \qquad
    p_{\text{YMC}} = \frac{E^{2}}{2}\left(\frac{\epsilon}{3} - b\right).
\end{equation}
Substituting the dimensionless variable in Eq.~(\ref{dimensionless}), one finds
\begin{equation}
    \rho_{\text{YMC}} = \frac{1}{2}b\kappa^{2}(y+1)e^{y}, \qquad
    p_{\text{YMC}} = \frac{1}{6}b\kappa^{2}(y-3)e^{y}.
\end{equation}
The YMC EoS parameter is therefore
\begin{equation}
    w_{\text{YMC}} = \frac{y-3}{3(y+1)}.
\end{equation}
Imposing the FHDE correspondence \(w^{(\text{fr})}_{\text{DE}}\leftrightarrow w_{\text{YMC}}\) gives
\begin{equation}
    \frac{y-3}{3(y+1)} = -1 + \frac{(3\alpha-2)(1-\Omega^{(\text{fr})}_{\text{DE}})}
    {2\alpha - \Omega^{(\text{fr})}_{\text{DE}}(3\alpha-2)},
\end{equation}
which yields
\begin{equation}\label{YMC-field}
    E^{2} = \kappa^{2}\exp\!\left(3 - \frac{12(\alpha-2)}
    {\alpha + 3\alpha\Omega^{(\text{fr})}_{\text{DE}} - 2(3+\Omega^{(\text{fr})}_{\text{DE}})}\right).
\end{equation}
Equation~(\ref{YMC-field}) describes the late-time evolution of the YMC configuration, which in this approximation is purely electric. This candidate is structurally distinct from the scalar-field models, since it is governed by a spin-1 gauge field in a cosmological background. The evolution of \(E^{2}\) and the corresponding EoS parameter is displayed in Figures (\ref{Figure 10}) and (\ref{Figure 11}).

Figures (\ref{Figure 10}) and (\ref{Figure 11}) show the following:
\begin{itemize}
    \item The electric field squared \(E^{2}\) decays asymptotically to zero as \(z\to-1\). For larger values of \(\alpha\), such as \(1.8\) and \(1.6\), the evolution is more rapid, whereas smaller values such as \(1.4\) and \(1.2\) lead to a slower and more gradual decay over \(-1<z\leq2\).

    \item The EoS parameter \(w_{\text{YMC}}\) approaches \(-1\) asymptotically as fractional effects become stronger, indicating that the YMC field increasingly mimics \(\Lambda\)CDM behaviour in the DE-dominated era. This is qualitatively consistent with the behaviour found for quintessence in Figure~\ref{quintEoS}.
\end{itemize}

For the gauge-field sector, Figure (\ref{Figure 10}) likewise indicates that larger values of \(\alpha\) are less favoured, since the corresponding rapid evolution of \(E^{2}\) is difficult to reconcile with a slowly varying DE component.

This completes the reconstruction of FHDE in terms of several spin-0 and one spin-1 effective field configurations. All of the candidates considered here display promising dynamical behaviour that can, in principle, be constrained by current and forthcoming observations~\cite{desicollaboration2024desi2024vicosmological,DESI:2025zgx,adame2024desi3}. In this framework, the observational burden falls primarily on a single parameter, namely the L\'evy index \(\alpha\), whose fractional origin may encode effects such as non-locality or memory and thereby provide a consistent mechanism for the dynamical nature of DE.

\section{Conclusion}

The central aim of this work has been to advance a research programme exploring the interplay between the HP and FC~\cite{FC,Miller1993AnIT,Grigoletto2013FractionalVO}. This framework, termed FHDE, was introduced in~\cite{Trivedi:2024inb}, where several concrete applications were also presented. The guiding question throughout the present study has been whether FHDE can be regarded as a realistic and physically well-motivated candidate for explaining the late-time accelerated expansion of the Universe. To address this question, we systematically examined a class of effective field frameworks capable of reproducing the dynamical features of the FHDE model established in~\cite{Trivedi:2024inb}.

In this chapter, we have developed a new HDE scenario, which we refer to as FHDE. This terminology reflects the fact that the model is motivated by corrections arising from FC and fractional quantum mechanics, thereby importing ideas from a non-trivial interface between mathematics and physics into gravitational and cosmological settings. The defining feature of the model is a modified HDE density that smoothly reduces to the standard HDE expression in the limit \(\alpha \to 2\). Here, the parameter \(\alpha\) encodes the fractional-calculus structure that emerges in fractional classical and quantum mechanics, as well as its physical implications.

More specifically, the FHDE energy density was derived from the holographic inequality after incorporating corrections to the Bekenstein--Hawking entropy through the FWDW equation. In this sense, the model provides a natural extension of conventional HDE while opening new directions for investigation. We then studied the cosmological evolution of the model by adopting the Hubble horizon as the IR cutoff. The density parameters for DM and DE, together with the deceleration parameter and the DE EoS parameter, were analysed in detail in order to characterise the cosmic evolution implied by the model. The resulting behaviour is particularly encouraging, since FHDE yields a viable late-time cosmology even with the Hubble horizon cutoff, despite the well-known difficulties of this choice in conventional HDE. In particular, we showed that observationally acceptable evolution is obtained for smaller values of \(\alpha\). We also examined further aspects of the model by studying its classical stability through the squared sound speed and by analysing the \(w^{(\text{fr})}_{\text{DE}}-w'^{(\text{fr})}_{\text{DE}}\) plane over an extended redshift interval. Taken together, these results indicate that fractional effects can provide a consistent HDE scenario even in the case of the Hubble horizon cutoff.

Several natural extensions of the present work suggest themselves. For instance, it would be worthwhile to investigate the model using the future event horizon or the particle horizon rather than the Hubble horizon. One may also consider more sophisticated infrared cutoffs, such as the Granda--Oliveros or generalised Nojiri--Odintsov prescriptions, and study the corresponding FHDE dynamics. Further possibilities include extending the model to non-GR cosmologies and improving observational constraints on the parameter \(\alpha\) using data from BBN, BAO, cosmic chronometers, and related probes; see, for example,~\cite{Hernandez-Almada:2021aiw,Barrow:2020kug,Anagnostopoulos:2020ctz}.

Our analysis of late-time cosmological evolution proceeded along two complementary lines. On the one hand, we employed effective field configurations built from both spin-\(0\) and spin-\(1\) degrees of freedom. On the other hand, we incorporated specific ingredients of FC that have already found applications in several gravitational contexts~\cite{Garcia-Aspeitia:2022uxz,
Micolta-Riascos:2023mqo,LeonTorres:2023ehd,Calcagni:2009kc}.

Two closely related questions shaped the discussion. First, can FHDE~\cite{Trivedi:2024inb} be established as a promising theoretical framework through suitable effective field realisations? Second, can one construct a consistent correspondence between such field frameworks and FHDE to obtain new insights into dynamical DE? Adopting the Hubble horizon as the infrared cutoff, we investigated the following effective field configurations: \((i)\) Quintessence, \((ii)\) Kinetic Quintessence (K-essence), \((iii)\) Dilaton, \((iv)\) Yang--Mills condensate, \((v)\) Dirac--Born--Infeld essence (DBI-essence), and \((vi)\) the Tachyon scalar field. For each model, we reconstructed two quantities: \((a)\) the kinetic term \(X_{i}\), and \((b)\) the potential \(V_{i}(\varphi)\), as functions of redshift, where \(i\) labels the effective field configuration. These quantities, together with the EoS parameter, were plotted for \(\alpha=1.2,1.4,1.6,\) and \(1.8\). The kinetic term, potential, and EoS parameter together control the cosmological behaviour of the model~\cite{copeland2006dynamics,Barreiro:1999zs,
bahamonde2018dynamical}: the potential determines the expansion dynamics and scaling properties, while the EoS parameter determines the late-time behaviour and its proximity to \(\Lambda\)CDM. In the models studied here, this limiting behaviour was realised through \(w(z)\to -1\) as \(z\to -1\). Our intention was not to replace a broader observational analysis, but rather to focus on the field variables most directly relevant for understanding the cosmological content of FHDE.

The main outcomes of the FHDE reconstruction across the chosen effective field configurations may be summarised as follows.

\paragraph{Kinetic energy \(X_{i}\)}

\begin{itemize}
    \item A common pattern emerges in Quintessence (Figure~\ref{Figure 1}), K-essence (Figure~\ref{Figure 4}), DBI-essence (Figure~\ref{Figure 12}), and the Tachyon field (Figure~\ref{Figure 15}): over the interval \( -1 < z \leq 2 \), a phenomenologically suitable evolution of \(X_{i}\) is obtained for \textit{smaller} values of \(\alpha\), such as \(\alpha=1.2\). In these cases, the kinetic term varies more gradually as redshift decreases, indicating that the field evolves toward stabilisation in a controlled manner. This is accompanied by the asymptotic decay of the corresponding potential \(V_{i}(\varphi)\) to zero as \(z\to -1\) in the DE-dominated regime.

    \item The Dilaton case (Figure~\ref{Figure 7}) provides a clear exception. Here, \textit{larger} values of \(\alpha\), such as \(\alpha=1.8\) and \(1.6\), produce a less dynamical evolution for \(z\lesssim1.4\), making them more favourable over the interval \( -1 < z \leq 1.5 \). This behaviour contrasts with that of the other field configurations, for which smaller values of \(\alpha\) are generally preferred as the Universe approaches DE domination in the asymptotic future.

    \item For the gauge-field configuration, namely YMC (Figure~\ref{Figure 10}), \textit{smaller} values of \(\alpha\) again yield a more appropriate evolution of the electric field component \(E^{2}\) as redshift decreases.
\end{itemize}

\paragraph{Potential \(V_{i}(\varphi)\) and coupling function \(f_{\text{kq}}(\varphi)\)}

\begin{itemize}
    \item Quintessence (Figure~\ref{Figure 2}), DBI-essence (Figure~\ref{Figure 13}), the Tachyon field (Figure~\ref{Figure 16}), and the K-essence coupling function \(f_{\text{kq}}(\varphi)\) (Figure~\ref{Figure 5}) display a common qualitative trend: the potential evolves less dynamically for \textit{larger} values of \(\alpha\), such as \(\alpha=1.8\) and \(1.6\), below certain redshift thresholds. For Quintessence this occurs for \(z\lesssim0.55\), for DBI-essence at \(z\lesssim0.99\), and for the Tachyon at \(z\lesssim1.8\). In the K-essence case, the coupling function evolves more slowly for \(\alpha=1.8\) throughout the entire range \( -1 < z \leq 2 \), without the crossover behaviour present in the other cases.

    \item The Dilaton potential \(\beta e^{\lambda\varphi}X_{\text{d}}\) (Figure~\ref{Figure 8}) behaves in the opposite manner: \textit{smaller} values of \(\alpha\), such as \(\alpha=1.2\) and \(1.4\), yield the less dynamical evolution. In this respect, the Dilaton differs from all the other reconstructed field configurations. Its potential nevertheless tends asymptotically toward \(\Lambda\)CDM-like behaviour as \(z\to -1\).
\end{itemize}

\paragraph{EoS parameter}

\begin{itemize}
    \item For Quintessence, K-essence, the Dilaton, YMC, and the Tachyon field, \textit{smaller} values of \(\alpha\), particularly \(\alpha=1.2\) and \(1.4\), drive the EoS parameter asymptotically toward \(w_{i}(z)\to -1\) in the far-future limit \(z\to -1\). In particular, for \(\alpha=1.2\), the present-day value of the EoS parameter at \(z=0\) is in good agreement with recent DESI constraints~\cite{desicollaboration2024desi2024vicosmological,
    adame2024desi1,adame2024DESI,DESI:2025zgx}.

    \item In the DBI-essence case (Figure~\ref{Figure 14}), the value of the EoS parameter at \(z=0\) is slightly larger than in the other reconstructed models, representing a mild but noteworthy deviation.
\end{itemize}

In closing, we have successfully reconstructed the kinetic and potential sectors of several DE field configurations within the fractional holographic framework of~\cite{Trivedi:2024inb}. Our results indicate that fractional modifications help prevent crossing into the phantom regime \(w(z)<-1\), while keeping the evolution within the quintessence domain. Smaller values of the fractional parameter \(\alpha\) consistently produce a cosmological evolution that approaches \(\Lambda\)CDM in the far-future limit \(z\to -1\). For the string-inspired configurations --- Dilaton, DBI-essence, and Tachyon --- the EoS parameter also supports viable late-time acceleration for small \(\alpha\), suggesting that fractional modifications can supply an effective dynamical description of DE in such settings. At the same time, since string theory has not yet been directly confirmed experimentally, the precise role of these models in DE physics remains open. Quintessence, K-essence, and Yang--Mills condensate appear especially promising, since they provide a viable and potentially testable framework when confronted with observations from CMB anisotropies, BAOs, and supernova data.

Building on the results obtained in~\cite{Trivedi:2024inb}, we regard the FHDE scenario as a viable framework worthy of further investigation, and several directions naturally follow.

A particularly important open question is whether FC leaves a distinctive observational imprint on DE physics. More concretely, one may ask whether there exists a late-time signature of \(\alpha\neq2\) that cannot be reproduced by other mechanisms. One possible avenue is the effect of FHDE on the Integrated Sachs--Wolfe (ISW) effect, which is sensitive to time-dependent gravitational potentials in the late Universe. A fractional DE component of the FHDE type may therefore induce measurable deviations in CMB--LSS cross-correlations, thereby providing an observational probe complementary to SNe Ia constraints.

Further work could also sharpen the observational constraints required to distinguish between the different effective field configurations reconstructed within FHDE. Additional directions include the dynamical stability of FHDE in a non-flat universe (\(k\neq0\)), interactions between DM and DE, and connections with modified gravity frameworks such as \(f(Q,C)\) and \(f(Q,\mathcal{L}_{m})\). In this connection, it is worth noting that a dynamical analysis of Barrow HDE models in~\cite{samaddar2024barrowholographicdarkenergy} --- including interacting and non-interacting sectors, as well as a Bianchi-type universe --- found that these models may asymptote toward \(\Lambda\)CDM-like late-time evolution, with \(w(z)\to -1\). This is qualitatively consistent with the behaviour found here for configurations \((i)\)--\((vi)\). A systematic dynamical-systems analysis of FHDE in both interacting and non-interacting settings, together with direct comparison to Barrow HDE and Tsallis HDE, would therefore help clarify both the similarities and the distinctive features of these models. The use of more general infrared cutoffs, such as the Granda--Oliveros and Nojiri--Odintsov prescriptions, also deserves detailed investigation within the FHDE framework.

\chapter{More Topics in Fractional Holographic Dark Energy Cosmology}\label{Chapter4}

Within the context of this chapter, we investigate whether future rip scenarios, such as the big rip, little rip, and pseudo rip, can arise with a specific choice of cutoff within the FHDE framework. As discussed previously, the standard HDE scenario is recovered in the limit $\alpha \to 2$, i.e., when fractional features progressively diminish. Our primary interest, however, lies in extracting novel insights from late-time cosmology 
in the regime where fractional effects dominate, i.e., for $\alpha \simeq 1.1$ or 
more generally throughout the range $1 < \alpha \leq 2$. In recent literature, 
the authors of~\cite{TrivediScherrer_2024,Brevik:2024ozg} presented intriguing results on future rip singularities within the conventional, Barrow, and Tsallis HDE frameworks, employing an ansatz-based approach in which the Hubble parameter $H(t)$ is prescribed so as to give rise to little and pseudo rip behaviours. In our theoretical investigation, we will focus only on the occurrence or avoidance of rip-like singularities. In the present chapter, we extend the investigations of~\cite{Trivedirip_2024} 
and~\cite{TrivediScherrer_2024} on rip scenarios to the FHDE setting, with particular emphasis on the non-local features induced by FC and their role in either facilitating or suppressing the occurrence of these late-time events.

\section{Late--Time Cosmic Singularities}\label{late-timecosmic}

A substantial body of literature has been devoted to the exploration of cosmological singularities that may arise in the far future of the 
Universe~\cite{Albarran_2016,Borislavov_Vasilev_2021,Bouhmadi-Lopez:2017ckh,
D_browski_2006,Albarran:2018mpg,Albarran:2015tga}, a programme that has been considerably propelled by the observational evidence for late-time 
acceleration. A particularly compelling class of such far-future events is the family of rip scenarios, in which the Universe undergoes progressive disintegration in various forms. Within the FHDE framework, two natural questions serve as our motivation: $(i)$ \textit{Which rip scenarios 
can occur within the FHDE model?} and $(ii)$ \textit{Can the non-local features inherent to the FHDE framework suppress or altogether prevent certain rip singularities?} By ``non-local features,'' we refer to the memory effect 
encoded in the fractional derivative, which influences the cosmological evolution of DE at late times. Within the FHDE framework and a broad cosmological perspective, the memory kernel of the fractional derivative parametrised by the L\'{e}vy index $\alpha$ controls how strongly past cosmic epochs affect the present behaviour of DE~\cite{Landim_2021,Rasouli_2024,Micolta-Riascos:2023mqo}. This is particularly relevant for rip singularities, since non-local terms can either amplify or inhibit singularity formation depending on how memory accumulates as the Universe approaches a finite-time future event. A recent study~\cite{TrivediScherrer_2024} examined all three rip singularities in the context of non-fractional HDE models, employing the GO cutoff~\cite{Granda_2008}. The principal finding is that the big and pseudo rip are the generic outcomes, while the little rip arises only for a very special class of IR cutoffs. For completeness, we recall the defining features of each rip scenario:

\begin{enumerate}
    \item \textbf{Big Rip (Type~I Singularity).} At a finite future time 
    $t \to t_{\text{finite}}$, the effective energy density and pressure 
    diverge, $\rho_{\text{eff}} \to \infty$ and $p_{\text{eff}} \to -\infty$, 
    and the Hubble parameter diverges, $H(t) \to \infty$~\cite{Caldwell_2003}. 
    The corresponding Hubble parameter ansatz is
    \begin{equation}\label{big rip ansatz}
        H(t) \approx \frac{H_{0}}{(t_{\text{rip}} - t)^{m}},
    \end{equation}
    where $m > 0$ is an arbitrary constant; throughout this paper we set 
    $m = 1$. In the approach to the big rip, the DE EoS parameter 
    remains phantom, satisfying $w < -1$ with a constant value or tending 
    to a limit strictly below $-1$ at finite time.

    \item \textbf{Little Rip.} In this scenario, the effective energy 
    density, pressure, and Hubble parameter all diverge, but only 
    asymptotically as $t \to \infty$. The little rip is therefore not a 
    finite-time singularity, in sharp contrast to the big 
    rip~\cite{Frampton_2011}. The Hubble parameter evolves as
    \begin{equation}\label{little rip ansatz}
        H(t) \approx H_{0}\exp(\lambda t).
    \end{equation}
    The DE EoS parameter remains phantom throughout, with the 
    energy density and Hubble parameter diverging only in the infinite-time 
    limit.

    \item \textbf{Pseudo Rip.} The Hubble parameter increases monotonically 
    but remains bounded above, approaching a finite asymptotic value 
    $H_{\infty} \equiv H(t \to \infty)$, so that $H(t) \to H_{\infty}$ as 
    $t \to \infty$. Bound structures are only partially disrupted before the 
    Universe asymptotes to a de~Sitter phase~\cite{Frampton_2012}. The 
    Hubble parameter takes the form
    \begin{equation}\label{pseudo rip ansatz}
        H(t) \approx H_{0} - H_{1}\exp(-\lambda t).
    \end{equation}
    The DE EoS parameter satisfies $w_{\text{DE}}(t) < -1$ 
    throughout the entire cosmic history, but approaches $-1$ from below at late times, $w_{\text{DE}}(t) \to -1^{-}$, so that the expansion asymptotes to de~Sitter rather than diverging. By construction, the  pseudo-rip occupies an intermediate position between a pure cosmological 
    constant ($w_{\text{DE}} = -1$) and a little rip ($w_{\text{DE}} < -1$ 
    with $\rho^{(\text{fr})}_{\text{DE}} \to +\infty$ as $t \to \infty$)~\cite{Frampton_2012}.
\end{enumerate}

The remainder of this section is organised as follows. In Section (\ref{Section 2}), we review the cosmological setup required for the interacting FHDE model. In Section (\ref{Rips}), we investigate whether all three rip scenarios --- big, little, and pseudo rip --- arise naturally within the interacting FHDE setting with the GO cutoff, and identify the range of L\'{e}vy index values $\alpha$ for which each singularity can occur. 
In Section (\ref{littleandpseudorip}), we adopt an ansatz-based approach, employing a well-known Hubble-parameter ansatz from the literature to characterise the EoS and SSS for the little and pseudo-rip within the Hubble cutoff.

\subsection{Cosmological Set-Up}\label{Section 2}

In this section, we establish the key expressions for the cosmological 
parameters that will play a central role in our investigation of future rip 
scenarios within the FHDE framework. We work throughout within the interacting 
dark sector. The starting point is the Friedmann equation for a flat ($k = 0$) 
FLRW Universe with an interacting dark sector. Defining the fractional density 
parameters $\Omega^{(\text{fr})}_{\text{DE}} = \rho^{(\text{fr})}_{\text{DE}}/3H^{2}$ 
and $\Omega^{(\text{fr})}_{\text{DM}} = \rho^{(\text{fr})}_{\text{DM}}/3H^{2}$, 
the Friedmann constraint reads
\begin{equation}\label{friedmann1}
    \Omega^{(\text{fr})}_{\text{DE}} + \Omega^{(\text{fr})}_{\text{DM}} = 1,
\end{equation}
The FHDE DE density parameter and the corresponding DE and DM density parameter are given by
\begin{equation}
    \Omega^{(\text{fr})}_{\text{DE}} = c^{2}
    \frac{L^{\frac{2-3\alpha}{\alpha}}_{\text{IR}}}{H^{2}}, \qquad 
    \Omega^{(\text{fr})}_{\text{DM}} = 1 - c^{2}
    \frac{L^{\frac{2-3\alpha}{\alpha}}_{\text{IR}}}{H^{2}}.
\end{equation}

We also require the continuity equations governing the interacting dark sector, 
in which energy exchange between DE and DM is encoded through 
an interaction term $Q$. The inclusion of this coupling is motivated by the 
question of whether, and to what extent, the interaction between the dark 
components --- via Eq.~(\ref{contd}) for DE and Eq.~(\ref{contm}) 
for DM --- can amplify or otherwise modify the occurrence of rip 
singularities.\footnote{For the impatient reader: the $Q$ term plays a 
particularly decisive role in producing a big rip within the GO cutoff 
framework, as will be seen in Section (\ref{littleandpseudorip}).} We consider two 
distinct forms for the interaction term:

\begin{enumerate}
    \item \textbf{Linear interaction.} The interaction term takes the form
    \begin{equation}\label{linear Q}
        Q = 9H^{3}\tilde{\beta}\left(\Omega^{(\text{fr})}_{\text{DE}} 
        + \Omega^{(\text{fr})}_{\text{DM}}\right).
    \end{equation}

    \item \textbf{Non-linear interaction.} The interaction term takes the form
    \begin{equation}\label{nonlinear Q}
        Q = 3H\tilde{\beta}\left(\frac{\Omega^{(\text{fr})}_{\text{DE}}}
        {1 - \Omega^{(\text{fr})}_{\text{DE}}}\right).
    \end{equation}
\end{enumerate}

Here, $\tilde{\beta}$ is the coupling constant governing the strength of the DE--DM interaction. It is observationally constrained to be 
very small, $\tilde{\beta} \approx 0$, but we will demonstrate in Sections (\ref{littleandpseudorip}) that even small variations in its value can have a significant influence on the occurrence of rip scenarios.

To complete the setup, we write the time derivative of the FHDE energy 
density (see Eq.~(\ref{fracrho})) as
\begin{equation}
    \dot{\rho}^{(\text{fr})}_{\text{DE}} = \rho^{(\text{fr})}_{\text{DE}}
    \left(\frac{2-3\alpha}{\alpha}\right)\frac{\dot{L}_{\text{IR}}}{L_{\text{IR}}}.
\end{equation}
Substituting this into the DE continuity equation~(\ref{contd}), 
one obtains the EoS parameter $w_{\text{DE}}(t)$ for a general IR cutoff $L_{\text{IR}}$:
\begin{equation}\label{general EoS}
    w_{\text{de}}(t) = -1 - \frac{1}{3H\Omega^{(\text{fr})}_{\text{DE}}}
    \left[\frac{Q}{3H^{2}} + \Omega^{(\text{fr})}_{\text{DE}}
    \left(\frac{2-3\alpha}{\alpha}\right)\frac{\dot{L}_{\text{IR}}}{L_{\text{IR}}}\right].
\end{equation}
The squared sound speed parameter is correspondingly given by
\begin{equation}\label{general vs}
    v^{2}_{s}(t) = w_{\text{de}} + \dot{w}_{\text{de}}
    \left(\frac{\alpha}{2-3\alpha}\right)\frac{L_{\text{IR}}}{\dot{L}_{\text{IR}}}.
\end{equation}
The subsequent sections investigate all three rip scenarios --- big, little, 
and pseudo rip --- as natural consequences of phantom DE domination 
at late times, supported by numerical plots of the EoS and squared sound 
speed parameters defined in Eqs.~(\ref{general EoS}) and~(\ref{general vs}) 
for both the GO and Hubble cutoffs.

\subsection{Rip Scenarios in Granda-Oliveros Cutoff}\label{Rips}

During DE domination, i.e., when the Universe transitions into the 
late-time accelerated expansion phase, the Friedmann equation takes the form
\begin{equation}\label{Friedmann11}
    H^{2} = \frac{\rho^{(\text{fr})}_{\text{DE}}}{3} = c^{2}L^{\frac{2-3\alpha}{\alpha}}_{\text{IR}}.
\end{equation}

In this section, the IR cutoff is chosen so as to permit a comprehensive investigation of all rip scenarios. The appropriate choice is the GO cutoff, which includes the time derivative $\dot{H}$, thereby enabling an analytic treatment of the rip conditions. It is worth recalling that, in general, HDE models do not admit the standard 
parametrisations discussed in~\cite{Frampton_2011,Frampton_2012}, because the Friedmann equation does not take the standard perfect-fluid form in these settings. This necessitates a distinct approach to the analysis of both little and big rip scenarios.

In HDE models, the choice we need to make is the IR cutoff, because some cutoff definitions can lead to late-time cosmic singularities, while others may not. For preliminary investigations, we use the GO cutoff. The corresponding FHDE energy density within the GO cutoff is written as
\begin{equation}
    \rho_{\text{GO}} = 3c^{2}\left(\gamma H^{2} + \delta\dot{H}
    \right)^{\frac{3\alpha-2}{2\alpha}},
\end{equation}
and the associated fractional density parameter 
$\Omega_{\text{GO}} = \rho_{\text{GO}}/3H^{2}$ is
\begin{equation}
    \Omega_{\text{GO}} = \frac{c^{2}\left(\gamma H^{2} + \delta\dot{H}
    \right)^{\frac{3\alpha-2}{2\alpha}}}{H^{2}}.
\end{equation}
Substituting the GO cutoff description into the Friedmann equation~(\ref{Friedmann}), one obtains the evolution equation for the 
Hubble parameter:
\begin{equation}\label{Hdot}
    \dot{H} = \frac{1}{\delta}\left[\left(\frac{H}{c}\right)^{n} 
    - \gamma H^{2}\right], \qquad n = \frac{4\alpha}{3\alpha-2},
\end{equation}
which integrates to
\begin{equation}\label{Hint}
    \int_{H_{i}}^{H_{f}}\frac{\delta}{\left(H/c\right)^{n} - \gamma H^{2}}
    \,dH = \int_{t_{i}}^{t_{f}}dt.
\end{equation}
Here $c$, $\gamma$, and $\delta$ are positive constants. 
Equation~(\ref{Hdot}) admits two branches: $\dot{H} > 0$ and $\dot{H} < 0$. 
The negative branch does not give rise to any rip singularity and is 
discarded. We therefore focus exclusively on the positive branch 
$\dot{H} > 0$. In this regime, a big rip occurs if $H(t)$ diverges at a 
finite time $t_{f}$, while a pseudo rip occurs if $H(t)$ increases 
monotonically but saturates to a finite asymptotic value $H_{\infty}$ as 
$t \to \infty$. The asymptotic behaviour of Eq.~(\ref{Hdot}) is controlled 
by the sign and power of the dominant term, which determines whether $H(t)$ 
diverges, saturates, or decays.

To determine the rip conditions, we examine the convergence properties of 
the integral in Eq.~(\ref{Hint}) as $H \to \infty$. The integral converges 
for $n \geq 2$, implying that $H$ diverges at a \textit{finite} time --- a 
\textit{big rip} singularity. For $n < 2$, the integral diverges and $H(t)$ 
approaches a finite value only as $t \to \infty$, corresponding to a 
\textit{pseudo rip}. Translating these conditions into constraints on the 
L\'{e}vy index: $n \geq 2$ is equivalent to $\alpha \leq 2$, while $n < 2$ 
requires $\alpha > 2$. Since the L\'{e}vy index is restricted to the range 
$1 < \alpha \leq 2$, the pseudo rip condition $\alpha > 2$ lies outside the 
physically admissible domain of the FHDE framework with the GO cutoff. 
Consequently, the big rip is the only generic future singularity that can 
occur within this setting. Crucially, for fractional values of $\alpha$ in 
the range $1 < \alpha \leq 2$, the occurrence of the big rip reflects the 
role of non-local features --- equivalently, the memory effect encoded in the 
fractional derivative --- in governing the late-time accelerated expansion 
within FHDE~\cite{Trivedi:2024inb,Bidlan_2025}.

We now demonstrate that FHDE with the GO cutoff generically fails to produce 
a little rip, except for very special cutoff choices. Following the 
approach of~\cite{TrivediScherrer_2024} for conventional HDE, we treat the 
DE density as a free function of the cutoff: 
$\rho_{\text{DE}} = 3c^{2}[f(L)]^{-2\mathcal{A}}$, where 
$\mathcal{A} = (3\alpha-2)/2\alpha$. Substituting into the Friedmann 
equation~(\ref{Friedmann}) gives $f(L) = (H/c)^{-1/\mathcal{A}}$. With the 
GO cutoff, the integral relation between $H$ and $t$ becomes
\begin{equation}
    \int_{H_{i}}^{H_{f}}\frac{\delta}{\left\{f^{-1}\!\left[
    \left(\frac{H}{c}\right)^{-\frac{1}{\mathcal{A}}}\right]\right\}^{-2} 
    - \gamma H^{2}}\,dH = \int_{t_{i}}^{t_{f}}dt.
\end{equation}
A little rip requires this integral to diverge as $H \to \infty$, which 
occurs precisely when the quantity $\left\{f^{-1}\!\left[(H/c)^{-1/\mathcal{A}}\right]
\right\}^{-2} \sim \gamma H^{2} + g(H)$ with $\int dH/g(H)$ divergent. The 
implied cutoff form $L \sim (\gamma H^{2} + g(H))^{-1/2}$ belongs to the 
broader Nojiri--Odintsov (NO) class~\cite{Nojiri_2017,Nojiri_2006}, but 
represents a highly contrived and non-standard choice with no natural 
motivation in the HDE literature. We therefore conclude that the little rip 
does not arise within the FHDE framework with the GO cutoff, except for a 
very special and physically unmotivated class of IR cutoffs, consistent with 
the findings of~\cite{TrivediScherrer_2024}.

To examine the EoS evolution near the big rip, we compute $\dot{\rho}_{\text{GO}}$ 
from Eq.~(\ref{fracrho}):
\begin{equation}\label{rhodot-GO}
    \dot{\rho}_{\text{GO}} = \mathcal{A}\,\rho_{\text{GO}}
    \left(\frac{2\gamma H\dot{H} + \delta\ddot{H}}{\gamma H^{2} 
    + \delta\dot{H}}\right).
\end{equation}
Substituting into the continuity equation~(\ref{contd}), one obtains 
the EoS parameter within the GO cutoff:
\begin{equation}
    w_{\text{GO}}(t) = -1 - \frac{1}{3H}\left[\mathcal{A}
    \left(\frac{2\gamma H\dot{H} + \delta\ddot{H}}
    {\gamma H^{2} + \delta\dot{H}}\right) + 
    \frac{Q}{3H^{2}\Omega_{\text{GO}}}\right].
\end{equation}
For the big rip, the Hubble parameter obeys the ansatz~(\ref{big rip ansatz}), 
$H(t) \approx H_{0}(t_{\text{rip}} - t)^{-m}$ with $m > 0$, so that 
$H(t) \to \infty$ as $t \to t_{\text{rip}}$. The squared sound speed 
parameter within the GO cutoff is
\begin{equation}
    v^{2}_{\text{GO}} = w_{\text{GO}} + \frac{\dot{w}_{\text{GO}}}{\mathcal{A}}
    \left(\frac{\gamma H^{2} + \delta\dot{H}}
    {2\gamma H\dot{H} + \delta\ddot{H}}\right).
\end{equation}

\begin{figure*}[htbp]
    \centering
    \begin{subfigure}[b]{0.45\linewidth}
        \centering
        \includegraphics[width=\linewidth]{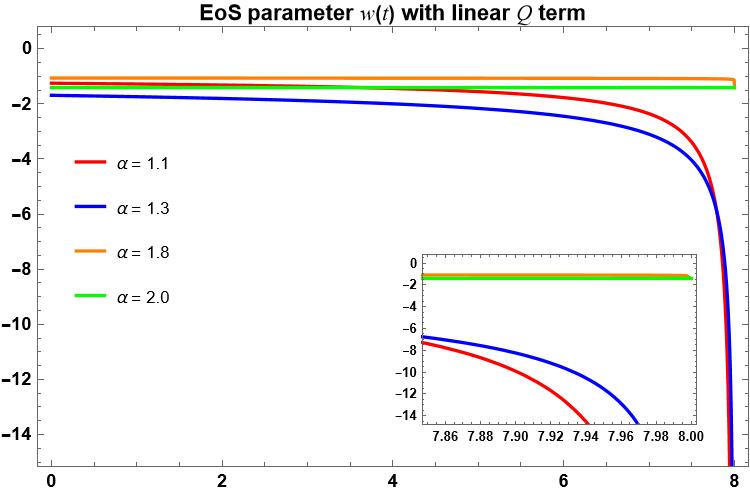}
        \caption{EoS parameter $w_{\text{GO}}(t)$ against cosmic time $t$ 
        for the big rip ansatz~(\ref{big rip ansatz}) with a linear 
        interaction term $Q$, for various values of the L\'{e}vy index 
        $\alpha$.}
        \label{Figure 1a}
    \end{subfigure}
    \hfill
    \begin{subfigure}[b]{0.45\linewidth}
        \centering
        \includegraphics[width=\linewidth]{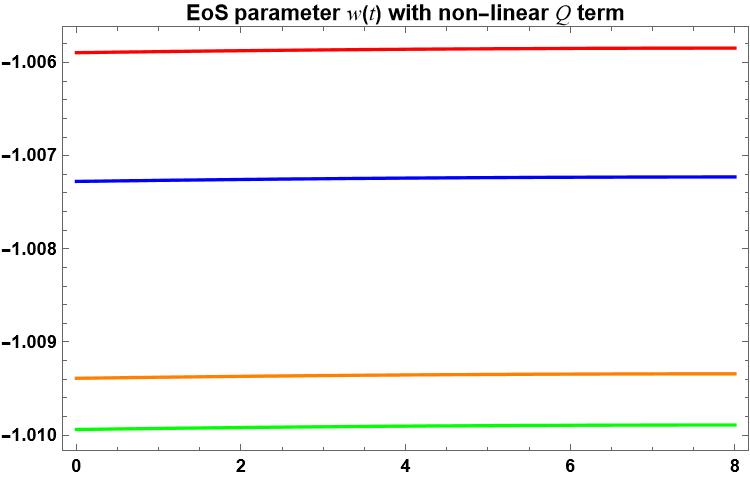}
        \caption{EoS parameter $w_{\text{GO}}(t)$ against cosmic time $t$ 
        for the big rip ansatz~(\ref{big rip ansatz}) with a non-linear 
        interaction term $Q$, for various values of the L\'{e}vy index 
        $\alpha$.}
        \label{Figure 1b}
    \end{subfigure}
    \caption{Redshift evolution of the EoS parameter $w_{\text{GO}}(t)$ 
    for the big rip scenario within the Granda--Oliveros cutoff.}
    \label{Figure 1}
\end{figure*}

\begin{figure*}[htbp]
    \centering
    \begin{subfigure}[b]{0.45\linewidth}
        \centering
        \includegraphics[width=\linewidth]{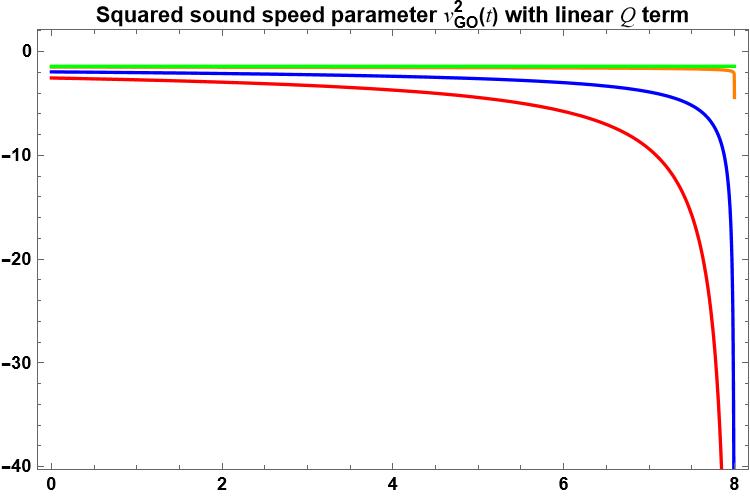}
        \caption{Squared sound speed parameter $v^{2}_{\text{GO}}(t)$ against 
        cosmic time $t$ for the big rip ansatz~(\ref{big rip ansatz}) with a 
        linear interaction term $Q$, for various values of the L\'{e}vy index 
        $\alpha$.}
        \label{Figure 2a}
    \end{subfigure}
    \hfill
    \begin{subfigure}[b]{0.45\linewidth}
        \centering
        \includegraphics[width=\linewidth]{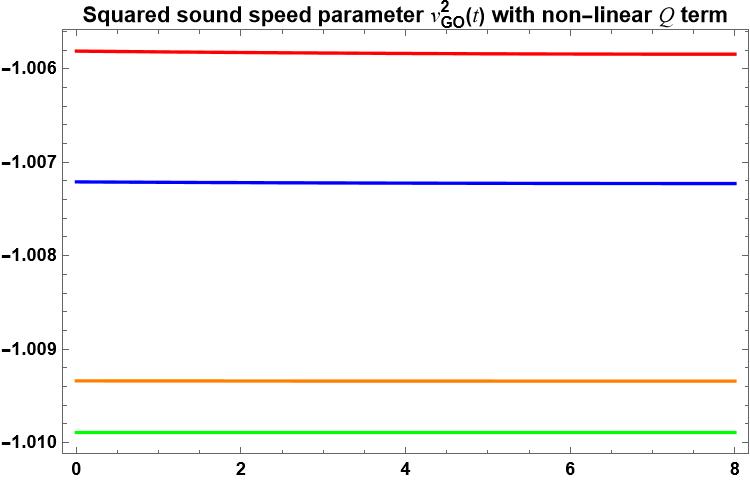}
        \caption{Squared sound speed parameter $v^{2}_{\text{GO}}(t)$ against 
        cosmic time $t$ for the big rip ansatz~(\ref{big rip ansatz}) with a 
        non-linear interaction term $Q$, for various values of the L\'{e}vy 
        index $\alpha$.}
        \label{Figure 2b}
    \end{subfigure}
    \caption{Redshift evolution of the squared sound speed parameter 
    $v^{2}_{\text{GO}}(t)$ for the big rip scenario within the 
    Granda--Oliveros cutoff.}
    \label{Figure 2}
\end{figure*}

We now present numerical plots of $w_{\text{GO}}(t)$ and $v^{2}_{\text{GO}}(t)$ 
for both linear and non-linear interaction terms, shown in Figures~\ref{Figure 1} 
and~\ref{Figure 2}. Throughout, we adopt the parameter values 
$\tilde{\beta} = 0.01$,\footnote{In the non-interacting limit $\tilde{\beta} = 0$, 
no big rip occurs. As the coupling strength increases towards $\tilde{\beta} \to 1$, 
the big rip becomes progressively more pronounced. We set $\tilde{\beta} = 0.01$ 
as the minimal coupling consistent with a physically plausible result.} 
$c = 0.495$,\footnote{Following the Planck+WP+BAO+HST+lensing constraints 
of~\cite{Li_2013}.} $H_{0} = 67.6$,\footnote{Consistent with recent 
determinations of the Hubble constant; see~\cite{guo2024newestmeasurementshubbleconstant}.} 
$\gamma = 0.9$, and $\delta = 0.5$~\cite{10.1093/mnras/stae2257}, for 
L\'{e}vy index values spanning the range $1 < \alpha \leq 2$.

The results are as follows:

\paragraph{Linear interaction (Figures~\ref{Figure 1a} and~\ref{Figure 2a}).}

\begin{enumerate}
    \item For fractional values $\alpha = 1.1$ and $1.3$, the EoS parameter 
    crosses the phantom divide ($w < -1$) and tends to $-\infty$ in the 
    late-time limit, consistent with the expected divergence of $w_{\text{GO}}$ 
    as the Universe approaches a big rip singularity.

    \item For larger values $\alpha = 1.8$ and $2.0$, the EoS parameter 
    remains close to $w \sim -1$, indicating the weakening or complete absence 
    of a big rip as fractional features diminish in the limit $\alpha \to 2$.

    \item The squared sound speed parameter is negative for all values of 
    $\alpha$, becoming increasingly negative as $\alpha \to 1.1$. This 
    signals classical instability throughout the approach to the big rip 
    singularity, a behaviour anticipated from the accumulation of phantom 
    energy.
\end{enumerate}

\paragraph{Non-linear interaction (Figures~\ref{Figure 1b} and~\ref{Figure 2b}).}

\begin{enumerate}
    \item Both $w_{\text{GO}}$ and $v^{2}_{\text{GO}}$ remain close to $-1$ 
    for all values of $\alpha$, corresponding to a nearly de~Sitter or 
    mildly phantom evolution without any finite-time divergence.

    \item This behaviour is qualitatively opposite to that observed for the 
    linear interaction term: the non-linear coupling suppresses the big rip 
    entirely across the parameter range considered.

    \item Consequently, the big rip singularity is realised exclusively in 
    the linear interaction scenario, and only for $\alpha$ values near 
    $\alpha \to 1.1$.
\end{enumerate}

In summary, for the linear interaction, fractional values $\alpha \to 1.1$ 
drive $w_{\text{GO}}(t)$ to large negative values with $v^{2}_{\text{GO}} < 0$, 
producing a classically unstable big rip evolution. For larger $\alpha$, and 
for all non-linear interaction cases, both $w_{\text{GO}}(t)$ and 
$v^{2}_{\text{GO}}$ remain near $-1$ without finite-time divergences, 
corresponding to a mild de~Sitter phantom behaviour.

\subsection{Little and Pseudo Rip in Hubble Cutoff}\label{littleandpseudorip}

We now motivate the ansatz-based approach to investigating late-time 
cosmological singularities within the Hubble cutoff in the FHDE framework. 
In recent work~\cite{Trivedirip_2024,TrivediScherrer_2024}, a methodology 
was developed for investigating the little rip and pseudo rip within 
conventional HDE and its extensions, including the Barrow and Tsallis 
variants. A key observation from~\cite{Trivedi:2024inb} is that a dynamical 
DE behaviour can be obtained within the Hubble cutoff $L = H^{-1}$, 
in contrast to conventional HDE where this cutoff yields a static EoS. 
Building on this result as our primary motivation, we import the 
well-established $H(t)$ ans\"{a}tze for the little and pseudo rip from the 
literature and use them to probe the EoS and squared sound speed (SSS) 
parameters within the Hubble cutoff in the FHDE framework.

The time derivative of the FHDE energy density within the Hubble cutoff 
$L = H^{-1}$ is
\begin{equation}
    \dot{\rho}_{\text{HH}} = 2\mathcal{A}\frac{\rho_{\text{HH}}\dot{H}}{H},
\end{equation}
where $\mathcal{A} = (3\alpha-2)/2\alpha$. Substituting this into the dark 
energy continuity equation~(\ref{contd}), the EoS parameter within the 
Hubble cutoff takes the form
\begin{equation}\label{EoSHH}
    w_{\text{HH}}(t) = -1 - \frac{1}{3H\Omega^{(\text{fr})}_{\text{DE}}}\left[
    \frac{Q}{3H^{2}} + 2\mathcal{A}\frac{\Omega^{(\text{fr})}_{\text{DE}}\dot{H}}{H}
    \right].
\end{equation}
Using Eq.~(\ref{general vs}) together with Eq.~(\ref{EoSHH}), the SSS 
parameter within the Hubble cutoff is
\begin{equation}\label{vsHH}
    v^{2}_{\text{HH}}(t) = w_{\text{HH}} + 
    \frac{\dot{w}_{\text{HH}}}{2\mathcal{A}}\frac{H}{\dot{H}}.
\end{equation}

\begin{figure*}[htbp]
    \centering
    \begin{subfigure}[b]{0.45\linewidth}
        \centering
        \includegraphics[width=\linewidth]{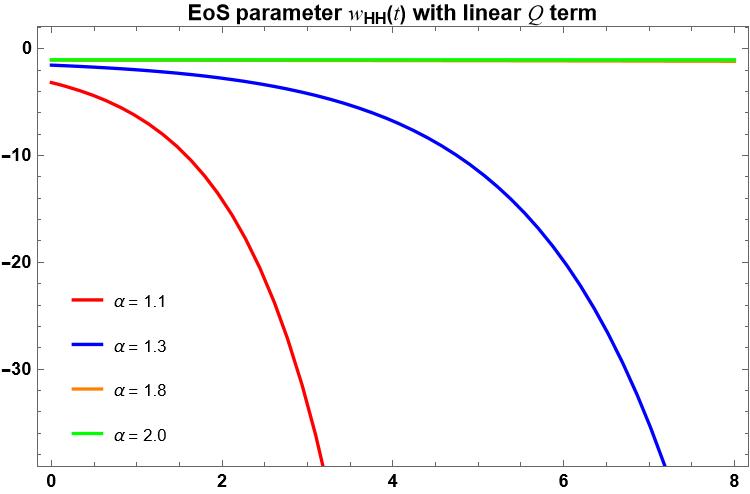}
        \caption{EoS parameter $w_{\text{HH}}(t)$ against cosmic time $t$ 
        for the little rip ansatz~(\ref{little rip ansatz}) with the linear 
        interaction term~(\ref{linear Q}), for various values of the 
        L\'{e}vy index $\alpha$.}
        \label{Figure 3a}
    \end{subfigure}
    \hfill
    \begin{subfigure}[b]{0.45\linewidth}
        \centering
        \includegraphics[width=\linewidth]{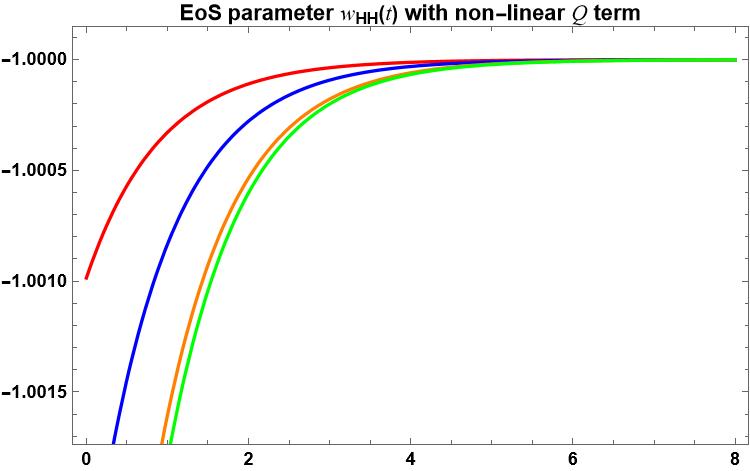}
        \caption{EoS parameter $w_{\text{HH}}(t)$ against cosmic time $t$ 
        for the little rip ansatz~(\ref{little rip ansatz}) with the 
        non-linear interaction term~(\ref{nonlinear Q}), for various values 
        of the L\'{e}vy index $\alpha$.}
        \label{Figure 3b}
    \end{subfigure}
    \caption{Redshift evolution of the EoS parameter $w_{\text{HH}}(t)$ 
    for the little rip scenario within the Hubble cutoff.}
    \label{Figure 3}
\end{figure*}

\begin{figure*}[htbp]
    \centering
    \begin{subfigure}[b]{0.45\linewidth}
        \centering
        \includegraphics[width=\linewidth]{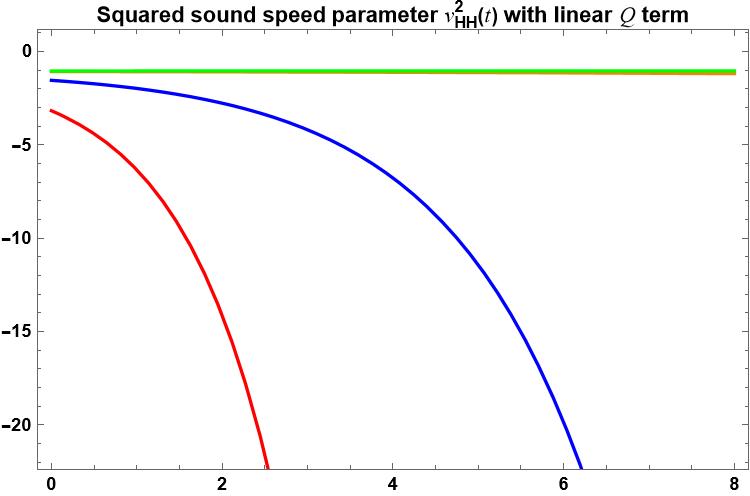}
        \caption{Squared sound speed parameter $v^{2}_{\text{HH}}$ against 
        cosmic time $t$ for the little rip ansatz~(\ref{little rip ansatz}) 
        with the linear interaction term~(\ref{linear Q}), for various 
        values of the L\'{e}vy index $\alpha$.}
        \label{Figure 4a}
    \end{subfigure}
    \hfill
    \begin{subfigure}[b]{0.45\linewidth}
        \centering
        \includegraphics[width=\linewidth]{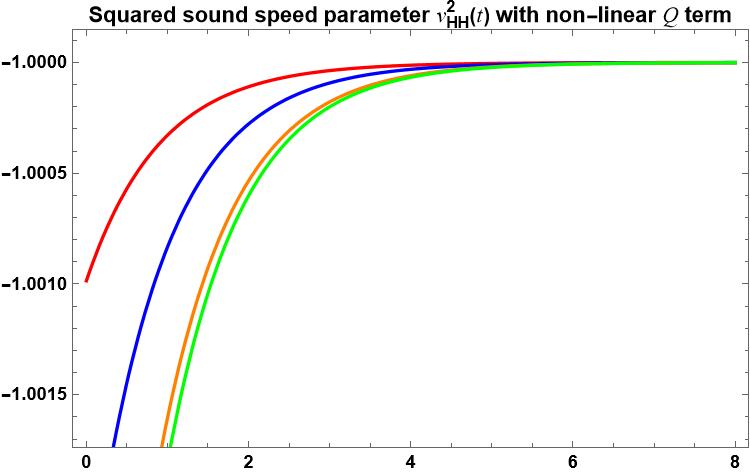}
        \caption{Squared sound speed parameter $v^{2}_{\text{HH}}$ against 
        cosmic time $t$ for the little rip ansatz~(\ref{little rip ansatz}) 
        with the non-linear interaction term~(\ref{nonlinear Q}), for 
        various values of the L\'{e}vy index $\alpha$.}
        \label{Figure 4b}
    \end{subfigure}
    \caption{Redshift evolution of the squared sound speed parameter 
    $v^{2}_{\text{HH}}$ for the little rip scenario within the Hubble 
    cutoff.}
    \label{Figure 4}
\end{figure*}

We now substitute the little rip ansatz~(\ref{little rip ansatz}) into 
Eqs.~(\ref{EoSHH}) and~(\ref{vsHH}), and present the resulting cosmological 
evolution in Figures~\ref{Figure 3} and~\ref{Figure 4} for both linear and 
non-linear interaction terms. The findings are as follows.

\paragraph{Linear interaction --- Figures~\ref{Figure 3a} and~\ref{Figure 4a}.}

\begin{enumerate}
    \item For fractional values $\alpha = 1.1$ and $1.3$, the EoS parameter 
    $w_{\text{HH}}(t)$ diverges to large negative values in the late-time 
    limit, closely analogous to the big rip behaviour observed in 
    Figure~\ref{Figure 1a}.

    \item For larger values $\alpha \to 2$, the EoS parameter remains 
    approximately constant throughout the late-time expansion, exhibiting 
    no little rip singularity behaviour.

    \item The SSS parameter $v^{2}_{\text{HH}}$ is negative throughout the 
    entire cosmic history, indicating classical instability under the little 
    rip ansatz~(\ref{little rip ansatz}) in the FHDE framework.

    \item Since both the EoS and SSS parameters diverge to large negative 
    values at a finite time, the linear interaction scenario corresponds to 
    a big rip type finite-time singularity rather than a genuine little rip.
\end{enumerate}

\paragraph{Non-linear interaction --- Figures~\ref{Figure 3b} and~\ref{Figure 4b}.}

\begin{enumerate}
    \item For all values of $\alpha$, the EoS parameter asymptotes to $-1$ 
    from below, $w_{\text{HH}} \to -1^{-}$, in the late-time limit --- the 
    characteristic behaviour expected of a little rip.

    \item The approach to $-1$ is more gradual for larger $\alpha \to 2$ 
    than for smaller values $\alpha \to 1.1$, where the evolution is 
    comparatively more rapid.

    \item The SSS parameter remains negative throughout the cosmic history, 
    indicating classical instability in this case as well.

    \item Since the EoS and SSS parameters asymptote to $-1$ from below 
    without any finite-time divergence, the non-linear interaction scenario 
    yields a genuine little rip evolution.
\end{enumerate}

In summary, for the linear interaction, $w_{\text{HH}}(t)$ diverges to 
large negative values for fractional $\alpha$ (e.g.\ $\alpha = 1.1$ and 
$1.3$), with $v^{2}_{\text{HH}} < 0$ throughout, signalling classical 
instability and a big rip type finite-time singularity rather than a little 
rip. In contrast, for the non-linear interaction, the EoS asymptotes to 
$-1$ from below for all values of $\alpha$, and the SSS parameter remains 
negative but finite, producing a classically unstable yet genuine little rip 
evolution without finite-time divergences.

\begin{figure*}[htbp]
    \centering
    \begin{subfigure}[b]{0.45\linewidth}
        \centering
        \includegraphics[width=\linewidth]{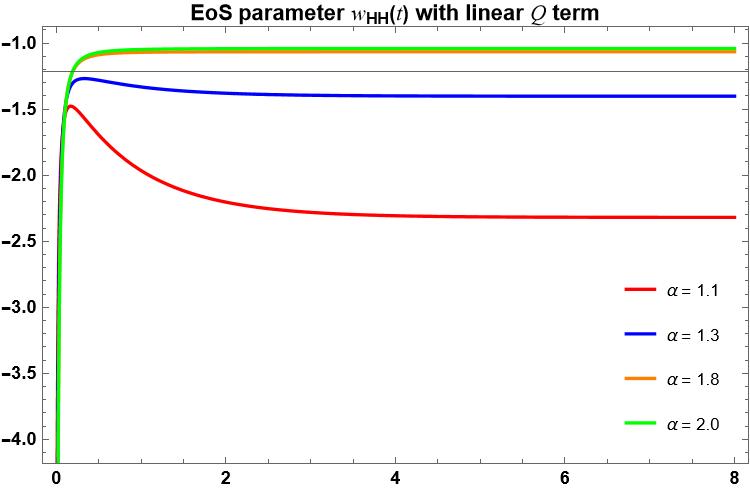}
        \caption{EoS parameter $w_{\text{HH}}(t)$ against cosmic time $t$ 
        for the pseudo rip ansatz~(\ref{pseudo rip ansatz}) with the linear 
        interaction term~(\ref{linear Q}), for various values of the 
        L\'{e}vy index $\alpha$.}
        \label{Figure 5a}
    \end{subfigure}
    \hfill
    \begin{subfigure}[b]{0.45\linewidth}
        \centering
        \includegraphics[width=\linewidth]{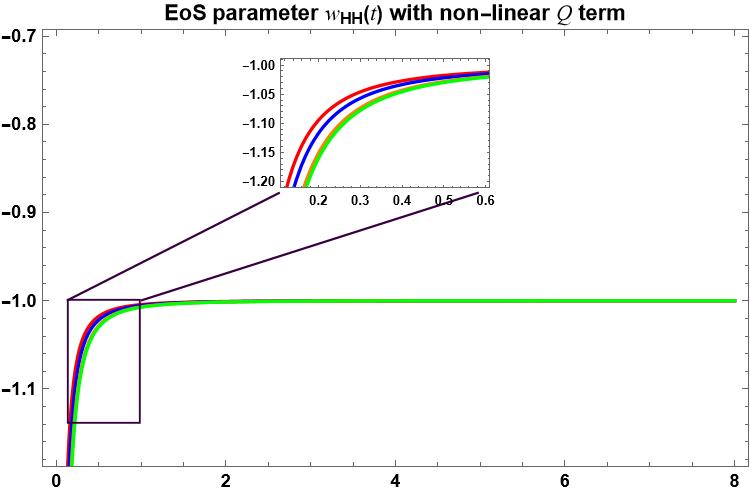}
        \caption{EoS parameter $w_{\text{HH}}(t)$ against cosmic time $t$ 
        for the pseudo rip ansatz~(\ref{pseudo rip ansatz}) with the 
        non-linear interaction term~(\ref{nonlinear Q}), for various values 
        of the L\'{e}vy index $\alpha$.}
        \label{Figure 5b}
    \end{subfigure}
    \caption{Redshift evolution of the EoS parameter $w_{\text{HH}}(t)$ 
    for the pseudo rip scenario within the Hubble cutoff.}
    \label{Figure 5}
\end{figure*}

\begin{figure*}[htbp]
    \centering
    \begin{subfigure}[b]{0.45\linewidth}
        \centering
        \includegraphics[width=\linewidth]{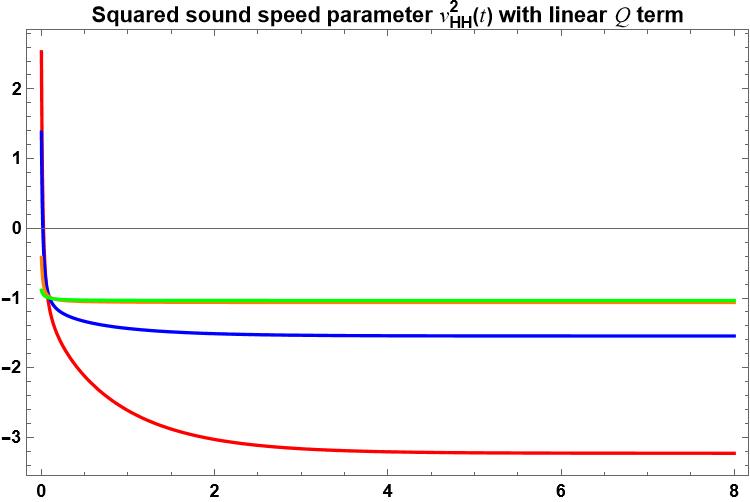}
        \caption{Squared sound speed parameter $v^{2}_{\text{HH}}$ against 
        cosmic time $t$ for the pseudo rip ansatz~(\ref{pseudo rip ansatz}) 
        with the linear interaction term~(\ref{linear Q}), for various 
        values of the L\'{e}vy index $\alpha$.}
        \label{Figure 6a}
    \end{subfigure}
    \hfill
    \begin{subfigure}[b]{0.45\linewidth}
        \centering
        \includegraphics[width=\linewidth]{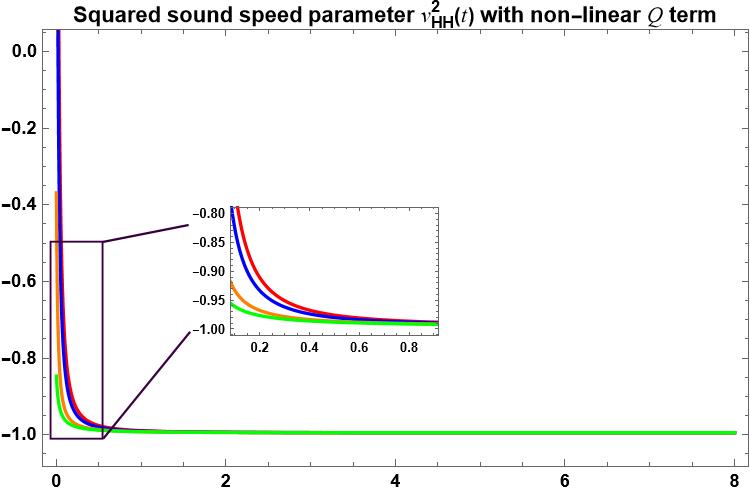}
        \caption{Squared sound speed parameter $v^{2}_{\text{HH}}$ against 
        cosmic time $t$ for the pseudo rip ansatz~(\ref{pseudo rip ansatz}) 
        with the non-linear interaction term~(\ref{nonlinear Q}), for 
        various values of the L\'{e}vy index $\alpha$.}
        \label{Figure 6b}
    \end{subfigure}
    \caption{Redshift evolution of the squared sound speed parameter 
    $v^{2}_{\text{HH}}$ for the pseudo rip scenario within the Hubble 
    cutoff.}
    \label{Figure 6}
\end{figure*}

We now turn to the pseudo rip, governed by the ansatz~(\ref{pseudo rip 
ansatz}). Under this ansatz, the Hubble parameter approaches a finite 
positive value as $t \to \infty$: $H(t) \to H_{0}$. We substitute this 
ansatz into Eqs.~(\ref{EoSHH}) and~(\ref{vsHH}) and present the resulting 
evolution in Figures~\ref{Figure 5} and~\ref{Figure 6}.

\paragraph{Linear interaction --- Figures~\ref{Figure 5a} and~\ref{Figure 6a}.}

\begin{enumerate}
    \item The EoS parameter asymptotes to a value close to $-1$ for large 
    $\alpha$, while for smaller $\alpha$ it approaches a value slightly 
    below $-1$. Across all values of $\alpha$, however, the evolution does 
    not correspond to a genuine pseudo rip.

    \item Among the values considered, only $\alpha \to 2$ yields a 
    pseudo rip-consistent behaviour, with $w_{\text{HH}} \to -1$ from 
    below and a sufficiently mild late-time evolution.

    \item The SSS parameter begins from a positive value for smaller $\alpha$ 
    but asymptotes to a large negative value in the late-time limit, 
    indicating classical instability throughout.

    \item The large negative values of $v^{2}_{\text{HH}}$ and the dramatic 
    late-time evolution are inconsistent with the smooth, asymptotic 
    character expected of a pseudo rip; the linear interaction is therefore 
    less suited to this scenario.
\end{enumerate}

\paragraph{Non-linear interaction --- Figures~\ref{Figure 5b} and~\ref{Figure 6b}.}

\begin{enumerate}
    \item The EoS parameter asymptotes to $-1$ from below, 
    $w_{\text{HH}} \to -1^{-}$, for all values of $\alpha$ during 
    late-time expansion, consistent with pseudo rip behaviour.

    \item Since this asymptotic approach holds for all $\alpha \in (1, 2]$, 
    the pseudo rip is a generic outcome of the non-linear interaction 
    scenario within the Hubble cutoff.

    \item The SSS parameter asymptotes smoothly to $-1$ from above for all 
    $\alpha$ in the late-time limit, indicating persistent but mild 
    classical instability.
\end{enumerate}

In summary, within the non-linear interaction regime, the EoS parameter 
approaches $-1$ from below at late times for all values of the L\'{e}vy 
index $\alpha$, producing pseudo rip behaviour generically, as shown in 
Figure~\ref{Figure 5b}. For the SSS parameter, the linear interaction can 
drive $v^{2}_{\text{HH}}$ from positive to large negative values, signalling 
a stronger classical instability, whereas the non-linear interaction yields 
a smoother asymptotic approach to $-1$ from above --- perturbatively unstable 
but without dramatic late-time divergences.

\section{Fractional HDE in Modified Gravity}\label{modified}

Gravity is a non-renormalisable theory and is therefore intrinsically sensitive to ultraviolet (UV) physics \cite{Kiefer:2004xyv}. Within the HDE framework, however, the UV cutoff is not an independent scale; rather, it is determined by the IR cutoff, which is fixed at cosmological distances in DE studies. As a result, any modification introduced at the IR scale necessarily affects the UV sector. In the absence of a well-defined renormalisation group equation for quantum gravity, a natural strategy is to consider modified gravity theories in which HDE provides the effective IR description. From this perspective, even a modified gravity model that is disfavoured by solar-system or galactic-scale constraints may still remain relevant when coupled to HDE, since the underlying physics at cosmological scales is qualitatively different.

In this regard, HDE offers a natural mechanism for alleviating this issue within such theories, thereby shifting attention toward questions of naturalness, the dynamics of DE, and consistency with observational data. From the observational point of view, HDE is characterised by only two free parameters, whereas most alternative DE models involve a larger parameter space, although the CC itself depends on only one parameter. Consequently, modified gravity models supplemented by HDE retain comparatively strong predictive power relative to many other dynamical DE scenarios \cite{wang2017holographic}.

A number of studies have already examined HDE and extended HDE models, including well-known examples such as Barrow and Tsallis HDE, under several choices of IR cutoff. FHDE, however, differs in an important way, as it carries features of the FWDW equation, which arises from the application of FC to the quantum-gravitational dynamics of a static, spherically symmetric BH. This structure provides a concrete mechanism through which non-local effects may enter late-time cosmological observables, particularly the EoS parameter and the deceleration parameter. The main objective of this section is therefore to derive these observables in the interaction-free case, namely \(Q=0\). Our approach is organised as follows:

\begin{enumerate}
    \item \textit{Scalar--Tensor Theory} -- Several constructions based on different subclasses of scalar--tensor gravity have already been explored. However, a direct formulation that incorporates non-local effects into the holographic description of DE with the simplest choice of cutoff, namely the Hubble cutoff, is still absent. To address this gap, we consider two representative scalar--tensor theories: \textit{Brans--Dicke} theory in Section (\ref{BD}) and \textit{Horndeski} theory in Section (\ref{Horndeski}).

    \item \textit{Vector--Tensor Theory} -- Another important modified gravity direction currently receiving considerable attention is the framework of massive vector fields coupled to gravity, commonly described by the \textit{Generalised Proca} (GP) class of theories. In a recent study \cite{Chiang:2025hrj}, a particular sub-class of GP theory, namely the \textit{Proca--Nuevo} theory, was investigated in a special cosmological setting motivated by contemporary cosmological tensions. Inspired by that construction, we extend the FHDE framework to the PN setting in Section (\ref{PN}).

    \item \textit{Braneworld Cosmology} -- Higher-dimensional theories in which the observable Universe is realised as a brane embedded in a higher-dimensional spacetime have attracted sustained interest in recent years. One of the best-known examples is the \textit{Dvali--Gabadadze--Porrati} (DGP) braneworld model. In \cite{Ghaffari:2014pxa}, a reconstruction of standard HDE was studied using the GO cutoff. In Section (\ref{Braneworld}), we extend this analysis to the FHDE framework by adopting the Hubble cutoff instead of the GO cutoff.
\end{enumerate}

To reiterate, the aim of this section is to derive, in a systematic and transparent manner, the late-time cosmological observables, specifically the EoS parameter and the deceleration parameter, in an interaction-free setting for several modifications of Einstein gravity.

\subsection{Brans--Dicke Theory}\label{BD}

The Jordan--Fierz--Brans--Dicke theory of gravitation, commonly known as Brans--Dicke (BD) theory, is one of the earliest and most influential modified gravity frameworks. BD gravity has remained both theoretically and observationally relevant, particularly through increasingly stringent bounds on the BD coupling parameter, with current constraints requiring \(\omega \gtrsim 40{,}000\) \cite{Faraoni:2004pi}. A central feature of the theory is its incorporation of Mach's principle through a time-dependent Newtonian gravitational constant,
\[
G_{\text{N}}(t)\simeq G_{0}\Tilde{\Phi}^{-1}(t),
\]
a property absent in general relativity \cite{Brans:1961sx}. In the literature, several HDE and extended HDE constructions have already been investigated in the BD framework for different choices of IR cutoff \cite{Sheykhi:2009dz, Setare:2006yj, Chattopadhyay:2014yda}. In its most general form, the BD action is written as
\begin{equation}\label{BDaction}
    \mathcal{S} = \frac{1}{2}\int d^{4}x\sqrt{-g}\left(\Tilde{\Phi} R 
    - \frac{\omega}{\Tilde{\Phi}}g^{\mu\nu}\nabla_{\mu}\Tilde{\Phi}\nabla_{\nu}\Tilde{\Phi} 
    - V(\Tilde{\Phi})\right) + \int d^{4}x\sqrt{-g}\,\mathcal{L}_{\text{matter}},
\end{equation}
where $\Tilde{\Phi}$, \(\omega\), \(R\), and \(V(\Tilde{\Phi})\) denote the BD scalar field, the BD coupling constant, the Ricci scalar, and the scalar self-interaction potential, respectively. For a spatially flat Universe containing dust-like DM and DE, the cosmological field equations in BD gravity are
\begin{equation}\label{BDEOM1}
    3H^{2} = \frac{\rho^{(\text{fr})}_{\text{DE}} + \rho^{(\text{fr})}_{\text{DM}}}{\Tilde{\Phi} M^{2}_{pl}} 
    + \frac{\omega\dot{\Tilde{\Phi}}^{2}}{2\Tilde{\Phi}^{2}} - \frac{3H\dot{\Tilde{\Phi}}}{\Tilde{\Phi}},
\end{equation}
and
\begin{equation}\label{BDEOM2}
    \frac{2\ddot{a}}{a} + \frac{\dot{a}^{2}}{a^{2}} = 
    -\frac{p^{(\text{fr})}_{\text{DE}}}{\Tilde{\Phi} M^{2}_{pl}} 
    - \frac{\omega\dot{\Tilde{\Phi}}^{2}}{2\Tilde{\Phi}^{2}} 
    - \frac{2\dot{a}\dot{\Tilde{\Phi}}}{a\Tilde{\Phi}} - \frac{\ddot{\Tilde{\Phi}}}{\Tilde{\Phi}}.
\end{equation}

An important departure from standard gravitational scenario is that the reduced Planck mass \(M_{pl}\) is no longer constant in the BD cosmological framework. This follows from the time dependence of Newton's gravitational constant, \(G_{\text{N}}(t)\simeq G_{0}\Tilde{\Phi}^{-1}(t)\), which implies
\[
M_{pl}^{2}(t)=\frac{\Tilde{\Phi}(t)}{8\pi G_{0}}.
\]
Substituting this relation into Eq. (\ref{BDEOM1}) and solving it together with the continuity equations, Eqs. (\ref{contd})--(\ref{contm}), yields
\begin{equation}\label{BD1}
    \frac{\dot{H}}{H^{2}} = -\frac{3}{2\Tilde{\Phi}^{2}}\left[8\pi G_{0}
    \left(1 + y + w^{(\text{BD})}_{\text{DE}}\right)\Omega^{(\text{fr})}_{\text{DE}} 
    - \frac{2\Tilde{\Phi}\dot{\Tilde{\Phi}}}{3H} + \frac{\omega\Tilde{\Phi}\ddot{\Tilde{\Phi}}}{9H^{3}} 
    - \frac{\Tilde{\Phi}\dot{\Tilde{\Phi}}\dot{H}}{3H^{3}} - \frac{\dot{\Tilde{\Phi}}^{2}}{3H^{2}} 
    - \frac{\Tilde{\Phi}\ddot{\Tilde{\Phi}}}{3H^{2}}\right],
\end{equation}
where \(y=\Omega^{(\text{fr})}_{\text{DM}}/\Omega^{(\text{fr})}_{\text{DE}}\). Taking the time derivative of the FHDE energy density in Eq. (\ref{fracrho}) while allowing the reduced Planck mass to vary with time gives
\begin{equation}
    \dot{\rho}^{(\text{fr})}_{\text{de}} = \rho^{(\text{fr})}_{\text{DE}}
    \left[\frac{\dot{\Tilde{\Phi}}}{\Tilde{\Phi}} + \left(\frac{3\alpha-2}{\alpha}\right)
    \frac{\dot{H}}{H}\right].
\end{equation}
Substituting this result into the DE continuity equation, Eq. (\ref{contd}), produces a second expression for \(\dot{H}/H^{2}\),
\begin{equation}\label{BD2}
    \frac{\dot{H}}{H^{2}} = -\frac{3\alpha}{3\alpha-2}\left[1 + w^{(\text{BD})}_{\text{DE}} 
    + \frac{Q}{9H^{3}\Omega^{(\text{fr})}_{\text{DE}}} 
    + \frac{\dot{\Tilde{\Phi}}}{3H\Tilde{\Phi}}\right].
\end{equation}

Equating Eqs. (\ref{BD1}) and (\ref{BD2}), one obtains the EoS parameter for FHDE in the BD cosmological background as
\begin{equation}\label{BDEoS}
    w^{(\text{BD})}_{\text{DE}} = -1 + 
    \frac{8\pi G_{0}(3\alpha-2)\Omega^{(\text{fr})}_{\text{DM}} 
    + (3\alpha-2)f(\Tilde{\Phi},\dot{\Tilde{\Phi}}) 
    - 2\alpha\Tilde{\Phi}^{2}\!\left(\dot{\Tilde{\Phi}}/3H\Tilde{\Phi}\right)}
    {2\alpha\Tilde{\Phi}^{2} - 8\pi G_{0}(3\alpha-2)\Omega^{(\text{fr})}_{\text{DE}}},
\end{equation}
where \(\Omega^{(\text{fr})}_{\text{DM}}\) is expressed in terms of \(\Omega^{(\text{fr})}_{\text{DE}}\) through Eq. (\ref{BDEOM1}), and the auxiliary function \(f(\Tilde{\Phi},\dot{\Tilde{\Phi}})\) is defined by
\begin{equation}
    f(\Tilde{\Phi},\dot{\Tilde{\Phi}}) = \frac{\omega\Tilde{\Phi}\ddot{\Phi}}{9H^{3}} 
    - \frac{2\Tilde{\Phi}\dot{\Tilde{\Phi}}}{3H} - \frac{\Tilde{\Phi}\dot{\Tilde{\Phi}}\dot{H}}{3H^{3}} 
    - \frac{\dot{\Tilde{\Phi}}^{2}}{3H^{2}} - \frac{\Tilde{\Phi}\ddot{\Tilde{\Phi}}}{3H^{2}}.
\end{equation}
Substituting the FHDE--BD EoS in Eq. (\ref{BDEoS}) into Eq. (\ref{BD2}), the corresponding deceleration parameter in the non-interacting case, \(Q=0\), becomes
\begin{equation}\label{BDq}
\begin{split}
    q^{(\text{BD})}_{\text{de}} = -1 + \frac{3\alpha}{3\alpha-2}
    \Bigg[\frac{(3\alpha-2)\left(8\pi G_{0}\Omega^{(\text{fr})}_{\text{DM}} 
    + f(\Tilde{\Phi},\dot{\Tilde{\Phi}})\right) 
    - 2\alpha\Tilde{\Phi}^{2}\!\left( \dot{\Tilde{\Phi}}/3H\Tilde{\Phi}\right)}
    {2\alpha\Tilde{\Phi}^{2} - 8\pi G_{0}(3\alpha-2)\Omega^{(\text{fr})}_{\text{DE}}} + \frac{\dot{\Tilde{\Phi}}}{3H\Tilde{\Phi}}\Bigg].
\end{split}
\end{equation}

Equations (\ref{BDEoS}) and (\ref{BDq}) therefore determine the EoS and deceleration parameters of FHDE in a BD cosmological background. In the interaction-free general relativistic limit, in which the FHDE framework was originally formulated \cite{Trivedi:2024inb}, one has \(\dot{\Tilde{\Phi}}\to0\) and \(Q\to0\). Under these conditions, both expressions reduce to the standard FHDE EoS and deceleration parameter derived in Section (\ref{FHDE}) of Chapter (\ref{Chapter2}); see Eqs. (\ref{FHDEEoS}) and (\ref{qeq}). This reduction confirms the consistency of the BD realisation with the original FHDE construction. More generally, the above results encode new qualitative features of FHDE in the BD cosmological setting, and a detailed numerical analysis of these observables is deferred to future work.

As a scalar--tensor theory, BD gravity provides a useful setting for exploring cosmological dynamics beyond Einstein gravity. Another major class within scalar--tensor theories is Horndeski gravity, which represents the most general second-order scalar--tensor framework. However, recent observational constraints have restricted its phenomenologically viable sector to a narrower subclass. In the next section, we therefore turn to one such restricted Horndeski model motivated by the constraints imposed by GW170817.

\subsection{Post-GW170817 Luminal Horndeski Theory}\label{Horndeski}

The joint observation of the binary neutron star merger GW170817 and its electromagnetic counterpart GRB170817A established that the propagation speed of gravitational waves, \(c_{g}\), is equal to the speed of light, \(c_{l}\), to a precision of \(|c_g/c_{l} - 1| \lesssim 10^{-15}\) \cite{Kase:2018aps}. This stringent luminal constraint significantly reduced the range of viable modified gravity models, particularly within Horndeski theory, which represents the most general class of scalar--tensor theories with second-order equations of motion. As a consequence, a large part of the Horndeski landscape was excluded as a DE alternative \cite{Ezquiaga:2017ekz}. To remain phenomenologically viable, Horndeski models must satisfy the luminal condition, which imposes strong restrictions on the quartic \(G_{4}\) and quintic \(G_{5}\) sectors. Under these conditions, the remaining viable models are essentially limited to standard quintessence-type scenarios or simple conformally coupled theories \cite{Ezquiaga:2017ekz, Euclid:2025vml}. Since reconstructions based on HDE within this restricted class of Horndeski theories have not yet been explored, we use this setting to develop a corresponding reconstruction in the FHDE framework. As in the previous cases, we adopt the Hubble cutoff for simplicity.

Restricting attention to Horndeski theories compatible with post-GW170817 luminal gravitational-wave propagation and excluding any explicit coupling to the Ricci scalar \(R\), the relevant scalar-field Lagrangian is
\begin{equation}
    \mathcal{L}_{\text{H}}=G_{2}(\Phi,X)+G_{3}(\Phi,X)\Box\Phi+\frac{M^{2}_{pl}R}{2},
\end{equation}
where the canonical kinetic term is defined by
\[
X=-\frac{1}{2}g^{\mu\nu}\partial_{\mu}\Phi\partial_{\nu}\Phi.
\]
Here, \(G_{i}\) are arbitrary functions of \(\Phi\) and \(X\), while \(G_{i,\Phi}\) and \(G_{i,X}\) denote partial derivatives with respect to these variables. Throughout the analysis, we set \(M_{pl}=1\). For a spatially flat FLRW background, the equations of motion are given by
\begin{equation}\label{HorndeskiEq1}
3H^{2}=\rho^{(\text{fr})}_{\text{DE}}+\rho^{(\text{fr})}_{\text{DM}}-G_{2}+2XG_{2,X}-2XG_{3,\Phi}+6H\dot{\Phi}XG_{3,X}
\end{equation}
and
\begin{equation}\label{HorndeskiEq2}
    2\dot{H}+3H^{2}=-p_{\text{DE}}-G_{2}+2X\left(G_{3,\Phi}+\ddot{\Phi}G_{3,X}\right).
\end{equation}

By solving Eqs. (\ref{HorndeskiEq1}) and (\ref{HorndeskiEq2}) algebraically together with the interaction-free continuity equations, Eqs. (\ref{contd}) and (\ref{contm}), one obtains the late-time cosmological observables in the form of the EoS parameter and the deceleration parameter. These are given by
\begin{equation}\label{HorndeskiEoS}
    w^{(\text{H})}_{\text{DE}}=\frac{(3\alpha-2)(1+y)\Omega^{(\text{fr})}_{\text{DE}}-2\alpha+(3\alpha-2)\Dot{\rho}_{\Phi}/6H^{3}}{2\alpha-(3\alpha-2)\Omega^{(\text{fr})}_{\text{DE}}}
\end{equation}
and
\begin{equation}\label{Horndeskiq}
    q^{(\text{H})}_{\text{DE}}=-1+\frac{3\alpha}{3\alpha-2}\Bigg[1+\frac{(3\alpha-2)(1+y)\Omega^{(\text{fr})}_{\text{DE}}-2\alpha+(3\alpha-2)\Dot{\rho}_{\Phi}/6H^{3}}{2\alpha-(3\alpha-2)\Omega^{(\text{fr})}_{\text{DE}}}\Bigg].
\end{equation}
Here,
\begin{equation}
    \rho_{\Phi}=-G_{2}+2XG_{2,X}-2XG_{3,\Phi}+6H\Dot{\Phi}XG_{3,X},
\end{equation}
and \(\Dot{\rho}_{\Phi}\) denotes the time derivative of \(\rho_{\Phi}\).

The observables defined in Eqs. (\ref{HorndeskiEoS}) and (\ref{Horndeskiq}) incorporate the characteristic ingredients of a restricted class of scalar--tensor gravity theories whose physical viability is dictated directly by observational constraints \cite{Ezquiaga:2017ekz}. In this sense, the BD and Horndeski frameworks considered here provide representative examples of scalar--tensor theories in which FHDE may offer a viable route toward describing dynamical DE with the Hubble radius as the IR cutoff. This line of investigation can be extended further to vector--tensor theories, which are of particular interest because of their richer non-linear structure. In the next section, we therefore turn to one such class of models, namely the Extended Proca--Nuevo theory.

\subsection{Extended Proca--Nuevo Theory}\label{PN}

We now present a reconstruction of FHDE in the framework of Extended Proca--Nuevo (EPN) theory, a non-linear vector--tensor theory describing a massive spin-\(1\) field \(A_{\mu}\) non-minimally coupled to gravity. The total gravitational action, including the Einstein--Hilbert contribution together with the EPN and standard matter sectors, is given by
\begin{equation}\label{EPNaction}
    S = \int d^{4}x\sqrt{-g}\left(\frac{M^{2}_{\text{pl}}}{2}R 
    + \mathcal{L}_{\text{EPN}} + \mathcal{L}_{\text{M}}\right),
\end{equation}
where \(M_{\text{pl}}\) denotes the reduced Planck mass, which will be set to unity throughout the analysis. The EPN Lagrangian takes the form
\begin{equation}\label{LEPN}
    \mathcal{L}_{\text{EPN}} = -\frac{1}{4}F^{\mu\nu}F_{\mu\nu} 
    + \Tilde{\Lambda}^{4}\left(\hat{\mathcal{L}}_{0} + \hat{\mathcal{L}}_{1} 
    + \hat{\mathcal{L}}_{2} + \hat{\mathcal{L}}_{3}\right),
\end{equation}
where the antisymmetric field strength tensor is defined by
\[
F_{\mu\nu} = \nabla_{\mu}A_{\nu} - \nabla_{\nu}A_{\mu}.
\]
The operator content of the theory is constructed from the building blocks
\begin{align}\label{Lhat}
    \hat{\mathcal{L}}_{0} &= \beta_{0}(X), \notag \\
    \hat{\mathcal{L}}_{1} &= \beta_{1}(X)\mathcal{L}_{1}[K] 
    + d_{1}(X)\frac{\mathcal{L}_{1}[\nabla A]}{\Tilde{\Lambda}^{2}}, \notag \\
    \hat{\mathcal{L}}_{2} &= \left(\beta_{2}(X) + d_{2}(X)\right)
    \frac{R}{\Tilde{\Lambda}^{2}} + \beta_{2,X}(X)\mathcal{L}_{2}[K] 
    + d_{2,X}(X)\frac{\mathcal{L}_{2}[\nabla A]}{\Tilde{\Lambda}^{4}}, \notag \\
    \hat{\mathcal{L}}_{3} &= \left(\beta_{3}(X)\mathcal{K}^{\mu\nu} 
    + d_{3}(X)\frac{\nabla^{\mu}A^{\nu}}{\Tilde{\Lambda}^{2}}\right)
    \frac{G_{\mu\nu}}{\Tilde{\Lambda}^{2}} - \frac{1}{6}\beta_{3,X}(X)
    \mathcal{L}_{3}[K] - \frac{1}{6}d_{3,X}(X)
    \frac{\mathcal{L}_{3}[\nabla A]}{\Tilde{\Lambda}^{6}},
\end{align}
which are written in terms of the matrix fields
\[
\mathcal{K}^{\mu}{}_{\nu} \equiv 
\left(\sqrt{g^{-1}f[A]}\right)^{\mu}{}_{\nu} - \delta^{\mu}_{\nu},
\qquad
(\nabla A)^{\mu}{}_{\nu} \equiv g^{\mu\alpha}\nabla_{\alpha}A_{\nu}.
\]

The symmetric tensor \(f_{\mu\nu}\) appearing in the matrix square root is defined as
\begin{equation}\label{fmunu}
    f_{\mu\nu}[A] = g_{\mu\nu} + 2\frac{\nabla_{(\mu}A_{\nu)}}{\Tilde{\Lambda}^{2}} 
    + \frac{\nabla_{\mu}A_{\alpha}g^{\alpha\beta}\nabla_{\nu}A_{\beta}}
    {\Tilde{\Lambda}^{4}},
\end{equation}
while the antisymmetric Levi--Civita contractions associated with a generic matrix \(M\) are given by
\begin{equation}\label{Ln}
    \mathcal{L}_{n}[M] = -\frac{1}{(4-n)!}\,
    \epsilon^{\mu_{1}\cdots\mu_{4}}\epsilon_{\nu_{1}\cdots\nu_{4}}\,
    M^{\nu_{1}}{}_{\mu_{1}}\cdots M^{\nu_{n}}{}_{\mu_{n}}\,
    \delta^{\nu_{n+1}}{}_{\mu_{n+1}}\cdots\delta^{\nu_{4}}{}_{\mu_{4}}.
\end{equation}
The Lorentz scalar controlling the self-interaction strength is
\begin{equation}\label{X}
    X \equiv -\frac{1}{2\Tilde{\Lambda}^{2}}A_{\mu}A^{\mu},
\end{equation}
where $\Tilde{\Lambda}$ sets the characteristic energy scale associated with the non-linearity of the theory. The operators in Eq. (\ref{Lhat}) contain non-minimal couplings to the Ricci scalar \(R\) and the Einstein tensor \(G_{\mu\nu}\), with their structure governed by the arbitrary coefficient functions \(\beta_{n}(X)\) and \(d_{n}(X)\). Here we use the shorthand \(\beta_{n,X}\equiv \partial\beta_{n}/\partial X\). The sector proportional to \(d_{n}\) coincides with the Generalised Proca (GP) class \cite{Heisenberg:2017mzp}, whereas the \(\beta_{n}\)-sector defines the Proca--Nuevo (PN) operators \cite{deRham:2020yet}.

A notable feature of the EPN construction is that the coefficient functions can be chosen so that all non-minimal gravitational couplings cancel algebraically. In \cite{Chiang:2025hrj}, this possibility was explored through a ``special'' EPN model obtained by imposing the conditions \(\beta_{2}+d_{2}=0\) and \(\beta_{3}+d_{3}=0\). Under these constraints, the Lagrangian operators reduce to
\begin{align}\label{special EPN}
    \hat{\mathcal{L}}^{s}_{0} &= \beta_{0}(X), \notag \\
    \hat{\mathcal{L}}^{s}_{1} &= \beta_{1}(X)\mathcal{L}_{1}[K] 
    + d_{1}(X)\frac{\mathcal{L}_{1}[\nabla A]}{\Tilde{\Lambda}^{2}}, \notag \\
    \hat{\mathcal{L}}^{s}_{2} &= \beta_{2,X}(X)\left(\mathcal{L}_{2}[K] 
    - \frac{\mathcal{L}_{2}[\nabla A]}{\Tilde{\Lambda}^{4}}\right), \notag \\
    \hat{\mathcal{L}}^{s}_{3} &= -\frac{1}{6}\beta_{3,X}(X)
    \left(\mathcal{L}_{3}[K] 
    - \frac{\mathcal{L}_{3}[\nabla A]}{\Tilde{\Lambda}^{6}}\right).
\end{align}

We now specify the cosmological sector of the theory. The analysis is restricted to a spatially flat FLRW background. Spatial homogeneity and isotropy constrain the vector field to the purely temporal configuration
\begin{equation}\label{vector field}
    A_{\mu}dx^{\mu} = -\phi(t)\,dt,
\end{equation}
where \(\phi(t)\) is a time-dependent scalar function. Varying the action in Eq. (\ref{EPNaction}) with respect to the metric and the vector field yields the full set of cosmological field equations,
\begin{align}
    H^{2} &= \frac{1}{3}\left(\rho^{(\text{fr})}_{\text{DE}}+\rho^{(\text{fr})}_{\text{DM}} 
    + \rho_{\text{EPN}}\right), \label{Friedmann} \\[6pt]
    \dot{H} + H^{2} &= -\frac{1}{6M^{2}_{\text{Pl}}}
    \left(\rho^{(\text{fr})}_{\text{DE}} +\rho^{(\text{fr})}_{\text{DM}} + \rho_{\text{EPN}} 
    + 3p^{(\text{fr}}_{\text{DE}}) + 3p_{\text{EPN}}\right), 
    \label{Raychaudhuri} \\[6pt]
    0 &= \beta_{0,X} + 3\left(\beta_{1,X} + d_{1,X}\right)
    \frac{H\phi}{\Tilde{\Lambda}^{2}} 
    + 6\left[\left(\beta_{2,X} + d_{2,X}\right) 
    + \left(\beta_{2,XX} + d_{2,XX}\right)
    \frac{\phi^{2}}{\Tilde{\Lambda}^{2}}\right]\frac{H^{2}}{\Tilde{\Lambda}^{2}} 
    \notag \\
    &\quad - \left[3\left(\beta_{3,X} + d_{3,X}\right) 
    + \left(\beta_{3,XX} + d_{3,XX}\right)
    \frac{\phi^{2}}{\Tilde{\Lambda}^{2}}\right]
    \frac{H^{3}\phi}{\Tilde{\Lambda}^{4}}, \label{vector EOM}
\end{align}
where the Hubble parameter is defined by \(H\equiv \dot{a}/a\), and an overdot denotes differentiation with respect to cosmic time \(t\). A key point is that Eq. (\ref{vector EOM}) is algebraic in \(\phi\), rather than a dynamical evolution equation. This feature follows directly from the non-polynomial structure of the EPN Lagrangian.

The effective EPN energy density and pressure are given explicitly by
\begin{align}
\rho_{\text{EPN}} &= \Tilde{\Lambda}^4 \Biggl\{
  -\beta_0 + \beta_{0,X}\frac{\phi^2}{\Tilde{\Lambda}^2}
  + 3\left(\beta_{1,X} + d_{1,X}\right)\frac{H\phi^3}{\Tilde{\Lambda}^4}
  \notag \\
  &\quad + 6\left[
    -\left(\beta_2 + d_2\right)
    + 2\left(\beta_{2,X} + d_{2,X}\right)\frac{\phi^2}{\Tilde{\Lambda}^2}
    + \left(\beta_{2,XX} + d_{2,XX}\right)\frac{\phi^4}{\Tilde{\Lambda}^4}
  \right]\frac{H^2}{\Tilde{\Lambda}^2}
  \notag \\
  &\quad - \left[
    5\left(\beta_{3,X} + d_{3,X}\right)
    + \left(\beta_{3,XX} + d_{3,XX}\right)\frac{\phi^2}{\Tilde{\Lambda}^2}
  \right]\frac{H^3\phi^3}{\Tilde{\Lambda}^6}
\Biggr\},
\label{eq:2.13}
\end{align}
\begin{align}
p_{\text{EPN}} &= \Tilde{\Lambda}^4 \Biggl\{
  \beta_0
  - \left(\beta_{1,X} + d_{1,X}\right)\frac{\phi^2\dot{\phi}}{\Tilde{\Lambda}^4}
  + 2\left(\beta_2 + d_2\right)\frac{3H^2 + 2\dot{H}}{\Tilde{\Lambda}^2}
  \notag \\
  &\quad - 2\left(\beta_{2,X} + d_{2,X}\right)
    \frac{\phi\!\left(3H^2\phi + 2H\dot{\phi} + 2\dot{H}\phi\right)}{\Tilde{\Lambda}^4}
  - 4\left(\beta_{2,XX} + d_{2,XX}\right)
  \frac{H\phi^3\dot{\phi}}{\Tilde{\Lambda}^6}
  \notag \\
  &\quad + \left[
    \left(\beta_{3,X} + d_{3,X}\right)
      \frac{2H^2\phi + 3H\dot{\phi} + \dot{H}\phi}{\Tilde{\Lambda}^3}
    + \left(\beta_{3,XX} + d_{3,XX}\right)
      \frac{H\phi^2\dot{\phi}}{\Tilde{\Lambda}^5}
  \right]\frac{H\phi^2}{\Tilde{\Lambda}^3}
\Biggr\}.
\label{rhoEPN}
\end{align}

Differentiating Eq. (\ref{rhoEPN}) with respect to time and dividing by \(9H^{3}\), we obtain
\begin{equation}\label{EPNrhodot}
\begin{split}
    \frac{\dot{\rho}_{\text{EPN}}}{9H^{3}}=\Tilde{\Lambda}^{4}\Bigg\{\frac{\beta_{0,X}}{\Tilde{\Lambda}^{2}}\frac{2\Phi\Dot{\Phi}}{9H^{3}}+\frac{(\beta_{1,X}+d_{1,X})}{\Tilde{\Lambda}^{4}}\left(\frac{\Phi^{3}\dot{H}+3\Phi^{2}\Dot{\Phi}H}{3H^{3}}\right)-\left[(\beta_{3,XX}+d_{3,XX})\frac{2\Phi^{4}\Dot{\Phi}}{9\Tilde{\Lambda}^{8}}\right]\\
+\frac{2}{3H\Tilde{\Lambda}^{2}}\Bigg[\frac{4(\beta_{2,X}+d_{2,X})}{\Tilde{\Lambda}^{2}}\Phi\Dot{\Phi}+\frac{4(\beta_{2,XX}+d_{2,XX})}{\Tilde{\Lambda}^{4}}\Phi^{3}\Dot{\Phi}\Bigg]\\
-\Bigg[5(\beta_{3,X}+d_{3,X})+(\beta_{3,XX}+d_{3,XX})\frac{\Phi^{2}}{\Tilde{\Lambda}^{2}}\Bigg]\frac{\Phi^{2}(\Dot{H}\Phi+H\Dot{\Phi})}{3H\Tilde{\Lambda}^{6}}\Bigg\}\\
+\frac{4\Dot{H}\Tilde{\Lambda}^{2}}{3H^{2}}\Bigg[-(\beta_{2}+d_{2})+2(\beta_{2,X}+d_{2,X})\frac{\Phi^{2}}{\Tilde{\Lambda}^{2}}+(\beta_{2,XX}+d_{2,XX})\frac{\Phi^{4}}{\Tilde{\Lambda}^{4}}\Bigg].
\end{split}
\end{equation}

By solving Eq. (\ref{Friedmann}) together with Eq. (\ref{EPNrhodot}) and the DM continuity equation, Eq. (\ref{contm}), one obtains
\begin{equation}
    \frac{\Dot{H}}{H^{2}}=\frac{3}{2}\left[\frac{\Tilde{\Lambda}^{4}\zeta(\Phi,\Dot{\Phi})+\rho_{\text{EPN}}/3H^{2}-1-\Omega^{(\text{fr})}_{\text{DE}}w^{(\text{fr})}_{\text{DE}}}{1-2\Tilde{\Lambda}^{2}\chi(\Phi)}\right].
\end{equation}
Here, the auxiliary functions \(\chi(\Phi)\) and \(\zeta(\Phi,\Dot{\Phi})\) are defined by
\begin{equation}
    \chi(\Phi)=-(\beta_{2}+d_{2})+2(\beta_{2,X}+d_{2,X})\frac{\Phi^{2}}{\Tilde{\Lambda}^{2}}+\left(\beta_{2,XX}+d_{2,XX}\right)\frac{\Phi^{4}}{\Tilde{\Lambda}^{4}}
\end{equation}
and
\begin{equation}
\begin{split}
 \zeta(\Phi,\Dot{\Phi})=\Tilde{\Lambda}^{4}\Bigg\{\frac{\beta_{0,X}}{\Tilde{\Lambda}^{2}}\frac{2\Phi\Dot{\Phi}}{9H^{3}}+\frac{(\beta_{1,X}+d_{1,X})}{\Tilde{\Lambda}^{4}}\left(\frac{\Phi^{3}\dot{H}+3\Phi^{2}\Dot{\Phi}H}{3H^{3}}\right)-\left[(\beta_{3,XX}+d_{3,XX})\frac{2\Phi^{4}\Dot{\Phi}}{9\Tilde{\Lambda}^{8}}\right]\\
-\Bigg[5(\beta_{3,X}+d_{3,X})+(\beta_{3,XX}+d_{3,XX})\frac{\Phi^{2}}{\Tilde{\Lambda}^{2}}\Bigg]\frac{\Phi^{2}(\Dot{H}\Phi+H\Dot{\Phi})}{3H\Tilde{\Lambda}^{6}}\\
+\frac{2}{3H\Tilde{\Lambda}^{2}}\Bigg[\frac{4(\beta_{2,X}+d_{2,X})}{\Tilde{\Lambda}^{2}}\Phi\Dot{\Phi}+\frac{4(\beta_{2,XX}+d_{2,XX})}{\Tilde{\Lambda}^{4}}\Phi^{3}\Dot{\Phi}\Bigg]\Bigg\}.
 \end{split}
\end{equation}

Comparison of the resulting expression for \(\dot{H}/H^{2}\) with the corresponding FHDE relation yields the reconstructed EoS and deceleration parameters in EPN cosmology,
\begin{equation}
    w^{(\text{EPN})}_{\text{DE}}=\frac{(2-3\alpha)\left(\Tilde{\Lambda}^{4}\zeta(\Phi,\Dot{\Phi})-1+\rho_{\text{EPN}}/3H^{2}\right)-2\alpha(1-2\Tilde{\Lambda}^{2}\chi(\Phi))}{2\alpha\left(1-2\Tilde{\Lambda}^{2}\chi(\Phi)\right)+\Omega^{(\text{fr})}_{\text{DE}}(2-3\alpha)}
\end{equation}
and
\begin{equation}
\begin{split}
    q^{(\text{EPN})}_{\text{DE}}=-1-\frac{3\alpha}{2-3\alpha}\Bigg\{1+\frac{(2-3\alpha)\left(\Tilde{\Lambda}^{4}\zeta(\Phi,\Dot{\Phi})-1+\rho_{\text{EPN}}/3H^{2}\right)}{2\alpha\left(1-2\Tilde{\Lambda}^{2}\chi(\Phi)\right)+\Omega^{(\text{fr})}_{\text{DE}}(2-3\alpha)}-\\
\frac{2\alpha(1-2\Tilde{\Lambda}^{2}\chi(\Phi))}{2\alpha\left(1-2\Tilde{\Lambda}^{2}\chi(\Phi)\right)+\Omega^{(\text{fr})}_{\text{DE}}(2-3\alpha)}\Bigg\}.
\end{split}
\end{equation}

The above relations define the reconstructed cosmological observables, namely the EoS parameter and the deceleration parameter, in the combined FHDE--EPN framework. A central feature of this reconstruction is the explicit appearance of the fractional parameter \(\alpha\), through which non-local effects are incorporated into the cosmological dynamics. In this way, the EPN setup provides a concrete modified-gravity realisation of FHDE. Having established this reconstruction, we next consider another modified cosmological setting, namely the DGP braneworld scenario.

\subsection{Dvali--Gabadadze--Porrati Braneworld Theory}\label{Braneworld}

The braneworld scenario with large extra dimensions, in which the observable Universe is described as a four-dimensional hypersurface or \textit{brane} embedded in a higher-dimensional bulk spacetime, has received sustained attention as a framework for modelling DE dynamics \cite{Euclid:2025vml}. In this setting, cosmological evolution on the brane is governed by a modified Friedmann equation that incorporates the gravitational backreaction of the bulk geometry. Among the best-known realisations of this idea is the Dvali--Gabadadze--Porrati (DGP) model \cite{Dvali:2000hr}, where the four-dimensional FLRW Universe is represented as a brane evolving in a five-dimensional Minkowski bulk of infinite volume. The recovery of four-dimensional Einstein gravity at observable scales is achieved by adding to the brane action an induced Einstein-Hilbert term constructed from the intrinsic curvature of the brane. The self-accelerating branch of the DGP model allows late-time cosmic acceleration to emerge as a purely geometric effect, without the need for an explicit DE component or exotic matter \cite{Deffayet:2000uy, Deffayet:2001pu}. However, this branch is known to suffer from ghost instabilities and cannot independently realise a crossing of the phantom divide; such behaviour requires the presence of at least one additional energy component on the brane. By contrast, the normal branch does not generate acceleration on its own, but it can exhibit a phantom-like effective EoS through the mechanism of dynamical gravitational screening localised on the brane.

In the DGP braneworld framework, the EoS parameter evolves dynamically due to bulk-induced effects. This feature is consistent with the broader interest in models beyond \(\Lambda\)CDM, particularly those that allow for a time-dependent DE sector. Observational results from WMAP9 and Planck-2013 for CMB anisotropies, BAO distance measurements from galaxy surveys, and magnitude--redshift data for distant SNe Ia from the SNLS3 and Union2.1 compilations have frequently been interpreted as motivating dynamical DE scenarios over a strictly constant CC description. In this context, the authors of \cite{Ghaffari:2014pxa} proposed, for the first time, an HDE model with a GO cutoff in the DGP braneworld theory. The GO cutoff is particularly useful because it naturally supports a dynamical DE sector. Nevertheless, in light of DESI DR2 observations, the EoS value predicted in that model falls outside the most recent observationally preferred range. More specifically, the model yields a present-day value of approximately \(w_{\text{DE}}\approx -0.62\), whereas current combined constraints with supernova data favour a value in the range \(w_{\text{DE}}\sim -0.75\) to \(-0.84\).

We now turn to the FHDE model and, for simplicity, adopt the Hubble cutoff instead of the GO cutoff. In the DGP braneworld model, the modified Friedmann equation is given by
\begin{equation}
    H^{2}=\left(\sqrt{\frac{\rho}{3}}+\frac{1}{4r^{2}_{c}}+\frac{\epsilon}{2r_{c}}\right)^{2}
\end{equation}
where \(\epsilon=\pm1\) labels the two branches of the DGP solution and \(r_{c}\) denotes the crossover length scale separating the short- and long-distance gravitational regimes, defined by \(r_{c}=G_{5}/2G_{4}\). In the limit \(r_{c}\gg1\), one recovers the standard Friedmann equation of four-dimensional cosmology. For a spatially flat FLRW Universe, the same equation can be written in the form
\begin{equation}\label{DGPFriedmann}
    H^{2}-\frac{\epsilon}{r_{c}}H=\frac{\rho}{3}.
\end{equation}

The parameter \(\epsilon\) plays a central role in distinguishing the two branches of the DGP model. For \(\epsilon=+1\), and in the absence of any matter or energy content on the brane, that is, \(\rho=0\), the model admits a de Sitter solution with constant Hubble parameter \(H=r_{c}^{-1}\). This gives rise to an accelerating Universe with a constant EoS parameter, similar to the CC. On the other hand, for \(\epsilon=-1\) and \(r_{c}\ll H^{-1}\), the term \(H^{2}\) may be neglected in the modified Friedmann equation, leading to
\begin{equation}
    H^{2}=\frac{\rho^{2}}{36M^{6}_{5}}.
\end{equation}
This is precisely the Friedmann equation of the spatially flat RS II braneworld model. It immediately shows that the model does not possess a self-accelerating solution, thereby requiring the introduction of a DE component on the brane. In the present work, this role is played by the FHDE density.

Substituting the FHDE density, with the Hubble cutoff, into the modified Friedmann equation (\ref{DGPFriedmann}), we obtain expressions for the EoS parameter and the deceleration parameter in terms of the L\'evy index \(\alpha\) and the remaining parameters of the FHDE--DGP setup. In the absence of interaction, the EoS parameter is found to be
\begin{equation}
w^{(\text{DGP})}_{\text{DE}}=\frac{-2\alpha +\left(3\alpha-2\right)\left[(1+y)\Omega^{(\text{fr})}_{\text{DE}}-\epsilon\left[\Dot{H}/3H^{3}r_{c}-\Dot{r}_{c}/3H^{2}r^{2}_{c}\right]\right]}{2\alpha-(3\alpha-2)\Omega^{(\text{fr})}_{\text{DE}}}
\end{equation}
and the corresponding deceleration parameter is
\begin{equation}
q^{(\text{DGP})}_{\text{DE}}=-1+\frac{3\alpha}{3\alpha-2}\Bigg\{1+\frac{-2\alpha +\left(3\alpha-2\right)\left[(1+y)\Omega^{(\text{fr})}_{\text{DE}}-\epsilon\left[\Dot{H}/3H^{3}r_{c}-\Dot{r}_{c}/3H^{2}r^{2}_{c}\right]\right]}{2\alpha-(3\alpha-2)\Omega^{(\text{fr})}_{\text{DE}}}\Bigg\}.
\end{equation}

These expressions encode the cosmological behaviour of FHDE in the DGP braneworld when the Hubble radius is chosen as the IR cutoff. A natural extension of the present analysis would be to examine the numerical evolution of these quantities for different values of \(\alpha\in(1,2]\), although such an investigation is deferred to future work. We emphasize that these modified cosmological parameters may lead to phenomenological predictions that agree more closely with recent DESI observations than the results reported in \cite{Ghaffari:2014pxa}. If confirmed, this would point to an interesting phenomenological connection between FC-motivated HDE and higher-dimensional braneworld cosmology, with the Hubble horizon playing the role of the IR cutoff.

\section{Conclusion}

The first half of this chapter, namely Section (\ref{late-timecosmic}), examined the interplay between FC and the HDE framework during the late-time expansion of the Universe. The main objective was to determine how the non-local features introduced by FC influence the emergence or avoidance of rip singularities in the FHDE model. To carry out this analysis, we adopted an explicit IR cutoff prescription and focused on the GO and Hubble cutoffs.

We began with the GO cutoff,
\[
L=(\gamma H^{2}+\delta\dot{H})^{-\frac{1}{2}},
\]
because its dependence on both \(H(t)\) and \(\dot{H}\) makes it particularly suitable for studying late-time cosmic behaviour. In comparison with more conventional prescriptions, such as the Hubble, particle horizon, and event horizon cutoffs \cite{Trivedirip_2024, TrivediScherrer_2024}, the GO cutoff provides a more flexible framework for tracking the onset of rip singularities. An additional advantage of this choice is that big-rip and pseudo-rip behaviours can be investigated without imposing an \textit{ansatz} for the Hubble parameter. Owing to its extended structure, the GO cutoff provides a broader framework for analysing the presence or absence of rip singularities. In particular, it allowed us to study all rip scenarios through the analysis of Eq. (\ref{Hint}).

Our results show that a big-rip singularity \textit{does} occur for fractional values of the L\'evy index within the physically admissible interval \(1<\alpha\leq2\). This conclusion is supported by the numerical results displayed in Figures (\ref{Figure 1}) and (\ref{Figure 2}), which illustrate the late-time evolution of the EoS parameter and the squared sound speed (SSS), respectively. By contrast, the analysis suggests that a pseudo-rip singularity may appear only when \(\alpha>2\). Since the L\'evy index is restricted to values not exceeding \(2\), this possibility is excluded from the present framework. This result indicates that pseudo-rip behaviour is not supported by fractional, non-local effects, whereas the big-rip singularity is directly associated with such features.

We further note that a little-rip singularity may arise in FHDE only for a specific class of IR cutoffs. In the present setting, its realisation is highly restrictive and requires a cutoff of the form
\[
L\sim(\gamma H^{2}+g(H))^{-\frac{1}{2}},
\]
which belongs to the broader class of NO cutoffs. Having clarified the rip structure associated with the GO cutoff, we then turned to Section (\ref{littleandpseudorip}), where the occurrence of little-rip and pseudo-rip singularities was studied by substituting the relevant \textit{ansatz} for the Hubble parameter \(H(t)\). The corresponding expressions are given in Eq. (\ref{little rip ansatz}) and Eq. (\ref{pseudo rip ansatz}) for the little-rip and pseudo-rip cases, respectively.

More precisely, we inserted the chosen forms of \(H(t)\) into the EoS parameter \(w_{\text{HH}}(t)\), Eq. (\ref{EoSHH}), and the SSS parameter \(v^{2}_{\text{HH}}(t)\), Eq. (\ref{vsHH}), derived in the FHDE framework for the Hubble cutoff \(L=H^{-1}\). Their late-time evolution was then analysed numerically. The results for the little-rip scenario are shown in Figures (\ref{Figure 3}) and (\ref{Figure 4}), while those for the pseudo-rip scenario are presented in Figures (\ref{Figure 5}) and (\ref{Figure 6}). The choice \(L=H^{-1}\) was made deliberately to show that late-time rip singularities can still emerge from fractional effects even when one adopts the simplest cutoff, despite the known limitations of the Hubble cutoff in standard HDE. The main outcomes of this analysis are listed in Table (\ref{Table}).

\begin{table}[ht]
    \centering
    \caption{Occurrence and avoidance of rip scenarios in the ansatz-based analysis.}
    \label{Table}
    \begin{tabular}{|c|c|c|}
        \hline
        Rip Scenario & Linear \(Q\) & Non-linear \(Q\) \\
        \hline
        B.R (\ref{big rip ansatz}) & Occurs for \(\alpha \to 1.1\) & No B.R \\
        L.R (\ref{little rip ansatz}) & No L.R & Occurs for all \(\alpha \in (1,2]\) \\
        P.R (\ref{pseudo rip ansatz}) & No P.R & Occurs for all \(\alpha \in (1,2]\) \\
        \hline
    \end{tabular}
\end{table}

To summarise the EoS behaviour for the different rip scenarios, we first consider the big-rip case. Within the GO cutoff, we studied the late-time evolution of \(w_{\text{GO}}(t)\) in both the linearly interacting regime, shown in Figure (\ref{Figure 1a}), and the non-linearly interacting regime, shown in Figure (\ref{Figure 1b}). For linear interaction, \(w_{\text{GO}}(t)\) diverges to negative infinity at late times, that is,
\[
w_{\text{GO}}(t)\rightarrow -\infty,
\]
for representative values such as \(\alpha=1.1\) and \(1.3\). This behaviour signals the appearance of a big-rip singularity when fractional effects become dominant. Although the values \(\alpha=1.1\) and \(1.3\) are illustrative, the essential result is that the same behaviour persists as \(\alpha\to1.1\), as indicated in Figure (\ref{Figure 1a}). In contrast, when the interaction term is taken to be non-linear, \(w_{\text{GO}}(t)\) remains approximately constant even at late times, and therefore does not support a big-rip singularity. We therefore conclude that, in the linearly interacting case, a big rip occurs when fractional effects dominate in the limit \(\alpha\to1.1\).

For the little-rip scenario, we analysed the late-time behaviour of \(w_{\text{HH}}(t)\) within the Hubble cutoff for both linear and non-linear interaction terms, as shown in Figures (\ref{Figure 3a}) and (\ref{Figure 3b}), respectively. In the linear case, the EoS parameter again tends to negative infinity as \(\alpha\to1.1\), with representative examples given by \(\alpha=1.1\) and \(1.3\). This behaviour is not characteristic of a little-rip singularity. By contrast, for values closer to \(\alpha=2\), such as \(\alpha=1.8\) and \(2.0\), the evolution tends to a constant value, as shown in Figure (\ref{Figure 3a}). However, the overall structure of the linear case is still more consistent with big-rip behaviour than with a little rip. In the non-linear interaction regime, the situation changes qualitatively: for all \(1<\alpha\leq2\), the EoS parameter approaches \(-1\) from below at late times. This asymptotic behaviour is precisely the one required for the realisation of a little-rip singularity. We therefore infer that a little rip occurs throughout the allowed interval \(1<\alpha\leq2\) in the non-linearly interacting case.

For the pseudo-rip scenario, we again examined the late-time EoS evolution within the Hubble cutoff in both interaction regimes, shown in Figures (\ref{Figure 5a}) and (\ref{Figure 5b}). In the linear regime, the EoS approaches \(-1\) for values such as \(\alpha=1.8\) and \(2.0\), whereas for \(\alpha=1.1\) and \(1.3\) it asymptotically approaches values smaller than \(-1\), as displayed in Figure (\ref{Figure 5a}). This indicates that the regime of strongest fractional effects, corresponding to \(\alpha\to1.1\), does not yield the asymptotic behaviour required for a pseudo rip. In the non-linear regime, however, the EoS evolves consistently with pseudo-rip behaviour: for every \(\alpha\in(1,2]\), it approaches \(-1\) from below during the late-time expansion of the Universe. Accordingly, we conclude that a pseudo-rip singularity is realised for all admissible values of \(\alpha\) in the non-linearly interacting case.

The squared sound speed remains negative for all rip scenarios considered in the FHDE framework during the late-time expansion of the Universe. This does not exclude the occurrence of these rip solutions. Rather, it indicates that the corresponding cosmological fluids are classically unstable at the perturbative level. In particular, any perturbation would generate a gradient instability, since negative \(v_s^2\) makes the system highly sensitive to even arbitrarily small disturbances. Such sensitivity to initial conditions, often associated with chaotic behaviour, suggests that the long-term evolution of rip scenarios in FHDE may be governed by nonlinear instabilities. These effects could be quantified through Lyapunov exponents or related diagnostics of dynamical instability \cite{GOLDHIRSCH1987311}. A rigorous treatment of these issues requires a dedicated perturbative analysis of the FHDE model, and several methods are available for this purpose; for a detailed review, see \cite{Astashenok_2024}. In particular, the framework developed in \cite{Astashenok_2024} could be extended by incorporating fractional effects alongside nonlinear dynamical indicators to determine stability thresholds more precisely within FHDE.

Although the present study is primarily theoretical, the results motivate a concrete observational programme in fractional cosmology. A substantial body of work has already shown that fractional and fractal features can be constrained through observational data. In FHDE, the DE density evolves as a function of the L\'evy index \(\alpha\), thereby modifying the late-time expansion rate relative to \(\Lambda\)CDM in a way controlled entirely by the fractional parameter. Recent analyses of rip cosmologies using combined CMB, BAO, and SN Ia data indicate that all rip scenarios remain statistically compatible with \(\Lambda\)CDM at the \(1\sigma\) level \cite{LazkozRip}. For FHDE, the central observational question is whether \(\alpha\) differs measurably from \(2\). A joint MCMC analysis using datasets such as DESI DR2 and Pantheon+ would therefore provide a direct route to constraining the fractional sector.

At the same time, the FHDE framework is not expected to leave a direct early-universe signature in the CMB temperature power spectrum, because its fractional holographic corrections are intrinsically IR and operate at late times. Nevertheless, one channel may still provide CMB sensitivity. The modified DE dynamics alter the late-time decay of gravitational potentials and thereby contribute to the CMB temperature anisotropies through the ISW effect; see \cite{Planck:2015fcm, Giannantonio_2012, St_lzner_2018}. In FHDE, values of \(\alpha\) close to \(1.1\) enhance the growth of DE density relative to the standard case \(\alpha=2\), which may amplify the ISW contribution on the Sachs--Wolfe plateau. Such an effect could, in principle, be detected through cross-correlations between CMB maps and large-scale structure surveys.

Another potentially informative observational window is provided by the GW spectrum. It has already been shown for Tsallis, Barrow, and R\'enyi HDE models that different entropy-based HDE constructions lead to qualitatively distinct GW transfer functions \cite{Maity:2024tkq}. Extending this analysis to FHDE would introduce an \(\alpha\)-dependent tensor spectral tilt, thereby opening a direct way to probe the relation between fractional effects and gravitational-wave propagation in HDE models.

The main results of this study are summarised in Table (\ref{Table}). Overall, the FHDE framework introduces qualitatively new features into the late-time singularity structure of DE cosmologies. In particular, non-locality, or memory effects, provides an additional mechanism that enriches the existing literature on rip singularities in HDE \cite{Brevik:2024ozg,TrivediScherrer_2024}. We stress, however, that the present analysis is entirely classical. An important direction for future work is a quantum cosmological treatment of the big-rip singularity in the GO cutoff.

In the second half of this chapter, particularly in Section (\ref{modified}), several reconstructions associated with the FHDE framework were investigated from a purely theoretical standpoint. We selected a set of established and comparatively less explored modified gravity models coupled to HDE and used them to derive late-time cosmological observables, including the EoS parameter and the deceleration parameter. The central aim was to embed non-local features into these modified-gravity settings to capture the dynamical character of DE through time- or redshift-dependent observables. We also note that numerical plots for different values of \(\alpha\in(1,2]\) were not included in this part of the analysis. This aspect is left for future work and may serve as a useful numerical extension for further study.

\chapter{Discussions and Outlook}\label{Chapter6}

The central contribution of this thesis is to place the DE problem within a framework that is explicitly sensitive to both holography and fractional operators. Rather than treating late-time cosmic acceleration as a purely phenomenological input, the thesis approaches it as a problem that may carry imprints of quantum-gravitational physics. The standard \(\Lambda\)CDM model remains extraordinarily successful at the observational level, yet its theoretical foundation is incomplete when one attempts to understand the smallness of the CC and the origin of the DE sector. In this respect, the present work does not merely propose another dynamical DE model; it seeks to motivate such a model from a deeper conceptual starting point, namely the \textit{interplay} between the HP and FC \cite{Li:2004rb,COPELAND_2006,Trivedi:2024inb}.

What makes the FHDE construction especially interesting is that its fractional correction is not introduced arbitrarily. Instead, it is traced back to BHT through a fractional modification of the Bekenstein--Hawking entropy, itself motivated by the FWDW framework and the broader programme of FQM \cite{Jalalzadeh_2021,moniz2020fractional,Varao:2024eig}. This is conceptually significant because it ties late-time cosmology to the quantum structure of spacetime through an identifiable chain of arguments: fractional quantum dynamics modifies black hole entropy, the entropy modifies the holographic bound, and the holographic bound, in turn, leads to a modified HDE density. This thesis, therefore, presents a coherent route from microscopic considerations to cosmological phenomenology. Whether this route is ultimately the correct one remains open, but it is internally motivated in a way that many DE parameterisations are not.

To this end, we would like to highlight the scientific publications from this thesis that correspond to the following chapters:
\begin{itemize}
    \item \textbf{Fractional Holographic Dark Energy} -- Available in published form; see \cite{Trivedi:2024inb}. This publication corresponds to Section (\ref{FHDE}) of Chapter (\ref{Chapter3}).
    \item \textbf{Reconstructing FHDE with Scalar and Gauge fields} -- Available in published form; see \cite{Bidlan_2025}. This publication corresponds to Section (\ref{ReconFHDE}) of Chapter (\ref{Chapter3}).
    \item \textbf{Future Rip Scenarios in Fractional Holographic Dark Energy} -- Currently under review; see \cite{Bidlan:2026xaa}. This paper corresponds to Section (\ref{late-timecosmic}) of Chapter (\ref{Chapter4}).
\end{itemize}

A particularly noteworthy aspect of the thesis is that the FHDE density reduces continuously to the standard HDE expression in the limit \(\alpha \to 2\). It ensures that the FHDE description includes the conventional HDE scenario as a special case, while allowing departures from it to be controlled by a single parameter, namely the L\'evy index \(\alpha\). In effect, \(\alpha\) plays the role of a parameter that quantifies the strength of fractional, non-local, or memory effects on the late-time cosmic dynamics. The theoretical appeal of this arrangement is that the model is neither disconnected from the standard HDE literature nor excessively bombarded with multiple parameters. It lies between conservative extensions and genuine conceptual constructions using FC in BHT.

From a cosmological perspective, the most non-trivial result of the thesis is the rehabilitation of the Hubble cutoff within the FHDE setting. In conventional HDE, the Hubble cutoff is well known to be problematic because it typically fails to produce a viable accelerating universe. The point is not simply that one more viable model has been added to the literature; rather, a cutoff that is usually regarded as too restrictive becomes viable once the entropy-area relation is modified by fractional effects. This suggests that the shortcomings of the Hubble cutoff in standard HDE may reflect the incompleteness of the underlying entropy prescription, rather than a fatal flaw of the cutoff itself. Furthermore, the detailed cosmological analysis supports this interpretation. The evolution of the DE density parameter, the DM density parameter, the deceleration parameter, and the EoS parameter shows that the FHDE model can reproduce a plausible late-time expansion, especially for smaller values of \(\alpha\). In these regimes, the DE sector becomes increasingly dynamical while remaining observationally consistent with DESI results, and the transition from matter domination to accelerated expansion occurs in a way qualitatively compatible with current cosmological expectations. That is precisely what makes it relevant in the present observational landscape in cosmology, where deviations from strict \(\Lambda\)CDM behaviour continue to attract attention \cite{desicollaboration2024desi2024vicosmological,DESI:2025zgx,adame2024desi3}.

At the same time, the thesis does not present FHDE as a ``completed theory''. The analysis of the \(w\)--\(w'\) plane and the squared sound speed reveals a more nuanced picture. On the one hand, the model tends to remain in the quintessence-like regime and, for suitable values of \(\alpha\), approaches \(w_{\text{DE}} \to -1\) in the far future without crossing deeply into phantom behaviour. On the other hand, the classical stability analysis indicates that the model is not free from difficulties, particularly in the asymptotic future, where instability may develop. It prevents the discussion from becoming triumphalist and instead positions FHDE as a promising but unfinished framework. In other words, the fractional correction improves the cosmological behaviour of HDE theory in an essential way, but it does not by itself solve every problem associated with the model.

One of the significant parts of the thesis is the reconstruction programme. By establishing correspondences between FHDE and several effective field configurations --- quintessence, K-essence, dilaton, DBI-essence, tachyon, and YMC --- the thesis moves beyond background cosmology and asks a deeper question: \textit{what kinds of effective field theories can reproduce the FHDE dynamics?} This is a useful strategy for two reasons. First, it embeds FHDE in a language that is familiar across the DE literature. Second, it allows one to diagnose which kinds of field structures are most naturally compatible with the fractional holographic scenario. The reconstructions reveal a pattern that is both encouraging and instructive. Across most of the models studied, smaller values of \(\alpha\) tend to produce smoother kinetic evolution and an EoS that approaches the \(\Lambda\)CDM limit in the far future. This repeated appearance of the same tendency suggests that the role of \(\alpha\) is not an artefact of one particular parametrisation, but a robust structural feature of the FHDE framework. In that sense, the thesis identifies a clear phenomenological message, i.e., a viable late-time cosmology is associated with a regime in which fractional effects are present but not extreme, and in which the deviation from the standard HDE limit is sufficient to generate acceleration without destabilising the dynamics too violently (see \cite{Rathore:2025ptu} for recent observational analysis of FHDE model).

More specifically, quintessence, K-essence, tachyon, and DBI-type reconstructions exhibit broad qualitative similarities, but the dilaton case behaves differently in a physically informative way. The dilaton prefers a somewhat different relationship between \(\alpha\) and the acceptable evolution of its kinetic and potential terms, indicating that the FHDE correspondence is sensitive to the structure of the underlying field theory. This implies that the significance lies not only in \textit{which} models work, but in \textit{how} and \textit{why} they differ. FHDE is therefore not merely flexible; it is discriminating. The inclusion of YMC adds another useful dimension to the FHDE reconstruction programme. Much of the DE literature is dominated by scalar-field models, so the extension to a spin-1 gauge-field configuration is conceptually rich. The YMC reconstruction shows that the qualitative tendency toward \(w \to -1\) and smoother late-time evolution for smaller \(\alpha\) is not confined to scalar sectors alone. This broadens the interpretive scope of the thesis. If a fractional holographic modification can be realised in both scalar and gauge-field languages, then the framework may be probing something more universal than a single model-building choice. At a minimum, it suggests that the FHDE background can be read as an effective cosmological signature compatible with multiple effective field descriptions.

Another important theme running through the thesis is the status of \(\alpha\) itself. Mathematically, it is a parameter inherited from the fractional framework; physically, it carries the features of non-locality, memory effects, or hidden geometric (fractal) structure associated with fractional dynamics. This makes \(\alpha\) more than just an arbitrary parameter. It becomes the central mediator between the microscopic and macroscopic sectors of the theory. Yet this also exposes one of the main conceptual challenges of the entire programme: although the thesis demonstrates that \(\alpha\) has rich cosmological consequences, its direct physical interpretation remains only partially understood. One may say that the thesis establishes \(\alpha\) as a meaningful phenomenological parameter, but not yet as a uniquely derived observable of quantum gravity. This is not a weakness of the thesis alone; it reflects the current state of the field. Still, the distinction matters. The next stage of the programme must move from ``\(\alpha\) works'' to ``\(\alpha\) means something \textit{definite}'' with future observations in cosmology.

The thesis also contributes by clarifying how FHDE compares with other entropy-corrected holographic models, especially Tsallis and Barrow scenarios. Structurally, these models share the feature that modifications to the entropy-area relation induce non-standard DE densities. However, the present work makes clear that FHDE differs in motivation. In the FHDE case, the correction is tied specifically to fractional quantum-mechanical and quantum-geometrodynamical considerations. For example, any two HDE models may have similar power-law structures at the effective level and yet reflect very different underlying physics. The value of the thesis lies partly in insisting on this distinction rather than collapsing all power-law corrections into a single phenomenological class.

The latter parts of the thesis, especially the extension to modified gravity settings and the analysis of future singularities, reinforce the idea that FHDE should be regarded as a flexible framework rather than a single isolated model. Once the DE density is modified to depend on \(\alpha\), it is natural to ask how this change propagates through non-GR cosmologies and whether it ameliorates or worsens late-time pathologies, such as rip singularities. Even when these investigations remain at the classical level, they are important because they test the stability of the central idea across different dynamical environments. A proposal as ambitious as FHDE cannot be assessed solely within a single background model; it must be stress-tested across multiple cosmological settings. The thesis takes meaningful steps in that direction.

What, then, is the broader significance of the work? In my view, it treats the opportunity to understand the elusive nature of DE as a possible window into the microscopic structure of spacetime. The HP already encourages us to believe that gravitational systems store and organise information in non-local ways, while FC offers a formal language for memory and non-local response. FHDE brings these ideas together in a cosmological setting and asks whether late-time acceleration may be one of their observable consequences. That is a strong conceptual manoeuvre because, even if the final theory of DE turns out to be different, this thesis shows that the conversation among holography, black hole thermodynamics, and fractional dynamics is indeed interesting. At the same time, the work points clearly toward the next set of questions. The most immediate is observational: can the parameter \(\alpha\) be constrained tightly enough using DESI, supernovae, CMB, BAO, DES, cosmic chronometers, and related probes to distinguish FHDE from \(\Lambda\)CDM and from other corrected HDE models? Closely related is the diagnostic question of whether FHDE leaves a distinctive imprint in late-time observables such as the ISW effect or structure-growth data. There is also the unresolved issue of future stability, which may require either alternative infrared cutoffs or a more complete interacting framework.

Since the introduction of FHDE in 2024 \cite{Trivedi:2024inb}, the framework has attracted increasing attention in both theoretical and observational cosmology. On the theoretical side, subsequent developments include the formulation of a fractional agegraphic DE model \cite{Naeem:2025ybg}, as well as FHDE reconstructions within \(f(Q)\) and \(f(T)\) teleparallel gravity \cite{Mazumdar:2025jfv,Meetei:2026eja}. In addition, a cosmographic connection between cosmological and Planckian scales, incorporating fractional entropy alongside Barrow and Tsallis entropies, has been explored in \cite{Bolotin:2026hot}. From the observational perspective, FHDE has also been examined in a range of cosmological settings \cite{Rathore:2025ptu,Robles-Barba:2025dkx, universe12050134}. Taken together, these developments indicate that FHDE has emerged as a serious and competitive DE candidate, with its relevance likely to grow as future observations place tighter constraints on dynamical DE models.

\section{Future Perspectives and Ongoing Works}

To conclude, several promising directions for the future development of the FHDE programme are worth highlighting. A natural next step, building directly on the results of this thesis, is to formulate a quantum cosmological treatment of the FHDE model itself. More specifically, one would aim to investigate the resolution of late-time cosmic singularities --- most notably the Big Rip --- by constructing a WDW equation for an effective DE component whose non-local structure is encoded in the fractional parameter \(\alpha\). And, indeed, we find that such a construction resolves the Big Rip singularity within appropriate cosmological settings in [Bidlan, Friedrich, Moniz; \textit{in preparation}]. Extending such a quantum treatment beyond the Big Rip would also be essential for understanding other finite- and infinite-time future singularities, including the Little Rip, the Pseudo Rip, and the Big Freeze. Another particularly interesting direction is to develop an inflationary realisation of FHDE, since the present thesis has focused primarily on its late-time cosmological implications (see \cite{Darabi:2016mjg} for a detailed review). In a related context, methods from fractional scalar-field cosmology may prove useful for studying spacetime singularities arising during the gravitational collapse of a matter cloud sourced by an FHDE component reconstructed through quintessence; see \cite{Saha:2023zos} for a related perspective. And, we find that gravitational memory features indeed contribute non-trivially in formation and avoidance of trapped surfaces in [Bidlan, Moniz; \textit{in prep.}]. Finally, confronting the FHDE model with observational probes such as the CMB spectrum through the ISW effect, together with DESI and DES data, remains an important open avenue for future work.

\appendix



\chapter{Implementing Fractional Operators in Gravity}\label{AppendixFC}

A systematic incorporation of fractional operators into gravitation and cosmology has been explored in the broader context of nonlocal and perturbative quantum field theories of gravity. As already noted in the introductory chapter, fractional calculus naturally introduces nonlocality. More precisely, the notion of ``nonlocality'' arises from the very definition of fractional operators, which may be represented as convolutions of a field \(\psi(x)\) with an integral kernel \(F(x-y)\) over the spacetime manifold. In this sense, several candidate quantum-gravity models possessing ultraviolet completions or infrared corrections can be understood as nonlocal theories, where the classical action and the associated equations of motion involve operators of the form \(\mathcal{F}(\Box)\), with infinitely many derivatives, and where \(\Box=\nabla_{\mu}\nabla^{\mu}\) denotes the covariant Laplace--Beltrami operator, or d'Alembertian. Symbolically, one may write
\begin{equation}
    \mathcal{F}(\Box)\psi(x)=\int d^{D}y\,F(y-x)\psi(y).
\end{equation}

Among the candidate frameworks for ultraviolet-complete nonlocal quantum gravity are string theory, nonlocal gravity with exponential or asymptotically polynomial form factors, and multifractional quantum field theories with fractional operators. Of these, the least explored are multifractional quantum field theories with fractional operators. For a detailed review of the scalar-field case, see \cite{Calcagni:2021ljs}. There, the authors investigate whether fractional form factors can improve the ultraviolet behaviour of quantum field theory by examining three broad classes of models, namely \((i)\) \(T[\partial^{\gamma}]\), \((ii)\) \(T[\partial+\partial^{\gamma}]\), and \((iii)\) \(T[\partial^{\gamma(\ell)}]\). The first contains a single fractional derivative of order \(\gamma\), the second mixes ordinary and fractional derivatives, and the third promotes the fractional order to a scale-dependent quantity distributed continuously over a length scale \(\ell\).

A central conclusion of \cite{Calcagni:2021ljs} is that perturbative scalar quantum field theories based on Lorentz-invariant or Lorentz-breaking nonlocal operators of fractional order, defined on spacetimes with scale-dependent spectral dimension, display a delicate interplay between unitarity and renormalisability. In general, these theories can be ghost-free or power-counting renormalisable, but not both simultaneously. Even so, some classes are one-loop unitary and finite, and may possibly remain so at higher orders. The extension of these ideas to gravity reveals an equally rich structure; see \cite{Calcagni:2021aap}. In the present chapter, we provide a concise overview of this interplay, largely following \cite{Calcagni:2021aap,Calcagni:2021ljs}. In particular, the classes introduced above admit natural covariant counterparts in terms of the d'Alembertian operator, namely \((i)\) \(T[\Box^{\gamma}]\), \((ii)\) \(T[\Box+\Box^{\gamma}]\), and \((iii)\) \(T[\Box^{\gamma(\ell)}]\), where the form factors involve non-integer powers of \(\Box\) while preserving covariance.

Before turning to each specific framework, it is useful to record some general variation formulae for a generic derivative operator \(\mathbf{D}\). Given \(\mathbf{D}\), the associated Levi--Civita connection, Ricci tensor, and Ricci scalar may be written as
\begin{equation}
\Gamma^{\sigma}_{\mu\nu}[\mathbf{D}]:=\frac{1}{2}g^{\rho\sigma}\left(\mathbf{D}_{\mu}g_{\nu\sigma}+\mathbf{D}_{\nu}g_{\mu\sigma}-\mathbf{D}_{\sigma}g_{\mu\nu}\right),
\end{equation}
\begin{equation}
    R_{\mu\nu}[\mathbf{D}]:=\mathbf{D}_{\sigma}\Gamma^{\sigma}_{\mu\nu}[\mathbf{D}]-\mathbf{D}_{\nu}\Gamma^{\sigma}_{\mu\sigma}[\mathbf{D}]+\Gamma^{\tau}_{\mu\nu}[\mathbf{D}]\Gamma^{\sigma}_{\sigma\tau}[\mathbf{D}]-\Gamma^{\tau}_{\mu\sigma}[\mathbf{D}]\Gamma^{\sigma}_{\nu\tau}[\mathbf{D}],
\end{equation}
with corresponding variations
\begin{equation}\label{varGamma}
    \delta\Gamma^{\rho}_{\mu\nu}[\mathbf{D}]=\frac{1}{2}g^{\rho\sigma}\left(\boldsymbol{\nabla}_{\mu}\delta g_{\nu\sigma}+\boldsymbol{\nabla}_{\nu}\delta g_{\mu\sigma}-\boldsymbol{\nabla}_{\sigma}\delta g_{\mu\nu}\right),
\end{equation}
\begin{equation}\label{varRicci}
    \delta R_{\mu\nu}[\mathbf{D}]=\boldsymbol{\nabla}_{\sigma}\delta\Gamma^{\sigma}_{\mu\nu}[\mathbf{D}]-\boldsymbol{\nabla}_{\nu}\delta\Gamma^{\rho}_{\mu\rho}[\mathbf{D}].
\end{equation}
One may also write\footnote{The covariant derivative \(\boldsymbol{\nabla}\) defined in terms of the generic derivative \(\mathbf{D}\) acts on a covariant rank-2 tensor as \(\boldsymbol{\nabla}_{\sigma}A_{\mu\nu}:=\mathbf{D}_{\sigma}A_{\mu\nu}-\Gamma^{\rho}_{\sigma\mu}[\mathbf{D}]A_{\rho\nu}-\Gamma^{\rho}_{\sigma\nu}[\mathbf{D}]A_{\mu\rho}\), so that \(\boldsymbol{\nabla}_{\sigma}g_{\mu\nu}=0\). For a contravariant vector, \(\boldsymbol{\nabla}_{\sigma}A^{\mu}=\mathbf{D}_{\sigma}A^{\mu}+\Gamma^{\mu}_{\sigma\nu}[\mathbf{D}]A^{\nu}\). Equation~(\ref{varGamma}) follows from the tensorial nature of \(\delta\Gamma^{\rho}_{\mu\nu}[\mathbf{D}]\), whose form in a local inertial frame can be promoted to an arbitrary frame by replacing \(\eta^{\rho\sigma}\) with \(g^{\rho\sigma}\) and ordinary derivatives with covariant derivatives.}
\begin{equation}
    \delta g^{\mu\nu}\mathcal{O}_{\mu\nu}:=g^{\mu\nu}\delta R_{\mu\nu}[\mathbf{D}]=\boldsymbol{\nabla}_{\sigma}\left(g^{\mu\nu}\delta\Gamma^{\sigma}_{\mu\nu}[\mathbf{D}]-g^{\mu\sigma}\delta\Gamma^{\rho}_{\mu\rho}[\mathbf{D}]\right).
\end{equation}
This expression can be rewritten in several equivalent ways; see \cite{Calcagni:2021aap}. In the conventional case, the tensor \(\mathcal{O}_{\mu\nu}\) vanishes in the absence of boundaries, or under boundary conditions with \(\delta g^{\mu\nu}=0\), and may otherwise be cancelled by the York--Gibbons--Hawking term. We now turn to the specific theories generated by different types of fractional operators and discuss their interplay with gravitation.

\section{$T[\partial^{\gamma}]$ and $T[\Box^{\gamma}]$ Theories}

In multifractional theories, the integro-differential operators encode a multiscale structure, implying that the spacetime manifold is endowed with one or more fundamental length scales. When such operators are incorporated into gravity, the natural starting point is an action resembling the Einstein--Hilbert action, with the usual integration measure \(d^{4}x\) preserved, while ordinary derivatives are replaced by fractional or multifractional ones. In the theory \(T[\partial^{\gamma}]\), one introduces a combination of Liouville and Weyl fractional derivatives through the operator \(\mathcal{D}^{\gamma}_{\pm\mu}\),
\begin{equation}
    \mathcal{D}^{\gamma}_{\pm\mu}:=\frac{1}{2}\left({}_{\infty}\partial^{\gamma}_{\mu}\pm {}_{\infty}\bar{\partial}^{\gamma}_{\mu}\right),
\end{equation}
where, suppressing the coordinate index \(\mu\),
\begin{equation}
    {}_{\infty}\partial^{\gamma}f(x):=-\frac{1}{\Gamma(m-\gamma)}\int_{-\infty}^{x}\frac{dx'}{(x-x')^{\gamma+1-m}}\partial_{x'}^{m}f(x'), \qquad m-1\leq\gamma<m,
\end{equation}
\begin{equation}
    {}_{\infty}\bar{\partial}^{\gamma}f(x):=\frac{1}{\Gamma(m-\gamma)}\int_{x}^{+\infty}\frac{dx'}{(x'-x)^{\gamma+1-m}}\partial_{x'}^{m}f(x'), \qquad m-1\leq\gamma<m,
\end{equation}
are, respectively, the Liouville and Weyl derivatives, with \(m=1,2,\dots\). Implementing these operators in the connection and curvature tensors, \cite{Calcagni:2021aap} arrives at the following geometric quantities for gravity in \(T[\partial^{\gamma}]\):
\begin{equation}
    \tilde{\Gamma}^{\rho}_{\mu\nu}:=\frac{1}{2}g^{\rho\sigma}\left(\mathcal{D}^{\gamma}_{\mu}g_{\nu\sigma}+\mathcal{D}^{\gamma}_{\nu}g_{\mu\sigma}-\mathcal{D}^{\gamma}_{\sigma}g_{\mu\nu} \right),
\end{equation}
\begin{equation}
    R^{(\gamma)\rho}_{\mu\sigma\nu}:=\mathcal{D}^{\gamma}_{\sigma}\tilde{\Gamma}^{\rho}_{\mu\nu}-\mathcal{D}^{\gamma}_{\nu}\tilde{\Gamma}^{\rho}_{\mu\sigma}+\tilde{\Gamma}^{\tau}_{\mu\nu}\tilde{\Gamma}^{\rho}_{\sigma\tau}-\tilde{\Gamma}^{\tau}_{\mu\sigma}\tilde{\Gamma}^{\rho}_{\nu\tau},
\end{equation}
\begin{equation}
    R^{(\gamma)}_{\mu\nu}:=R^{(\gamma)\sigma}{}_{\mu\sigma\nu}, \qquad R^{(\gamma)}:=g^{\mu\nu}R^{(\gamma)}_{\mu\nu}.
\end{equation}
The corresponding gravitational action is
\begin{equation}\label{fractionalderivativestheory}
    S=\frac{1}{16\pi G_{\text{N}}}\int d^{4}x\sqrt{|g|}\left[R^{(\gamma)}-2\Lambda\right],
\end{equation}
with equations of motion
\begin{equation}
    R^{(\gamma)}_{\mu\nu}-\frac{1}{2}g_{\mu\nu}R^{(\gamma)}+\Lambda g_{\mu\nu}+\mathcal{O}_{\mu\nu}=8\pi G_{\text{N}}T_{\mu\nu}.
\end{equation}

If \(T[\partial^{\gamma}]\) is to be regarded as a physical theory, then one is naturally led to the regime \(\gamma=1-\epsilon\lesssim1\), so as to preserve agreement with known gravitational phenomenology. However, there is at present no compelling theoretical reason to impose such a fine tuning, namely a value of \(\gamma\) extremely close to, but distinct from, unity. Furthermore, the symmetry structure of the theory is substantially modified: ordinary diffeomorphism invariance is deformed, and local inertial frames are described by fractional Lorentz transformations. In curved spacetime, the fractional covariant derivatives do not commute, but their commutator is no longer simply proportional to the fractional Riemann tensor,
\begin{equation}
    \left[\tilde{\nabla}_{\mu}, \tilde{\nabla}_{\nu}\right]A^{\rho}
    =R^{(\gamma)\rho}_{\lambda\mu\nu}A^{\lambda}+\text{(extra terms)},
\end{equation}
as discussed in \cite{Calcagni:2021aap}. These additional contributions strongly suggest that the underlying spacetime possesses a non-metric structure independent of the metric one. Together with the complicated Leibniz rule for fractional derivatives, this motivates a shift in attention from theories based directly on fractional derivatives to those constructed from fractional powers of the d'Alembertian.

We therefore turn to the covariant theory \(T[\Box^{\gamma}]\), where the action is built from form factors \(\mathcal{F}_{i}(\Box)\) acting on the standard curvature tensors. A generic nonlocal gravitational action takes the form
\begin{equation}\label{D'alemb}
    S=\frac{1}{16\pi G_{\text{N}}}\int d^{D}x\sqrt{|g|}\left[R-2\Lambda+R\mathcal{F}_{1}(\Box)R+G_{\mu\nu}\mathcal{F}_{2}(\Box)R^{\mu\nu}+R_{\mu\nu\rho\sigma}\mathcal{F}_{3}(\Box)R^{\mu\nu\rho\sigma}\right],
\end{equation}
supplemented by a matter action \(S_{m}\). For simplicity, following \cite{Calcagni:2021aap}, we set \(\mathcal{F}_{1}=\mathcal{F}_{3}=0\) and choose
\begin{equation}
    \mathcal{F}_{2}(\Box)=\frac{\left(-\ell_{\star}^{2}\Box\right)^{\gamma-1}-1}{\Box}
    \qquad\Rightarrow\qquad
    (-\Box)^{\gamma}h_{\mu\nu}=0.
\end{equation}
The action then becomes
\begin{equation}
    S=\frac{1}{16\pi G_{\text{N}}}\int d^{4}x\sqrt{|g|}\left[R-2\Lambda+G_{\mu\nu}\frac{\left(-\ell_{\star}^{2}\Box\right)^{\gamma-1}-1}{\Box}R^{\mu\nu}\right],
\end{equation}
with field equations
\begin{equation}
    \begin{split}
        \kappa^{2}T_{\mu\nu}={}&\left(-\ell_{\star}^{2}\Box\right)^{\gamma-1}G_{\mu\nu}+\Lambda g_{\mu\nu}
        +g_{\mu\nu}\nabla^{\sigma}\nabla^{\tau}\frac{\left(-\ell_{\star}^{2}\Box\right)^{\gamma-1}-1}{\Box}G_{\sigma\tau}
        -2\nabla^\sigma \nabla_{(\mu}\frac{\left(-\ell_*^2 \Box\right)^{\gamma-1}-1}{\Box}\,G_{\nu)\sigma} \\
        &-\frac{1}{2} g_{\mu\nu} G_{\sigma\tau}\frac{\left(-\ell_*^2 \Box\right)^{\gamma-1}-1}{\Box}\,R^{\sigma\tau}
        +2G^\sigma{}_{(\mu}\frac{\left(-\ell_*^2 \Box\right)^{\gamma-1}-1}{\Box}\,G_{\nu)\sigma} \\
        &+\frac{1}{2}\left[
        G_{\mu\nu}\frac{\left(-\ell_*^2 \Box\right)^{\gamma-1}-1}{\Box}\,R
        +R\frac{\left(-\ell_*^2 \Box\right)^{\gamma-1}-1}{\Box}\,G_{\mu\nu}
        \right]
        + \Theta_{\mu\nu}(R_{\alpha\beta},G^{\alpha\beta}).
    \end{split}
\end{equation}
Here \(\Theta_{\mu\nu}\) can be written explicitly once a suitable integral or series representation of the form factor is specified.

Since the free graviton equation, \( (-\Box)^{\gamma}h_{\mu\nu}=0 \), is structurally identical to the free massless scalar equation with a fractional d'Alembertian kinetic term, it is natural to expect that the unitarity and renormalisability properties of the scalar theory \(T[\Box^{\gamma}]\) carry over, at least qualitatively, to the corresponding gravitational model. As discussed in \cite{Calcagni:2021aap}, unitarity requires \(-2n<\gamma<1-2n\leq1\), where \(n\in\mathbf{N}\), and this can be reduced to the interval \(0<\gamma<1\) if one also demands a well-defined spectral dimension \(d_{s}=D/\gamma\).\footnote{The spectral dimension is the effective dimensionality of spacetime probed by a diffusing test particle, and is physically meaningful when positive semi-definite.} By contrast, power-counting arguments alone are insufficient to establish renormalisability, which must instead be checked order by order in perturbation theory. At one loop, no new divergences arise provided \(\gamma\neq D/4-n/2\), with \(n\in\mathbf{N}\). For \(D=4\), and within the above unitarity interval, this excludes \(\gamma=1/2\). For further quantum aspects of this fractional-gravity framework, see \cite{Calcagni:2021aap}.

\section{$T[\partial+\partial^{\gamma}]$ and $T[\Box+\Box^{\gamma}]$ Theories}

The multifractional theory \(T[\partial+\partial^{\gamma}]\) can be incorporated into the action in two distinct ways. In the first, one modifies the curvature tensors themselves by replacing the standard connection and Ricci tensor with their fractional counterparts \(\tilde{\Gamma}^{\rho}_{\mu\nu}\) and \(R^{(\gamma)}_{\mu\nu}\) from the \(T[\partial^{\gamma}]\) theory. The resulting action is then the sum of the Einstein--Hilbert term and its fractional analogue,
\begin{equation}\label{multifractionaltheory}
    S=\frac{1}{2\kappa^{2}}\int d^{D}x\sqrt{|g|}\left[R+\ell_{\star}^{2(\gamma-1)}R^{(\gamma)}-2\Lambda\right],
\end{equation}
with equations of motion
\begin{equation}
    \left[R_{\mu\nu}+R^{(\gamma)}_{\mu\nu}\right]-\frac{1}{2}g_{\mu\nu}\left[R+R^{(\gamma)}\right]+\Lambda g_{\mu\nu}+\mathcal{O}_{\mu\nu}=\kappa^{2}T_{\mu\nu}.
\end{equation}

The second realisation of \(T[\partial+\partial^{\gamma}]\) implements multiscaling at the level of each derivative, rather than only at the level of the action. One defines a multifractional Levi--Civita connection and the associated covariant derivatives. The corresponding generalisations of the connection and Ricci tensor are
\begin{equation}
    \bar{\Gamma}^{\rho}_{\mu\nu}:=\frac{1}{2}g^{\rho\sigma}\left(\mathcal{D}^{\gamma}_{\mu}g_{\nu\sigma}+\mathcal{D}^{\gamma}_{\nu}g_{\mu\sigma}-\mathcal{D}^{\gamma}_{\sigma}g_{\mu\nu} \right),
\end{equation}
\begin{equation}
    R_{\mu\nu}:=\mathcal{D}_{\sigma}\bar{\Gamma}^{\sigma}_{\mu\nu}-\mathcal{D}_{\nu}\bar{\Gamma}^{\sigma}_{\mu\sigma}+\bar{\Gamma}^{\tau}_{\mu\nu}\tilde{\Gamma}^{\sigma}_{\sigma\tau}-\tilde{\Gamma}^{\tau}_{\mu\sigma}\tilde{\Gamma}^{\sigma}_{\nu\tau}.
\end{equation}
Because of the multiscaling of the derivatives, there is no simple analogue of fractional Lorentz transformations even at the level of local inertial frames. In contrast with the first construction, this approach generates mixed terms of derivative order \(1+\gamma\). The action reads
\begin{equation}
    S=\frac{1}{2\kappa^{2}}\int d^{D}x\sqrt{|g|}\left(R-2\Lambda\right),
\end{equation}
and the corresponding equations of motion are
\begin{equation}
    R_{\mu\nu}-\frac{1}{2}g_{\mu\nu}R+\Lambda g_{\mu\nu}+\mathcal{O}_{\mu\nu}=\kappa^{2}T_{\mu\nu}.
\end{equation}

The d'Alembertian counterpart, \(T[\Box+\Box^{\gamma}]\), is constructed from a multiscaling graviton kinetic term involving both the ordinary and the fractional d'Alembertian. The form factor is chosen as
\begin{equation}
    \mathcal{F}_{2}(\Box)=\ell_{\star}^{2}\left(-\ell_{\star}^{2}\Box\right)^{\gamma-2}
    \qquad\Rightarrow\qquad
    \left[\Box+\ell_{\star}^{-2}\left(-\ell_{\star}^{2}\Box\right)^{\gamma}\right]h_{\mu\nu}=0.
\end{equation}
With this choice, the action (\ref{D'alemb}) reduces to
\begin{equation}
    S=\frac{1}{2\kappa^{2}}\int d^{D}x\sqrt{|g|}\left[R-2\Lambda+\ell_{\star}^{2}G_{\mu\nu}(-\ell_{\star}^{2}\Box)^{\gamma-2}R^{\mu\nu}\right],
\end{equation}
whose equations of motion are
\begin{equation}\label{multifractionalD'Alem}
     \begin{split}
        \kappa^{2}T_{\mu\nu}={}&\left[1-\left(-\ell_{\star}^{2}\Box\right)^{\gamma-1}\right]G_{\mu\nu}+\Lambda g_{\mu\nu}
        +\ell^{2}_{\star}g_{\mu\nu}\nabla^{\sigma}\nabla^{\tau}\left(-\ell_{\star}^{2}\Box\right)^{\gamma-2}G_{\sigma\tau}\\
        &-2\ell^{2}_{\star}\nabla^{\sigma}\nabla_{(\mu}\left(-\ell_{\star}^{2}\Box\right)^{\gamma-2}G_{\nu)\sigma}
        -\frac{1}{2}\ell^{2}_{\star}g_{\mu\nu}G_{\sigma\tau}\left(-\ell_{\star}^{2}\Box\right)^{\gamma-2}R^{\sigma\tau}\\
        &+2\ell^{2}_{\star}G^{\sigma}_{(\mu}\left(-\ell_{\star}^{2}\Box\right)^{\gamma-2}G_{\nu)\sigma}
        +\frac{1}{2}\ell_{\star}^{2}\left[G_{\mu\nu}(-\ell_{\star}^{2}\Box)^{\gamma-2}R+R(-\ell_{\star}^{2}\Box)^{\gamma-2}G_{\mu\nu}\right]\\
        &+\Theta_{\mu\nu}\left(R_{\alpha\beta},G^{\alpha\beta}\right).
    \end{split}
\end{equation}
According to \cite{Calcagni:2021aap,Calcagni:2021ljs}, this theory is likely non-renormalisable, much like Einstein gravity, because the ordinary \(\Box\) term dominates in the ultraviolet. Nevertheless, it remains potentially interesting as an effective framework generating classical and quantum modifications of gravity at large scales, with possible cosmological applications to dark matter and dark energy.

\section{$T[\partial^{\gamma(\ell)}]$ and $T[\Box^{\gamma(\ell)}]$ Theories}

The theory \(T[\partial^{\gamma(\ell)}]\) is obtained by promoting the fractional order to a scale-dependent quantity \(\gamma(\ell)\). Starting from the action (\ref{fractionalderivativestheory}), one introduces variable-order fractional derivatives \(\mathcal{D}_{\mu}^{\gamma(\ell)}\), leading to
\begin{equation}
    S=\frac{1}{2\kappa^{2}\ell_{\star}}\int_{0}^{\infty}d\ell\,\tau(\ell)\int d^{D}x\sqrt{|g|}\left[R^{(\gamma(\ell))}-2\Lambda\right].
\end{equation}
In compact form, the equations of motion become \cite{Calcagni:2021aap}
\begin{equation}
    R_{\mu\nu}^{(\gamma(\ell))}-\frac{1}{2}g_{\mu\nu}R^{(\gamma(\ell))}+\Lambda g_{\mu\nu}+\mathcal{O}^{(\ell)}_{\mu\nu}=\kappa^{2}T^{(\ell)}_{\mu\nu}.
\end{equation}

These equations are structurally identical to those of the fixed-order theory \(T[\partial^{\gamma}]\), with the replacement \(\gamma\to\gamma(\ell)\), but without integrating over \(\ell\) in the field equations themselves. The scale dependence is also inherited by the matter sector, whose action contains the same variable fractional exponents. Among the three derivative-based theories, \(T[\partial+\partial^{\gamma}]\) is arguably the simplest but also the least elegant, whereas \(T[\partial^{\gamma(\ell)}]\) preserves part of the simplicity of the fixed-order case while admitting a richer symmetry structure. All three theories, however, share the technical difficulty of evaluating composition rules and Leibniz rules for fractional derivatives, making the derivation of the field equations --- especially the tensor \(\mathcal{O}_{\mu\nu}\) --- rather involved.

The corresponding covariant theory with scale-dependent fractional powers of the d'Alembertian is \(T[\Box^{\gamma(\ell)}]\), whose graviton kinetic term is defined through the form factor
\begin{equation}\label{formfactorscale-dependent}
    \mathcal{F}_{2}(\Box)=\frac{(-\ell_{\star}^{2}\Box)^{\gamma(\ell)-1}}{\Box},
    \qquad
    (-\Box)^{\gamma(\ell)}h_{\mu\nu}=0.
\end{equation}
The action may then be written as
\begin{equation}
    S=\frac{1}{2\kappa^{2}\ell_{\star}}\int_{0}^{\infty}d\ell\,\tau(\ell)\int d^{D}x\sqrt{|g|}\left[R^{(\gamma(\ell))}-2\Lambda+G_{\mu\nu}\frac{(-\ell_{\star}^{2}\Box)^{\gamma(\ell)-1}}{\Box}R^{\mu\nu}\right],
\end{equation}
with equations of motion of the same form as Eq.~(\ref{multifractionalD'Alem}), after the replacement \(\gamma\to\gamma(\ell)\), and where the weight function \(\tau\) is normalised such that
\begin{equation}
    \int_{0}^{+\infty}d\ell\,\tau(\ell)=\ell_{\star}.
\end{equation}
These equations are to be understood as valid at each scale \(\ell\).

This scale-dependent fractional-d'Alembertian theory may be more promising as a candidate fundamental framework than either \(T[\Box^{\gamma}]\), which is unitary and one-loop finite but lacks multiscaling, or \(T[\Box+\Box^{\gamma}]\), where unitarity and renormalisability do not coexist. In particular, the conditions of unitarity and one-loop finiteness lead to
\begin{equation}
    \gamma(\ell)\in\left(0,\frac{1}{2}\right)\cup\left(\frac{1}{2},1\right),
\end{equation}
thereby selecting a class of theories that is well behaved in the ultraviolet while still allowing non-trivial cosmological signatures.

\bibliographystyle{unsrt}
\bibliography{main}

\end{document}